\documentclass[superscriptaddress,twocolumn]{revtex4-1}
\usepackage[T1]{fontenc}
\usepackage[pdftex]{graphicx}
\usepackage{graphicx}
\usepackage{subfig}
\usepackage[utf8]{inputenc}
\usepackage{amsmath}
\usepackage{braket}
\usepackage{color, colortbl}
\usepackage{xcolor}
\definecolor{Gray}{gray}{0.9}
\newcommand{\evo}{\text{evo}}

\begin{document}

\title{Many-body effects on the thermodynamics of closed quantum systems}
\author{A. H. Skelt}
\affiliation{Department of Physics, University of York, UK}
\author{K. Zawadzki}
\affiliation{Department of Physics, Northeastern University, Boston, Massachusetts 02115, USA}
\author{I. D'Amico}
\affiliation{Department of Physics, University of York, UK}
\affiliation{International Institute of Physics, Federal University of Rio Grande do Norte, Natal, Brazil}

\date{\today}

\begin{abstract}
Thermodynamics of quantum systems out-of-equilibrium is very important for the progress of quantum technologies, however, the effects of many body interactions and their interplay with temperature, different drives and dynamical regimes is still largely unknown. Here we present a systematic study of these interplays: we consider a variety of interaction (from non-interacting to strongly correlated) and dynamical (from sudden quench to quasi-adiabatic) regimes, and draw some general conclusions in relation to work extraction and entropy production. As treatment of many-body interacting systems is highly challenging, we introduce a simple approximation which includes, for the average quantum work, many-body interactions only via the initial state, while the dynamics is fully non-interacting. We demonstrate that this simple approximation is surprisingly good for estimating both the average quantum work and the related entropy variation, even when many-body correlations are significant.%
\end{abstract}


\maketitle

\section{Introduction}
Progress on applications of quantum technologies is linked to acquiring deeper understanding of the out-of-equilibrium thermodynamic properties of small quantum systems. Small quantum systems operating at finite temperature constitute the hardware for most of these technologies, so that out-of-equilibrium quantum thermodynamics guides, and ultimately may limit, these technologies \cite{Goold2016,Vinjanampathy2015,Millen2016,Parrondo2015,Liuzzo-Scorpo2016}.  Quantum thermodynamics has then become a rapidly expanding field, also supported by the advances in experimental techniques. Nonetheless, a remaining challenge is properly accounting for the effects of many-body interactions on quantum thermodynamics properties.
Many-body interactions in quantum systems  give rise to complex phenomena, such as collective behaviours and quantum phase transitions. From a practical perspective, these effects are challenging to calculate and often require approximations.  In this respect, there have been recent works studying out-of-equilibrium thermodynamics of many-body systems such as quantum harmonic oscillator chains and spin chains \cite{Silva2008,Dorner2012,Joshi2013,Mascarenhas2014,Sindona2014,Fusco2014,Zhong2015,Eisert2015,Bayat2016,Solano-Carrillo2016}, and a proposal for a density-functional-theory-based set of approximations which is in principle applicable to systems of high complexity \cite{Herrera2017,Herrera2018}.

Here we  present a systematic study of the out-of equilibrium thermodynamics of many-body quantum systems subject to a set of qualitatively different driving potentials. For each type of driving potential, we consider dynamical regimes from sudden quench, to finite times, to quasi-adiabatic; for each dynamical regime we consider different interaction strengths, from non-interacting to strongly correlated systems. For each driving potential, dynamical regime, and interaction strength, we consider  different temperatures: low, intermediate and high temperature. For all cases considered we calculate and discuss the average quantum work extracted and entropy produced in the dynamical trajectory. Our systematic study allows to uncover some important dependencies of work and entropy on the systems' correlation and dynamical regimes, which cut across the different applied drives.

Afterwards we consider two quite drastic approximations and compare their estimates with the exact results. The first is the completely non-interacting approximation, where many-body interactions are set to zero in all phases of the thermodynamic processes considered. The second approximation assumes knowledge of the initial interacting many-body state, but completely neglects interactions afterwards, during the driven dynamics. Our results show that including interactions just within the initial state provides surprisingly good accuracy. We provide an analytical analysis that explains this accuracy in the sudden quench and adiabatic regimes for a general Hamiltonian and driving potential.

\section{Theory}

\subsection{Hubbard model}
The one-dimensional Hubbard model can depict systems from weakly to strongly correlated and model numerous phases of matter and related phase transitions, such as metallic, antiferromagnetic, Mott-insulator, superconductivity, and FFLO transition \cite{Franca2006,Franca2012,dePicoli2018,Essler2005}.  It is being widely used to study many physical systems, from coupled quantum dots, to molecules, to chains of atoms \cite{Coe2010,Yang2011,Murmann2015,Coe2011,Brown2019,Nichols2019}. These are systems of importance, as hardware, to quantum technologies.  For small chains, the Hubbard model is numerically exactly solvable, yet still displays non-trivial behaviours, including, for repulsive interactions, the precursor to the metal-Mott insulator phase transition. Hence it is often the system of choice for exploring approximations to interacting quantum systems \cite{Herrera2018,Herrera2017,Carrascal2015,Fuks2014}.

For a fermionic system of $N$ sites, the Hamiltonian of the Hubbard model can be written as ($\hbar = 1$)
\begin{multline}
\label{eq:Hubbard_Hamiltonian}
{\hat{H}(t)} = -J \sum_{i,\sigma}^N \left( \hat{c}^{\dagger}_{i,\sigma} \hat{c}_{i+1,\sigma} + \hat{c}^{\dagger}_{i+1,\sigma} \hat{c}_{i,\sigma} \right) \\ + U \sum_i^N \hat{n}_{i,\uparrow} \hat{n}_{i,\downarrow} + \sum_i^N v_i(t) \hat{n}_i,
\end{multline}
where $J$ is the hopping parameter,  $\hat{c}^{\dagger}_{i,\sigma}$ ($\hat{c}_{i,\sigma}$) is the creation (annihilation) operator for a fermion with spin $\sigma$ ($\sigma= \uparrow$ or $\downarrow$) in site $i$,  $U$ is the strength of the on-site Coulomb interaction,  $\hat{n}_{i,\sigma} = \hat{c}^{\dagger}_{i,\sigma} \hat{c}_{i,\sigma}$ is the spin $\sigma$, $i$-site number operator, $\hat{n}_i = \hat{n}_{i,\uparrow} + \hat{n}_{i,\downarrow}$, and $v_i$ is the on-site potential.

We shall use the time-dependent, inhomogeneous, one-dimensional Hubbard model (i) to calculate the exact average quantum work and the entropy production for the set of system sizes and regimes described in section \ref{sec:systems} and (ii) within the approximations for these quantities described in section \ref{sec:approx}.

\subsection{Average quantum work and entropy variation}
\label{sec:q_work}
The average quantum work, much like its classical counter part, is described as the usable energy in a quantum system \cite{Vinjanampathy2015}.  In a closed system at temperature $T$, it can be calculated as \cite{Vinjanampathy2015}

\begin{equation}
\label{eq:q_work_trace}
\langle W \rangle = \mathrm{Tr}\left[ \rho_{f} \hat{H}_{f} \right] - \mathrm{Tr}\left[ \rho_0 \hat{H}_0 \right],
\end{equation}
with $\rho_{0(f)}$ and $\hat{H}_{0(f)}$ the initial (final) system state and Hamiltonian, respectively.

For a given dynamic process, the variation in thermodynamic entropy is defined using the average work and the change in the free energy of the system \cite{Batalhao2015,Goold2016},
\begin{equation}
\label{eq:entropy}
\Delta S =  \beta \left( \langle W \rangle - \Delta F \right)
\end{equation}
where $\beta = 1/k_B T$, and the free energy variation is
\begin{equation}
\label{eq:free_energy}
\Delta F = - \frac{1}{\beta} \ln \left( \frac{Z_f}{Z_0} \right),
\end{equation}
with $Z_{0(f)}$ the partition function at the beginning (end) of the dynamics, $Z_{0(f)} = \mathrm{Tr}\left[ \exp\left( -\beta \hat{H}_{0(f)} \right) \right]$.
This thermodynamic entropy can be considered a measure of the degree of irreversibility of the system dynamics: in fact it captures an uncompensated heat which would need to be dispersed to the environment for the system to return to thermodynamic equilibrium at the end of the driven process \cite{Herrera2017,Batalhao2015}.

\section{Systems, driving potentials, temperature range and dynamical regimes}
\label{sec:systems}

We shall calculate the work extracted and entropy produced for different systems, temperatures, driving potentials, and dynamical regimes.

We will consider Hubbard chains of 2, 4, and 6 sites, at half-filling and under open boundary conditions, and explore low ($T = 0.2J/k_B$), medium ($T = 2.5J/k_B$), and high ($T = 20J/k_B$) temperatures.

For each system size and temperature, we will explore regimes from non-interacting ($U=0J$) all the way to strongly correlated ($U=10$), and dynamics from sudden quench ($\tau = 0.5/J$, $\tau$ the overall driving time) all the way to quasi-adiabatic ($\tau=10/J$).

\begin{figure*}
\centering
\subfloat[Comb dynamics.]{\includegraphics[width=0.3\textwidth]{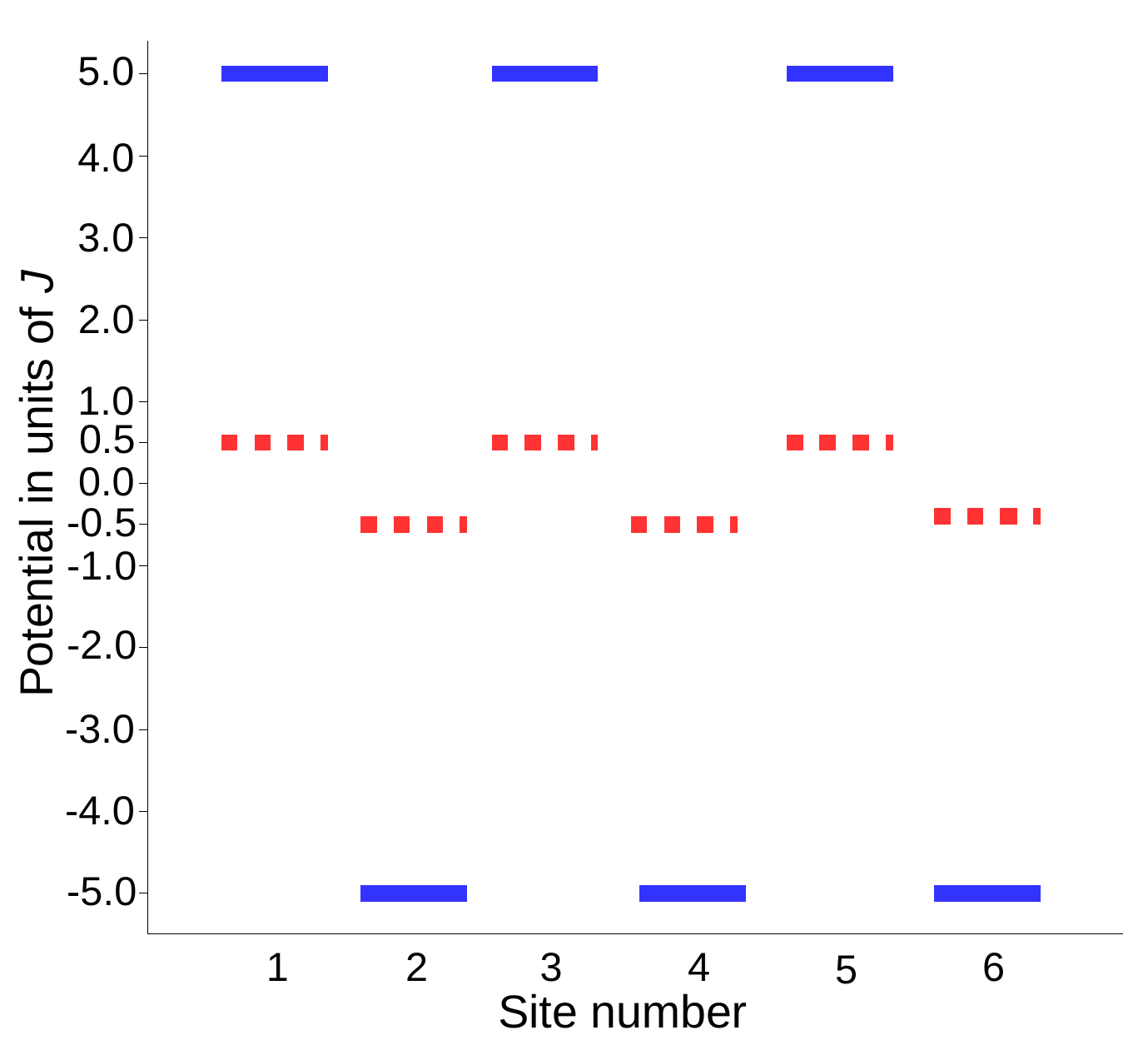}}
\subfloat[MI dynamics.]{\includegraphics[width=0.3\textwidth]{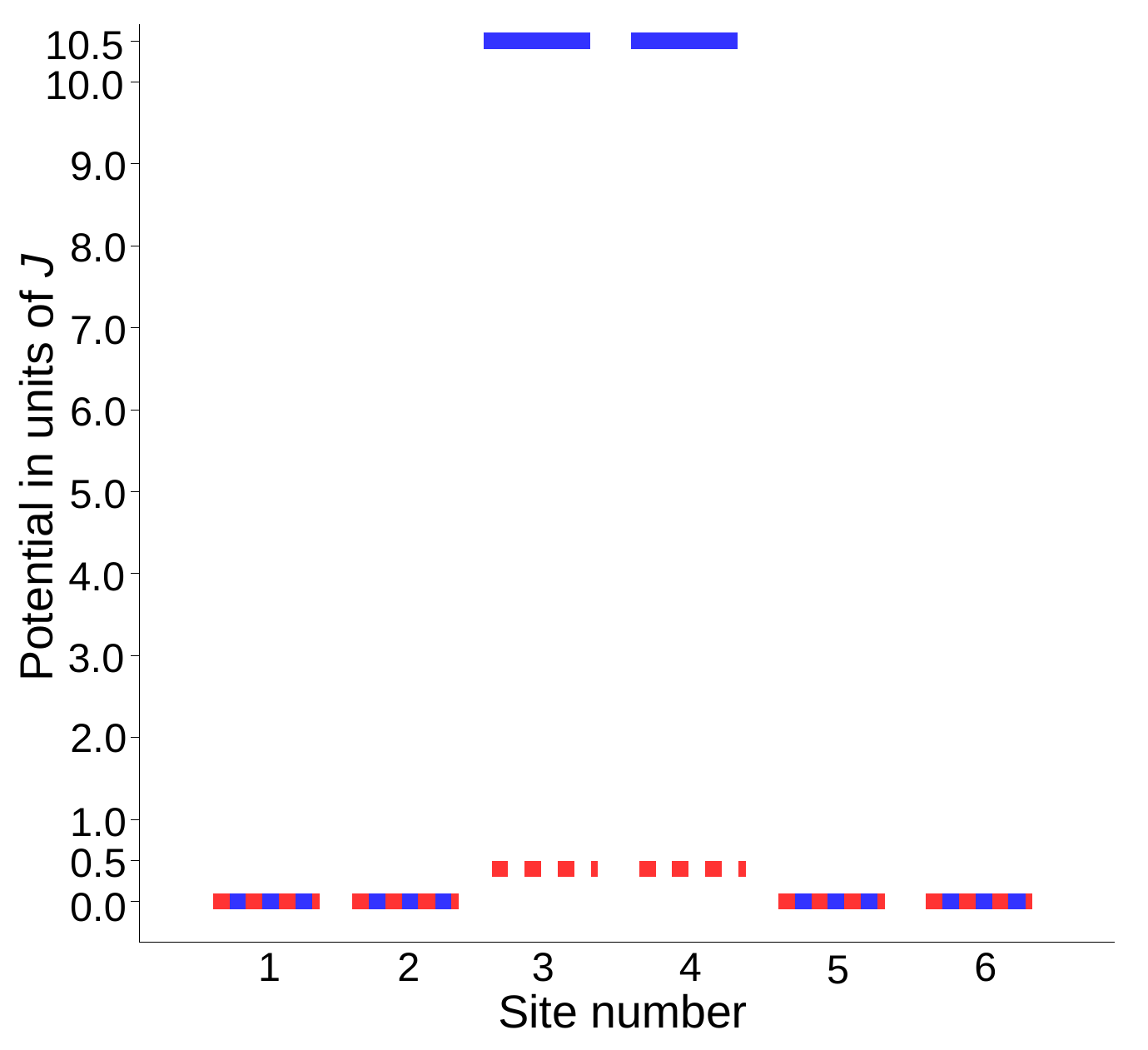}}
\subfloat[AEF dynamics.]{\includegraphics[width=0.3\textwidth]{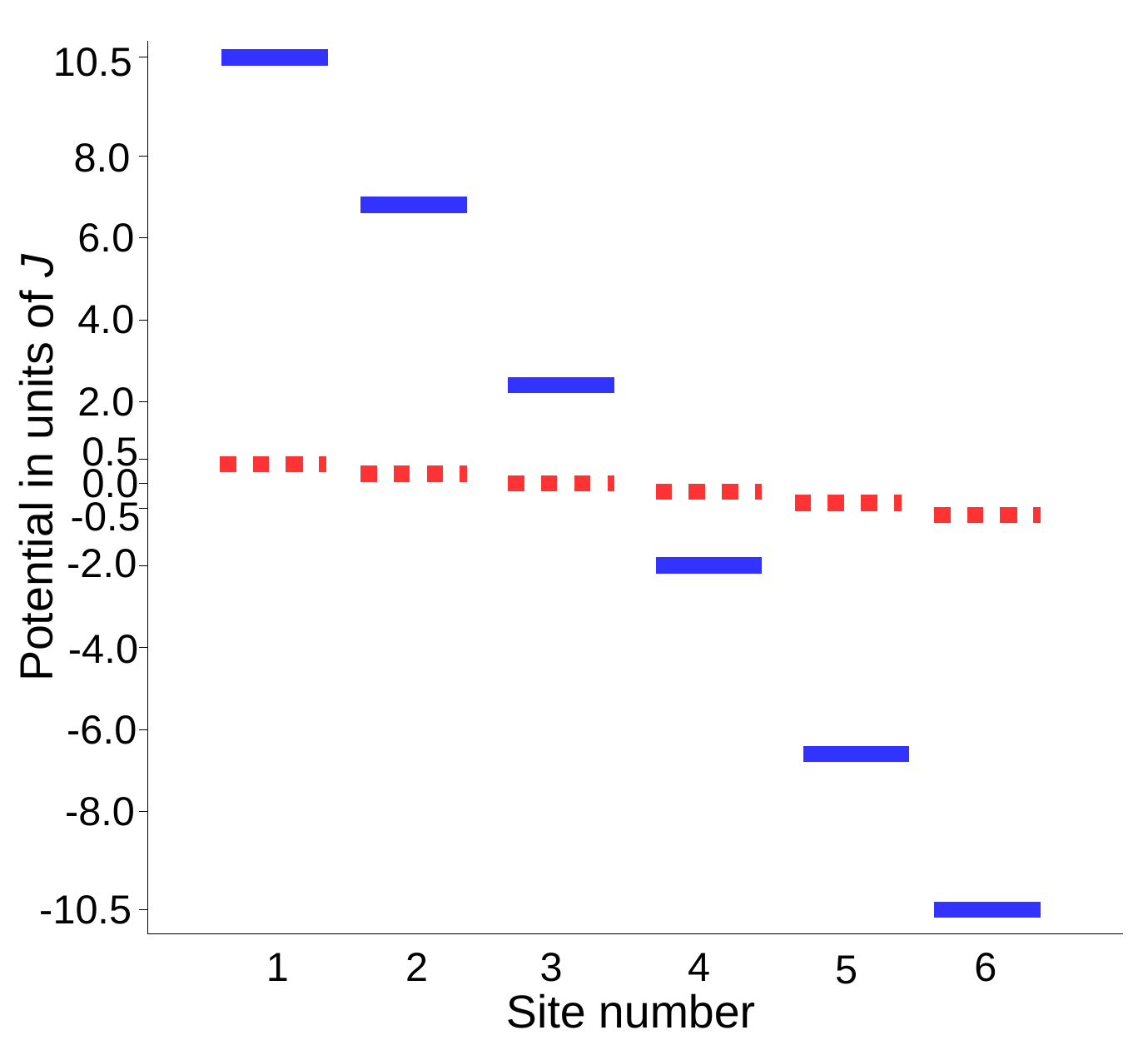}}
\caption{On-site driving potentials versus site number for a 6 site chain; red dashed lines show the potentials at $t=0$, and blue solid lines show the potentials at $t=\tau$.}
\label{fig:potentials}
\end{figure*}

For each parameter combination, we will consider three types of driving potentials \footnote{We note that for the 2 site chain the only relevant dynamics becomes the `Applied Electric Field' one.}, where each potential has a linear time dependency via $v_i (t) = \mu_i^0 + \mu_i^\tau t / \tau$, with $\mu_i^0$ and $\mu_i^{\tau}$ the time-independent coefficients for site $i$ at time $0$ and $\tau$ respectively.  With this choice, the {\it character} of the dynamics will depend on $\tau$, while the final Hamiltonian $\hat{H}_f$ will be independent of it. These driving potentials are:
\begin{itemize}
\item ``Comb'': for each site $i$, $\mu_i^0 = \mu_0 (-1)^i$ at $t=0$ and $\mu_i^{\tau} = \mu_{\tau} (-1)^i$ at $t=\tau$, where $\mu_0 = 0.5J$ and $\mu_{\tau} = 4.5J$.
\item ``Middle Island (MI)'': the inhomogeneity $\mu_i$ is driven only for the middle two sites of the chain;  $\mu_i^0 = 0$ for $i \neq L/2, ~(L/2)+1$ where $i$ goes from 1 to $L$, and $L$ is the chain length. For the middle two sites, $i = L/2, ~(L/2)+1$, $\mu_i^0 = 0.5J$ and $\mu_i^{\tau} = 10J$.
\item ``Applied Electric Field (AEF)'': this potential mimics the application of a potential difference between the extremes of the chain. The sites form a linear slope from $i=1$ to $i=L$ and are described using $\mu_i^0 = 2 \mu^0 / L \times i - \mu^0$ where $\mu^0 = 0.5J$, and $\mu_i^\tau = 2 \mu^\tau / L \times i - \mu^\tau$ with $\mu^{\tau} = 10J$.
\end{itemize}
The $t=0$ and $t=\tau$ form of the driving potentials for a six-site Hubbard chain are illustrated in figure \ref{fig:potentials}.

\section{Approximations}
\label{sec:approx}
\subsection{Average quantum work}
The type of approximations we consider for the average quantum work are of the form:
\begin{equation}
\label{eq:q_work_trace_approx}
\langle W^{is+\evo} \rangle = \mathrm{Tr}\left[ \rho_{f}^{is+\evo} \hat{H}_{f}^{\evo} \right] - \mathrm{Tr}\left[ \rho_0^{is} \hat{H}_0^{\evo} \right].
\end{equation}
Here $is$ (initial system) refers to the approximation used to derive the system state at $t=0$,  $\rho_0^{is} = \exp \left( -\beta \hat{H}_0^{is} \right)/ \mathrm{Tr}\left[ \exp \left( -\beta \hat{H}_0^{is} \right) \right]$, and $\evo$ is the approximation used for the evolution operator $\mathcal{U}_{\evo}= \mathcal{T} e^{-i \int_0^{\tau}\hat{H}_{t}^{\evo}(t) dt}$ where $\mathcal{T}$ is the time-ordered operator.   The final state is then  $\rho_{f}^{is+\evo} = \mathcal{U}_{\evo} \rho_0^{is} \mathcal{U}_{\evo}^{\dagger}$. We note that  $\hat{H}_{0}^{\evo}=\hat{H}_{t}^{\evo}(t=0)$.  In the approximation where $is$ and $\evo$ are the same, only one acronym shall be written. As (\ref{eq:q_work_trace_approx}) indicates, the Hamiltonians $\hat{H}_{f}$ and $\hat{H}_{0}$ explicitly entering (\ref{eq:q_work_trace}) and the evolution Hamiltonian $\hat{H}_{t}^{\evo}$ are to be taken in the same approximation: if this is not the case, we found that the mismatch in eigenstates leads to spurious oscillations in the work production (not shown).

We will consider two approximations, as described in Table~\ref{tab:approxes}.  The first, $\langle W ^ {NI} \rangle$, corresponds to a completely non-interacting system, the one obtained by setting $U=0$ in the Hubbard Hamiltonian.
The second approximation, $\langle W ^{exact + NI} \rangle$, uses the exact many-body initial state, but the approximated (non-interacting) Hamiltonian for the evolution of the system, according to the notation previously introduced.

\begin{table*}[ht!]
    \centering
    \begin{tabular}{|c|c|c|c|c|}
         \hline
         Acronym & Approximation & Hamiltonian & Initial State \\
         \hline
         NI & $\langle W ^ {NI} \rangle $ & $\displaystyle \hat{H}^{NI} = -J \sum_{i\sigma}^N \left( \hat{c}^{\dagger}_{i,\sigma} \hat{c}_{i+1,\sigma} + \hat{c}^{\dagger}_{i+1,\sigma} \hat{c}_{i,\sigma} \right) + \sum_i^N v_i {n}_i$ & $\rho^{NI}_0 = \exp \left( -\beta \hat{H}^{NI}_0 \right)/ Z^{NI}$ \\
         \hline
         exact + NI & $\langle W ^ {exact+NI} \rangle $ & $\displaystyle \hat{H}^{NI} = -J \sum_{i\sigma}^N \left( \hat{c}^{\dagger}_{i,\sigma} \hat{c}_{i+1,\sigma} + \hat{c}^{\dagger}_{i+1,\sigma} \hat{c}_{i,\sigma} \right) + \sum_i^N v_i {n}_i$ & $\rho^{exact}_0 = \exp \left( -\beta \hat{H}^{exact}_0 \right)/ Z^{exact}$ \\
         \hline
    \end{tabular}
    \caption{Types of approximations with their Hamiltonians and initial states.}
    \label{tab:approxes}
\end{table*}

\subsection{Entropy variation}
When approximating the entropy, we use $\langle W ^ {is+\evo} \rangle $, while the free energy is estimated in the same way as $\rho^{is}$. In the NI approximation, this implies that the free energy is constant.
For $\Delta S ^ {exact+NI}$, the exact free energy is then used  as this uses the same assumption made for calculating the initial thermal state, that the system Hamiltonian can be exactly -- or very accurately -- diagonalised.

\section{Work extraction}
The system is considered at equilibrium at time $t=0^-$, when the coupling with the thermal bath is switched off. Then the closed system is driven by a time-dependent external potential from the initial Hamiltonian $\hat{H}_{0}$ to the final Hamiltonian $\hat{H}_{f}$ in a time $\tau$, and the extracted work $\langle W_{ext}\rangle$ from this dynamics is calculated according to (\ref{eq:q_work_trace}), with $\langle W_{ext}\rangle=-\langle W\rangle$.

We stress that with each of the driven dynamics described in section~\ref{sec:systems}, the final Hamiltonian $\hat{H}_{f}(U)$ is the same for all $\tau$'s, so that the latter controls the rate of driving. Therefore, the larger $\tau$ is, the slower the system has evolved and hence more adiabatic the evolution.

For each of the many-body systems, their approximations, temperatures, and driving potentials described in section \ref{sec:approx}, we will consider the parameter space $0.5\le \tau \times J \le 10$ and $0\le U/J \le 10$. Due to the sheer number of results from all the combination of parameters, we will only explicitly show results for 6 site chains (all 6 site chains' exact results, and part of them for the approximations), and comment on the rest.

\subsection{Exact results}\label{exact:res}

\begin{figure*}
\centering
\subfloat[$T=0.2J/k_B$ with MI dynamics.]{\includegraphics[width=0.3\textwidth]{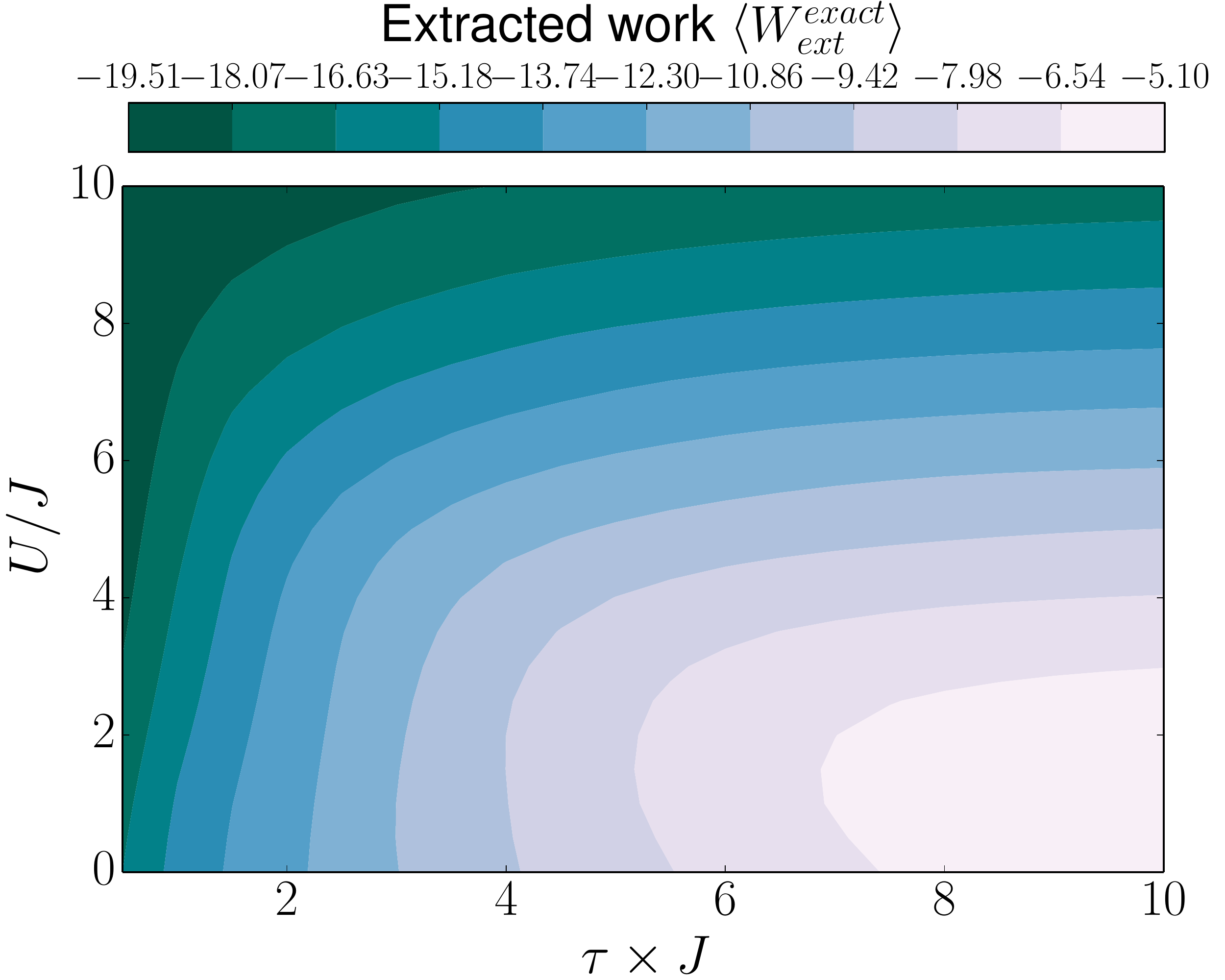}}
\subfloat[$T=0.2J/k_B$ with comb dynamics.]{\includegraphics[width=0.3\textwidth]{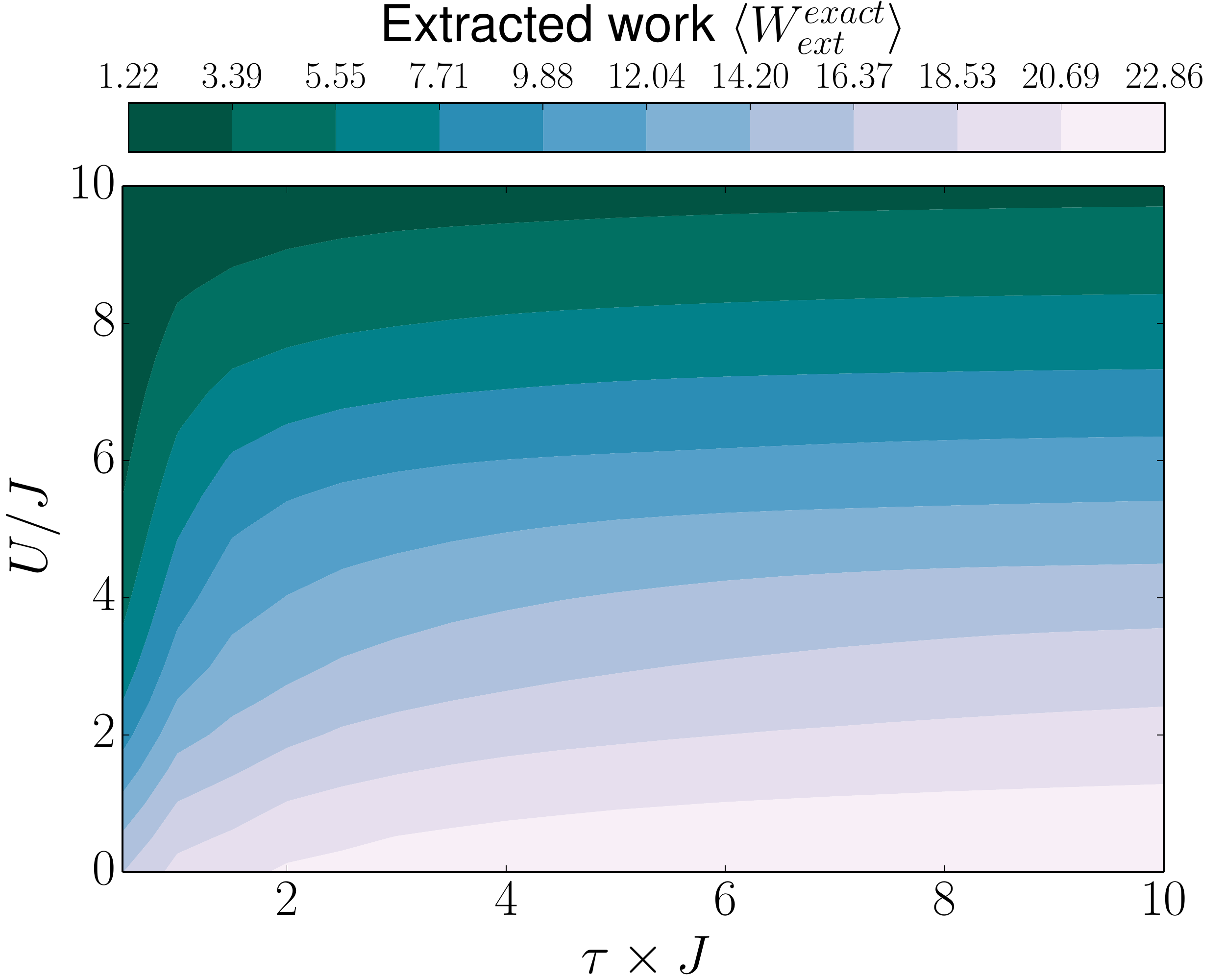}}
\subfloat[$T=0.2J/k_B$ with AEF dynamics.]{\includegraphics[width=0.3\textwidth]{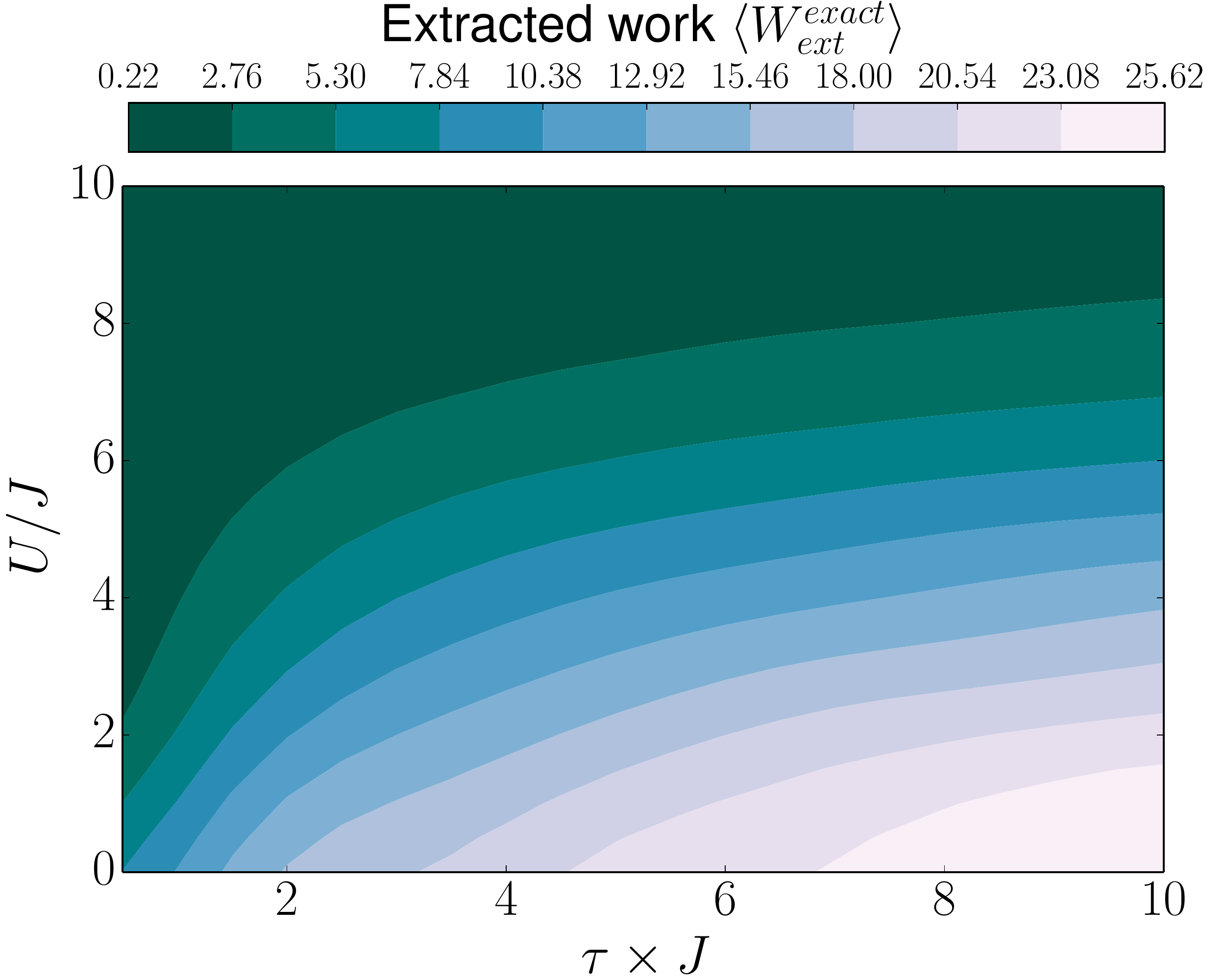}}
\par
\subfloat[$T=2.5J/k_B$ with MI dynamics.]{\includegraphics[width=0.3\textwidth]{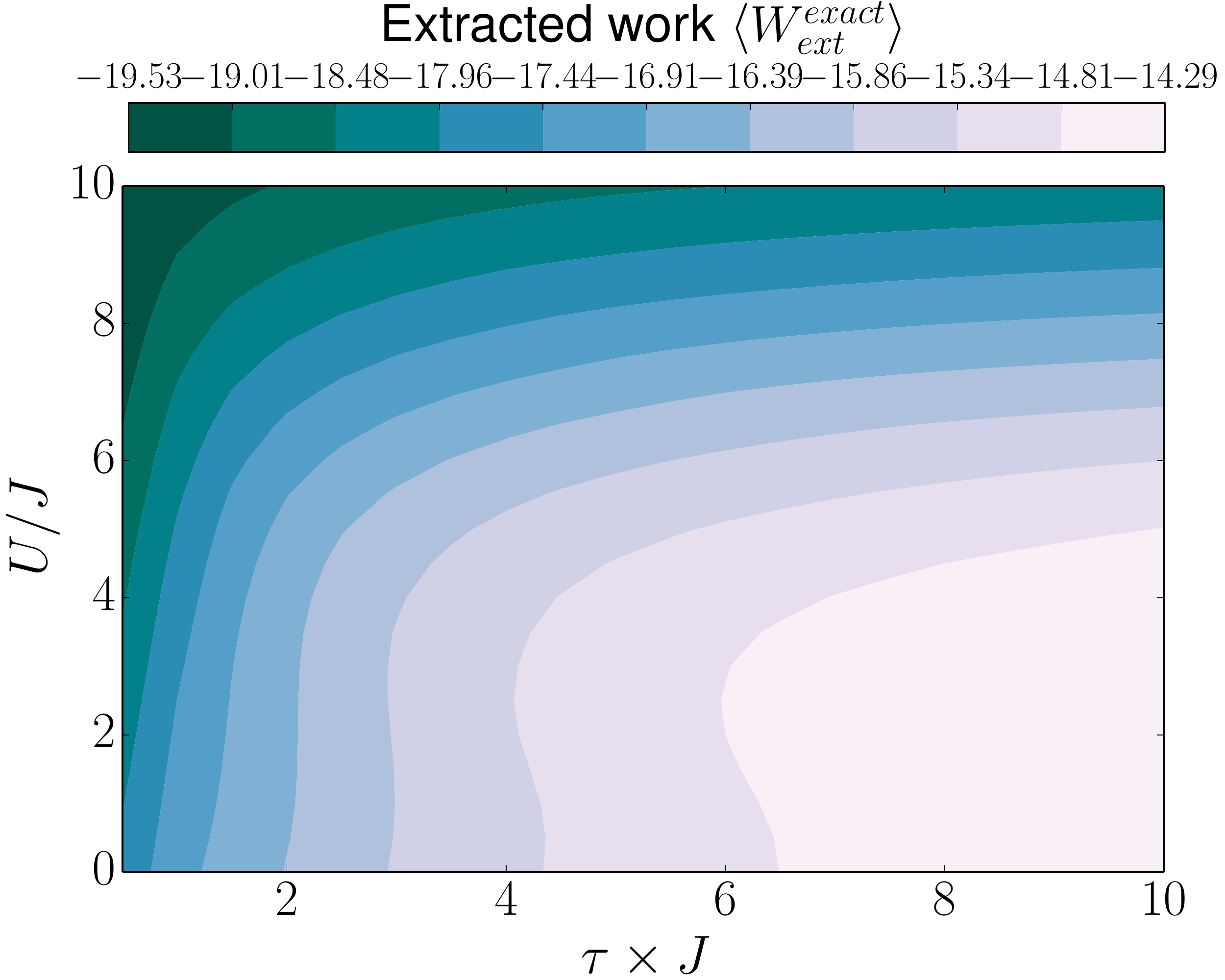}}
\subfloat[$T=2.5J/k_B$ with comb dynamics.]{\includegraphics[width=0.3\textwidth]{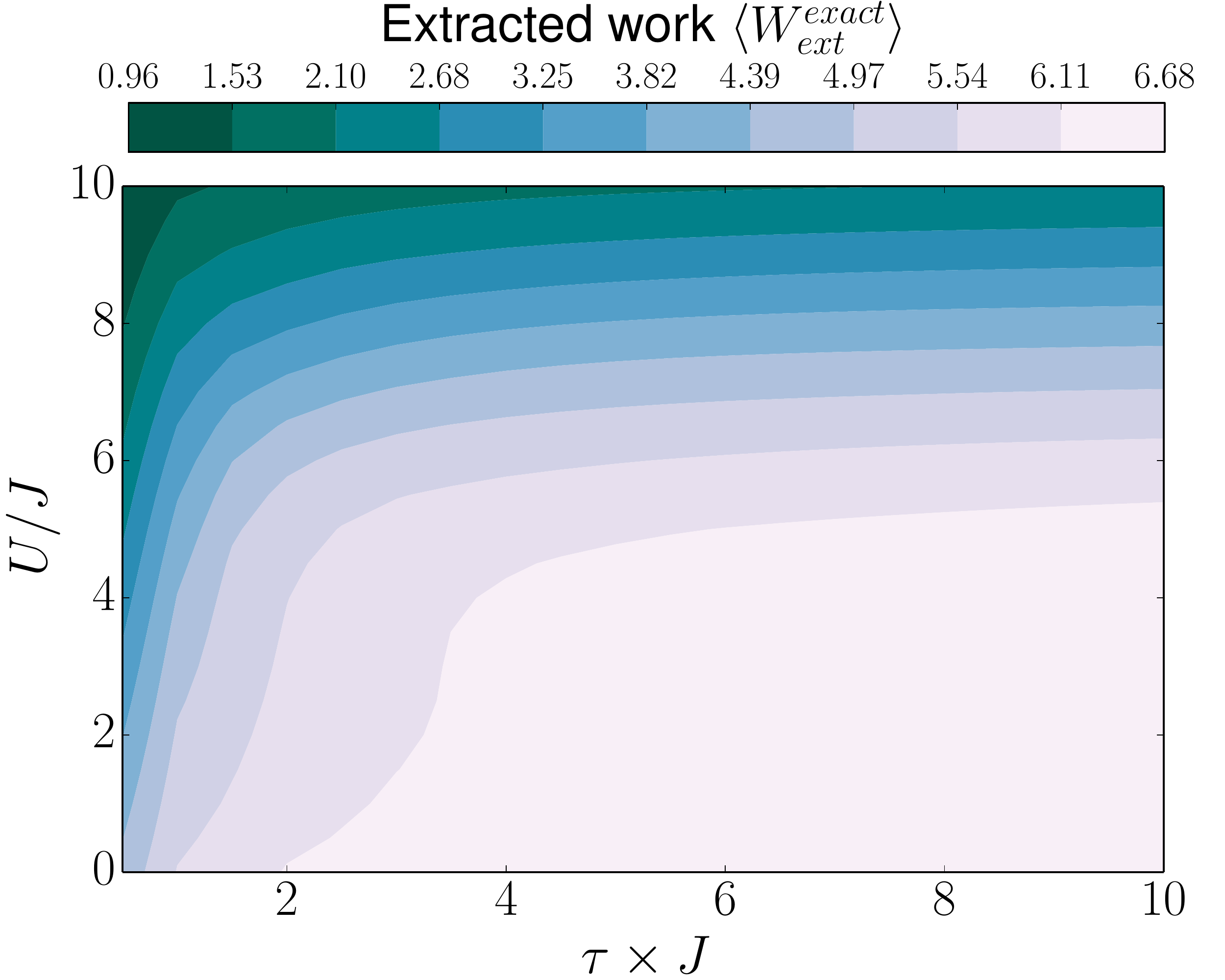}}
\subfloat[$T=2.5J/k_B$ with AEF dynamics.]{\includegraphics[width=0.3\textwidth]{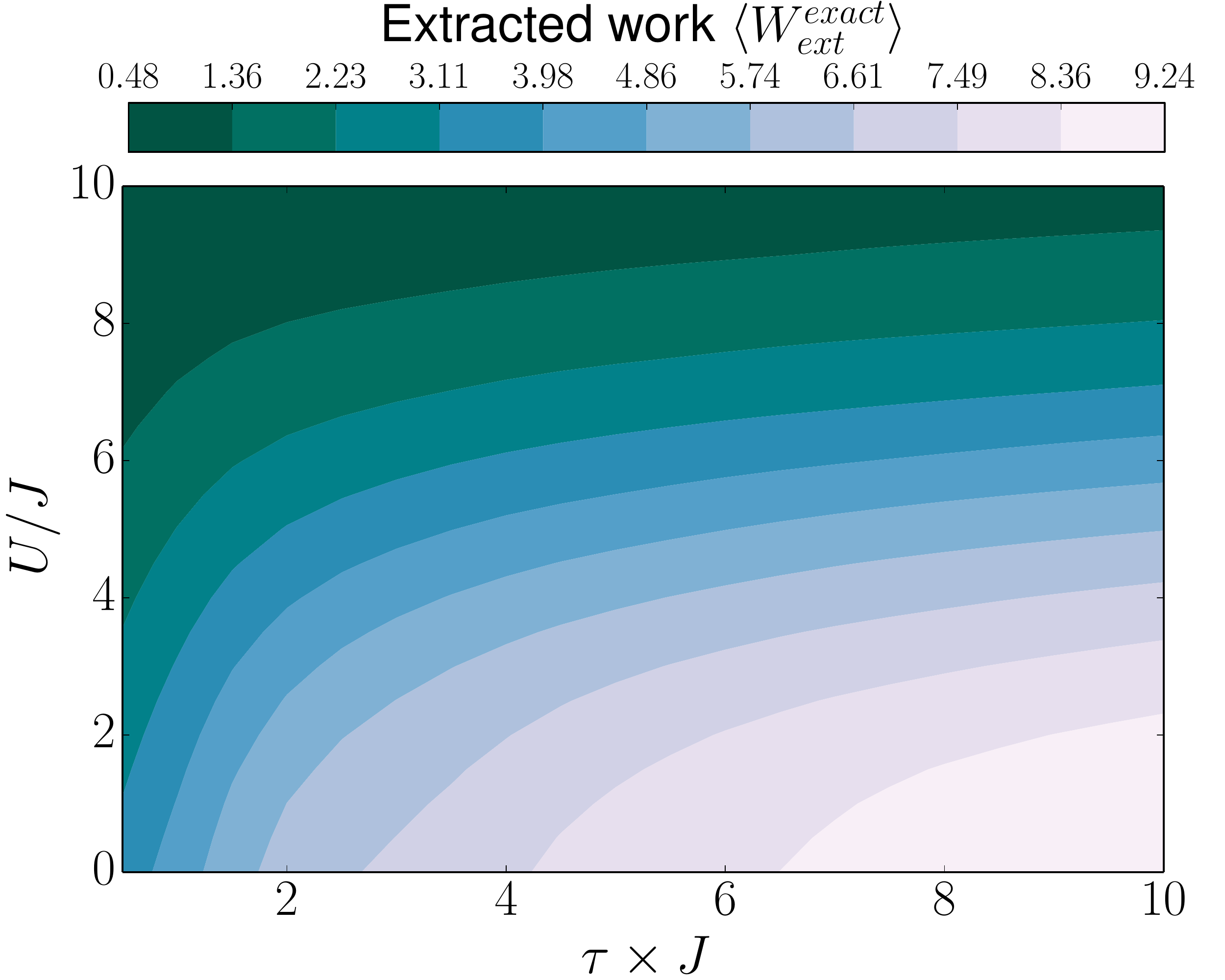}}
\par
\subfloat[$T=20J/k_B$ with MI dynamics.]{\includegraphics[width=0.3\textwidth]{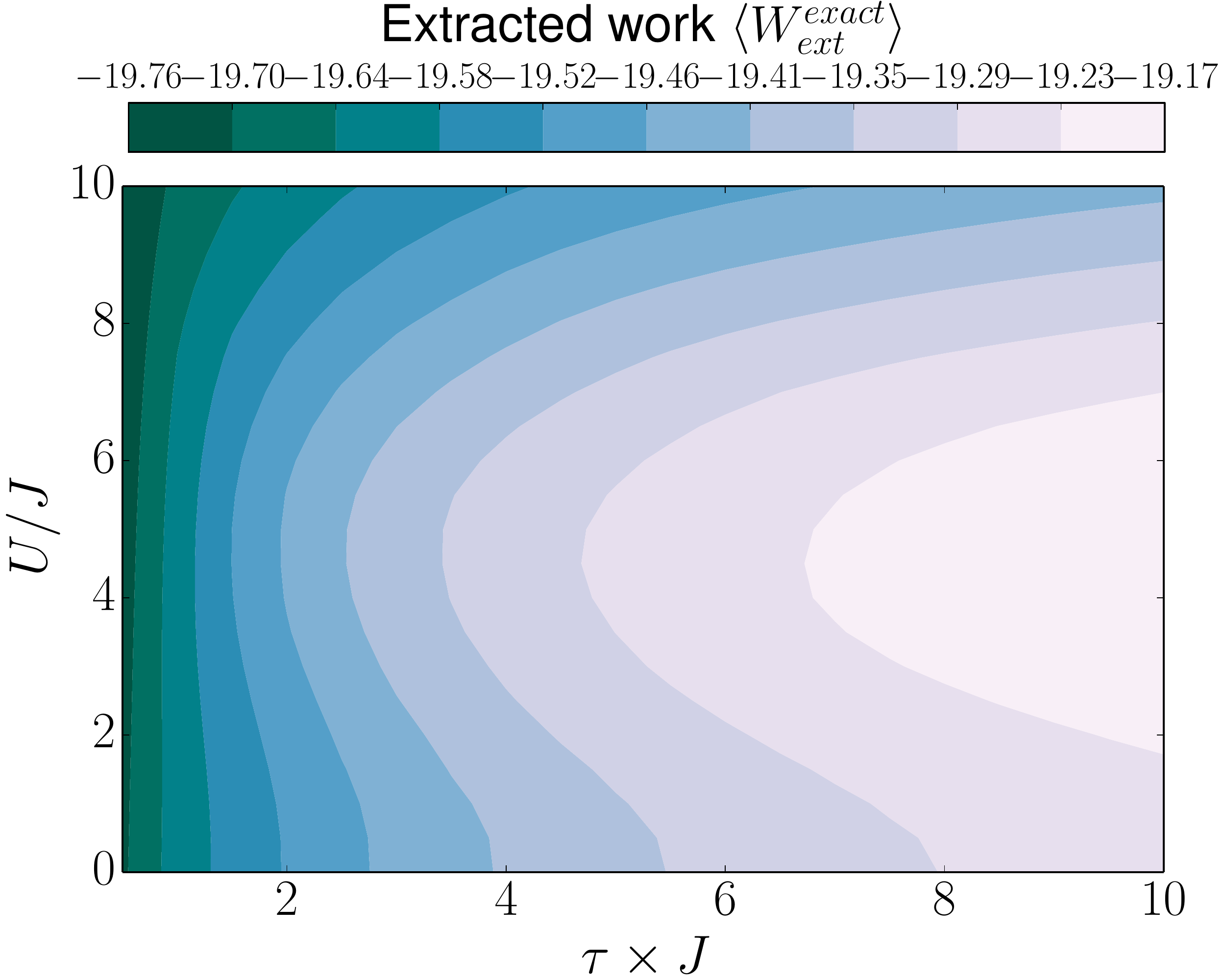}}
\subfloat[$T=20J/k_B$ with comb dynamics.]{\includegraphics[width=0.3\textwidth]{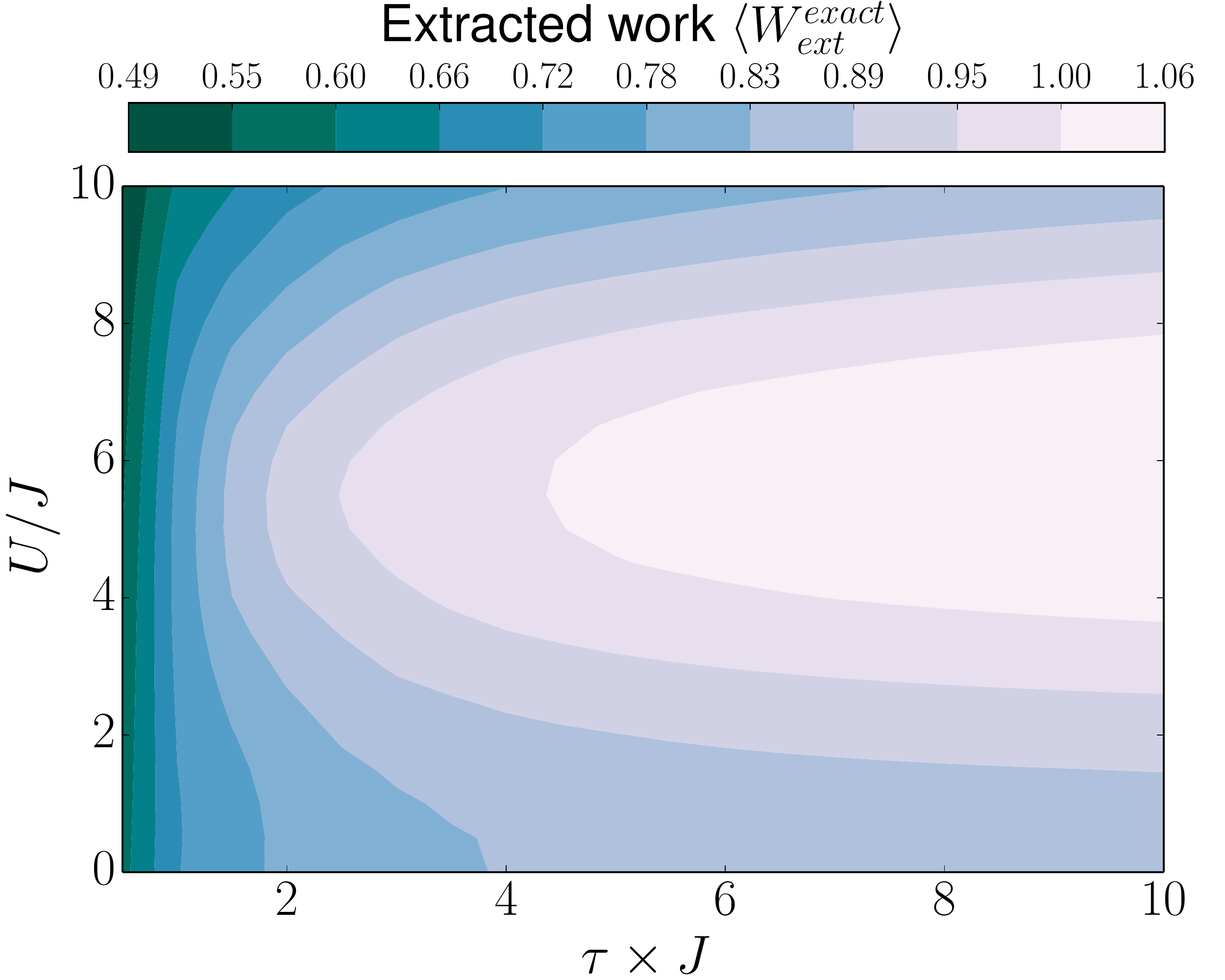}}
\subfloat[$T=20J/k_B$ with AEF dynamics.]{\includegraphics[width=0.3\textwidth]{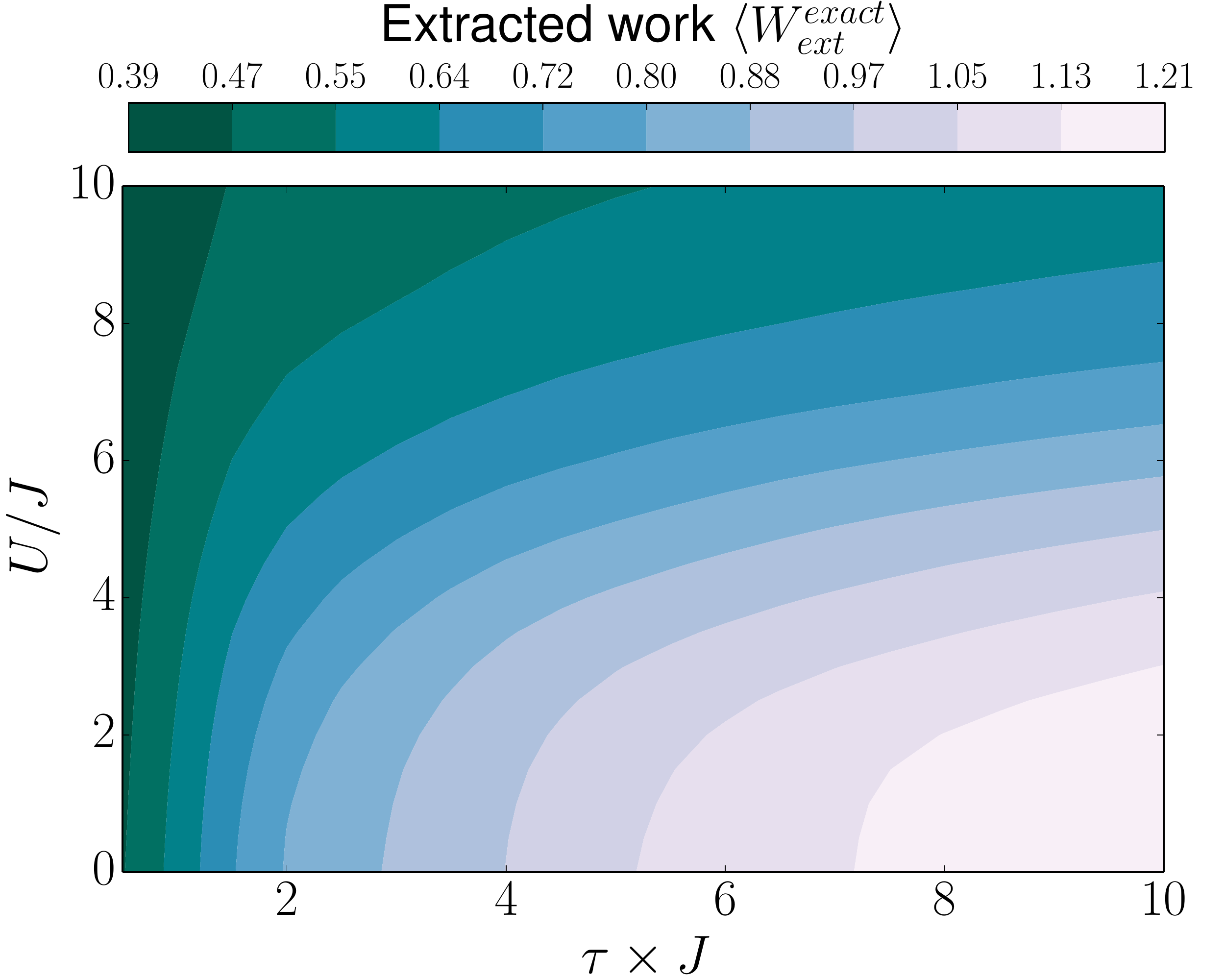}}
\caption{(a)-(i) Exact extracted average quantum work versus total dynamics time $\tau$ ($x$-axis)and interaction strength $U$ ($y$-axis). Data are presented for 6 site Hubbard chains driven by MI, comb, and AEF potentials, and at low, medium, and high temperatures, as indicated.  The lighter the colour shade, the more work is extracted, compatible with the respective work range indicated over each panel.}
\label{fig:ex_work}
\end{figure*}

Figure~\ref{fig:ex_work} shows the exact average quantum work extracted from a 6 site chain driven via ``applied electric field (AEF)'' (right column), ``comb'' (middle column), and ``middle island (MI)'' (left column) potentials at temperatures of $T=0.2J/k_B$ (first row), $T=2.5J/k_B$ (second row), and $T=20J/k_B$ (third row).
Each panel shows a wide range of regimes:  from non-interacting to strongly correlated systems as $U$ increases along the $y$-axis; and from sudden quench towards adiabaticity as $\tau$ increases along the $x$-axis. A lighter shade of colour corresponds to higher extracted work. 

Figure~\ref{fig:ex_work} presents a  variety of behaviours, with the extracted work varying over a wide range of values, and even from positive to negative. This confirms that the chosen dynamics are a good test-bed for understanding work extraction in systems representable via Hubbard chains, and hence a good test bed for related approximations.

At all temperatures, the largest work can be extracted via the AEF dynamics, while work needs to be done on the system in order to perform the MI dynamics. For all driving potentials, increasing temperatures {\it decreases} both the range of extracted/provided work and the maximum extractable work (the minimum work to be performed on the system in the case of MI). The maximum applied potential difference at $t=\tau$ is comparable to the highest temperature, so, as the temperature rises, the systems become less sensitive to the applied field. For all temperatures and dynamics the highest work that can be extracted (the lowest work performed on the system in the case of MI) is reached for large $\tau$'s, as the system gets closer to adiabaticity: here the dynamical state better adjusts to the driving force and, compatible with temperature and many-body interactions, the energetics favour low-potential chain sites.

In the case of AEF, at all temperatures, increasing $U$ hampers the transient current dynamics, so the highest work is achieved for zero to weak correlations. The maximum potential step between nearby sites is about $4J$ so, as $U$ increases, the AEF dynamics tends to freeze and lesser and lesser work can be extracted from the system.  At the highest temperature, the thermal energy is almost equal to the potential difference between the chain extremes at the end of the dynamics.

At low temperatures, `comb' dynamics  presents a work extraction pattern similar to AEF; however, as the temperature increases, maximum work extraction can be achieved for higher many-body interactions, and at high $T$ it is achieved only for relatively strong many-body interactions ($4 \lesssim U/J\lesssim 8$). This can be understood by realising that in this case the thermal energy $k_BT=20J$ is twice the potential barrier between even and odd chain sites, and hence a certain degree of repulsion is necessary to depopulate the high-energy sites completely and maximise work extraction.

Extracted work under MI dynamics is negative, meaning that work must be performed on the systems to achieve the final states. Indeed in this case the drive raises the potential of the central sites and, under the dynamics, the (closed) system cannot decrease its overall energy.
As even for $\tau=10$ the system is not fully adiabatic, a finite amount of repulsion helps to completely deplete the central island, even at low temperature, and more so as temperature increases. The $10.5J$ maximum barrier between high and low potential sites is about half of the maximum thermal energy considered.

\subsubsection{Adiabatic regime}
\label{adiab_ex}
From figure~\ref{fig:ex_work} we see that, as $\tau$ increases and we approach the adiabatic regime, the average work becomes strongly dependent on $U$ and weakly dependent on $\tau$. This behaviour can be easily explained.
For a closed system, $\langle W_{ext} \rangle $ can be also calculated from the work distribution function, and hence written as \cite{Vinjanampathy2015}
\begin{equation}\label{W_PW}
 \langle W_{ext} \rangle =\sum_{n,m} p_{n,m}^{\tau}E_m^f - \sum_n p_n^0 E_n^0,
\end{equation}
where $E_m^f$ ($E_n^0$) is the $m$-th ($n$-th) eigenvalue of the final (initial) Hamiltonian, and  $p_{n,m}^{\tau}=p_n^0p_{m|n}^\tau$ is the joint probability distribution  of arriving to the eigenstate $|\Psi_m^f\rangle$ of $H_f$, given the probability $p_n^0$ of initially being in the eigenstate $|\Psi_n^0\rangle$ of $H_0$.

In this formalism, the effect of the dynamics driven by the external potential (and hence the dependence on $\tau$) is contained in the conditional probabilities $p_{m|n}^\tau=|\langle\Psi_m^f| \mathcal{U}_{evo}(\tau)|\Psi_n^0\rangle|^2$.
In the adiabatic regime $\mathcal{U}_{evo}(\tau)|\Psi_n^0\rangle= |\Psi_n^{f}\rangle$, so that we obtain
\begin{eqnarray}
  \langle W_{ext} \rangle^{adiabatic}&=&
    \sum_{n,m}E_m^f p_n^0 |\langle\Psi_m^{f}|\Psi_n^{f,NI}\rangle |^2 - \sum_n p_n^0 E_n^0\nonumber \\
 & = & \sum_{n}p_n^0(E_n^f - E_n^0),
 \label{p_nm_ad_ex}
\end{eqnarray}
which indeed depends on the interaction coupling $U$ but not on $\tau$.
By looking at  figure~\ref{fig:ex_work} we can then note that the value of $\tau$ at which the different systems enter a nearly adiabatic behaviour depends both on the driving potential and on the temperature, with the `comb' driving potential getting closer to adiabaticity for smaller values of $\tau$, and increasing temperature seemingly decreasing adiabaticity.  The latter can be understood by realising that when temperature increases to the point of dominating by far inter-particle interactions, the system behaves as non-interacting, and hence the dependence of any quantity over $U$ will be lost (see also discussion in section \ref{NI_appr}).

\subsection{Approximated results}

We will firstly approximate the many-body system with a non-interacting system and use this to estimate quantum thermodynamic quantities. Afterwards we will extend this approach to include some memory of the electron-electron interaction through the system initial state.

\subsubsection{Non-interacting approximation}
\label{NI_appr}

\begin{figure*}
\centering
\subfloat[$T=0.2J/k_B$.]{\includegraphics[width=0.3\textwidth]{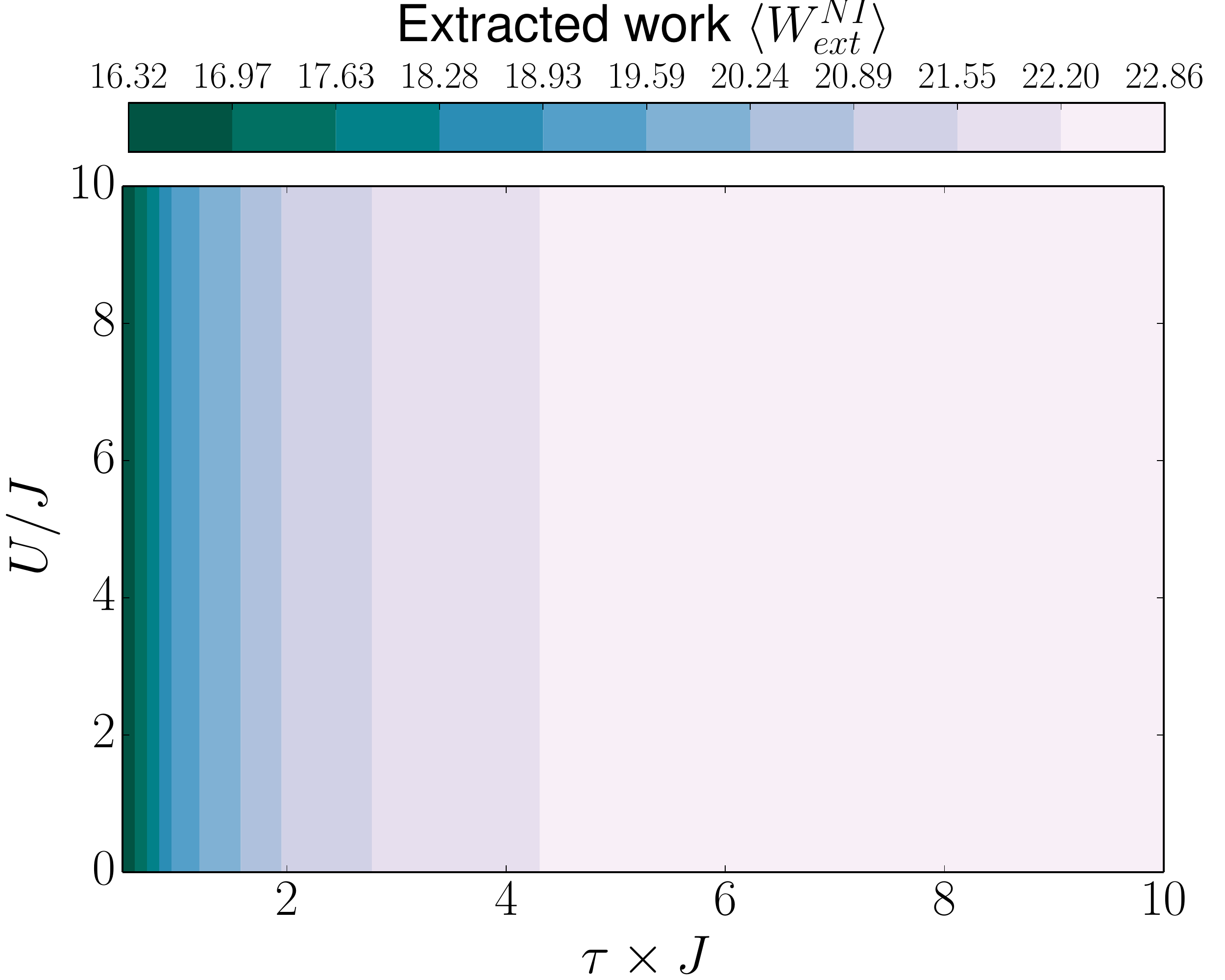}}
\subfloat[$T=2.5J/k_B$.]{\includegraphics[width=0.3\textwidth]{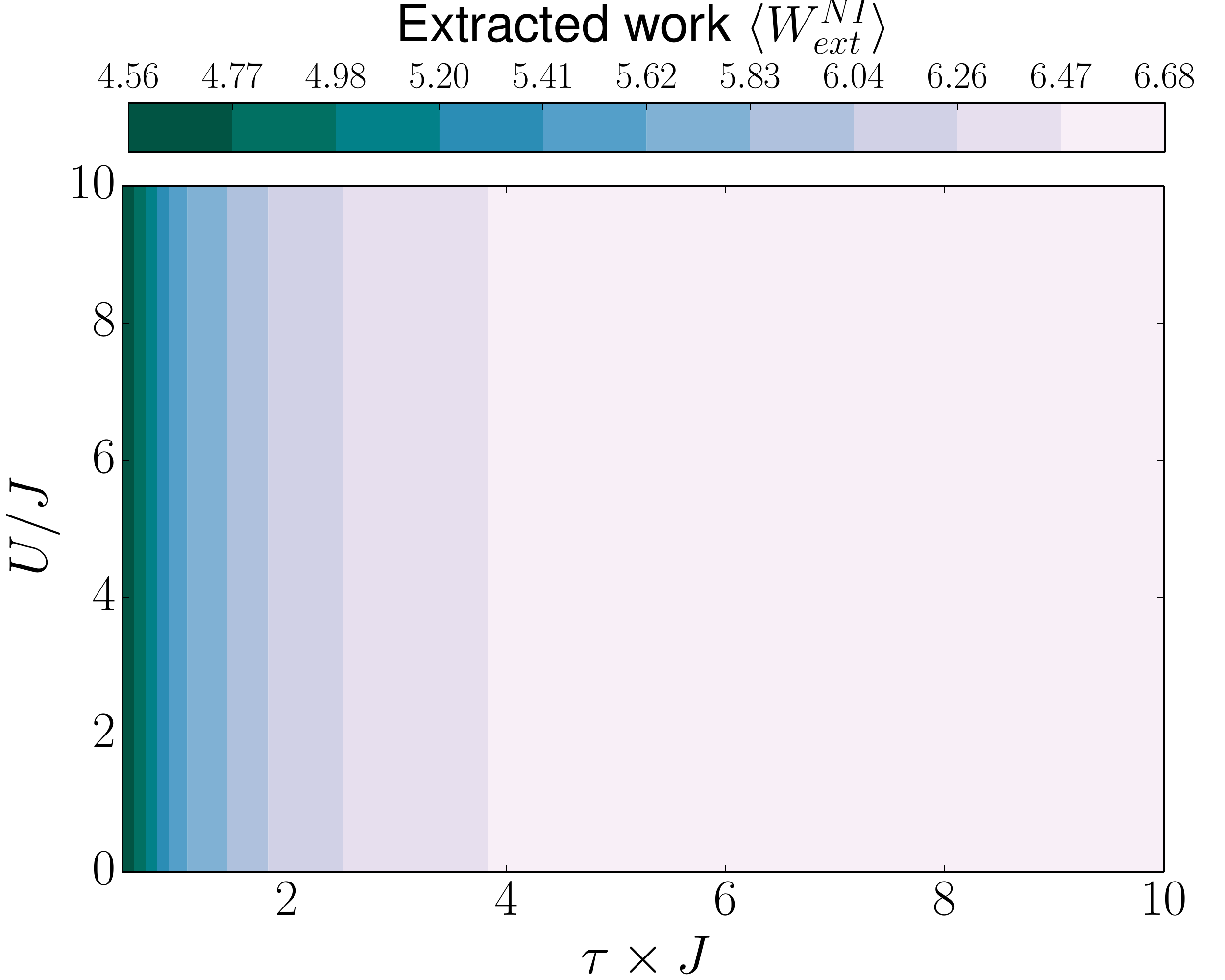}}
\subfloat[$T=20J/k_B$.]{\includegraphics[width=0.3\textwidth]{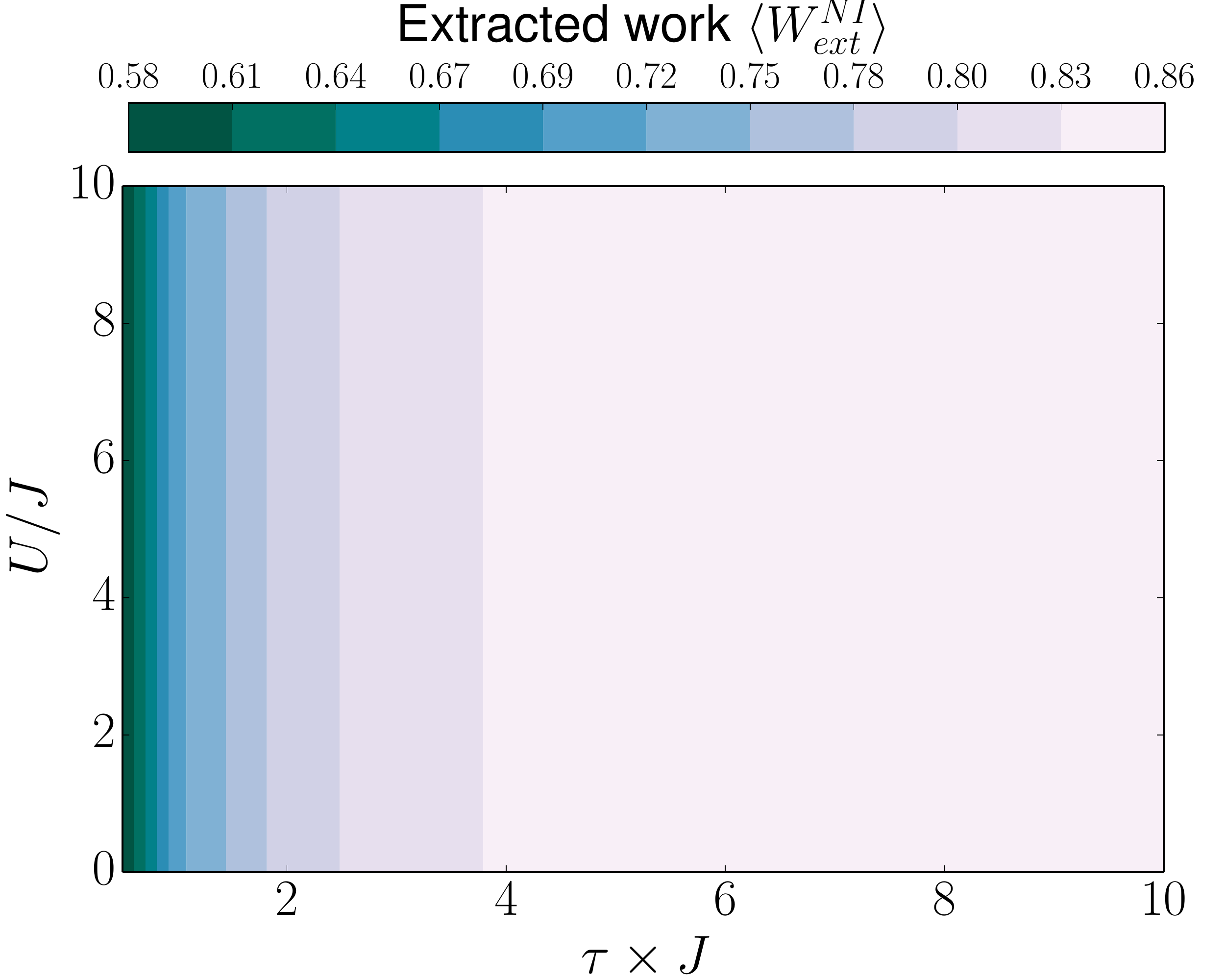}}
\label{fig:NI_work_extracted}
\centering
\subfloat[$T=0.2J/k_B$.]{\includegraphics[width=0.3\textwidth]{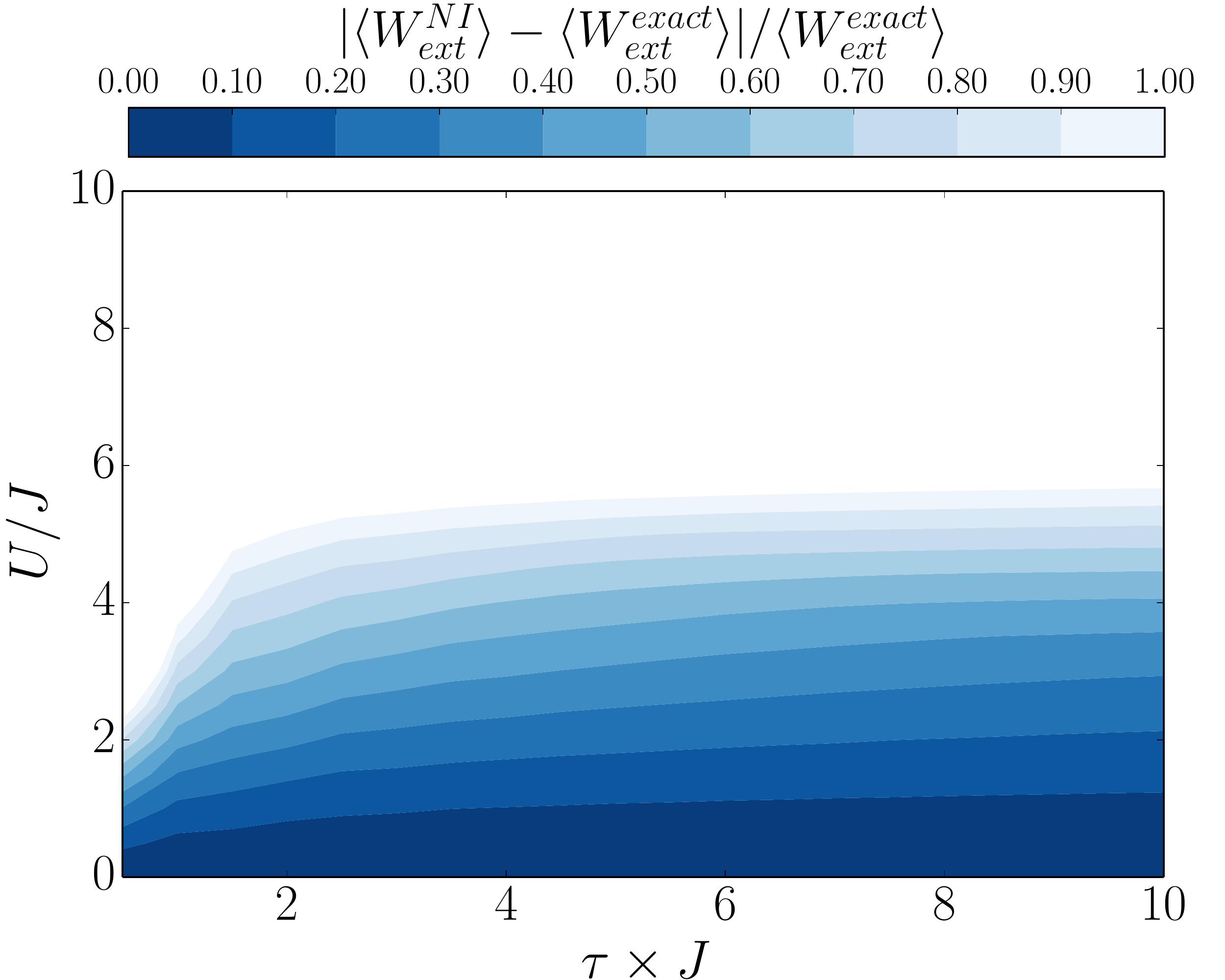}}
\subfloat[$T=2.5J/k_B$.]{\includegraphics[width=0.3\textwidth]{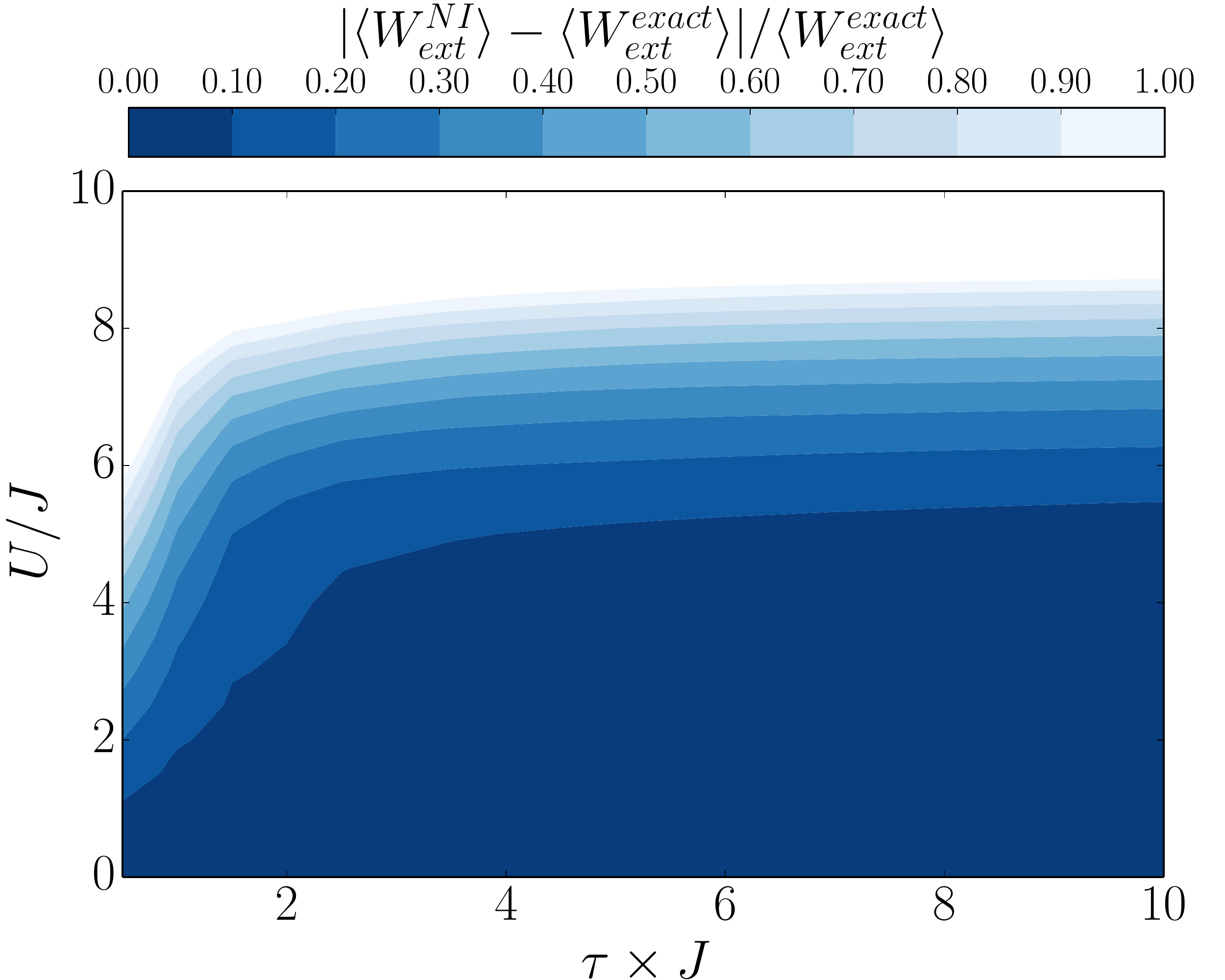}}
\subfloat[$T=20J/k_B$.]{\includegraphics[width=0.3\textwidth]{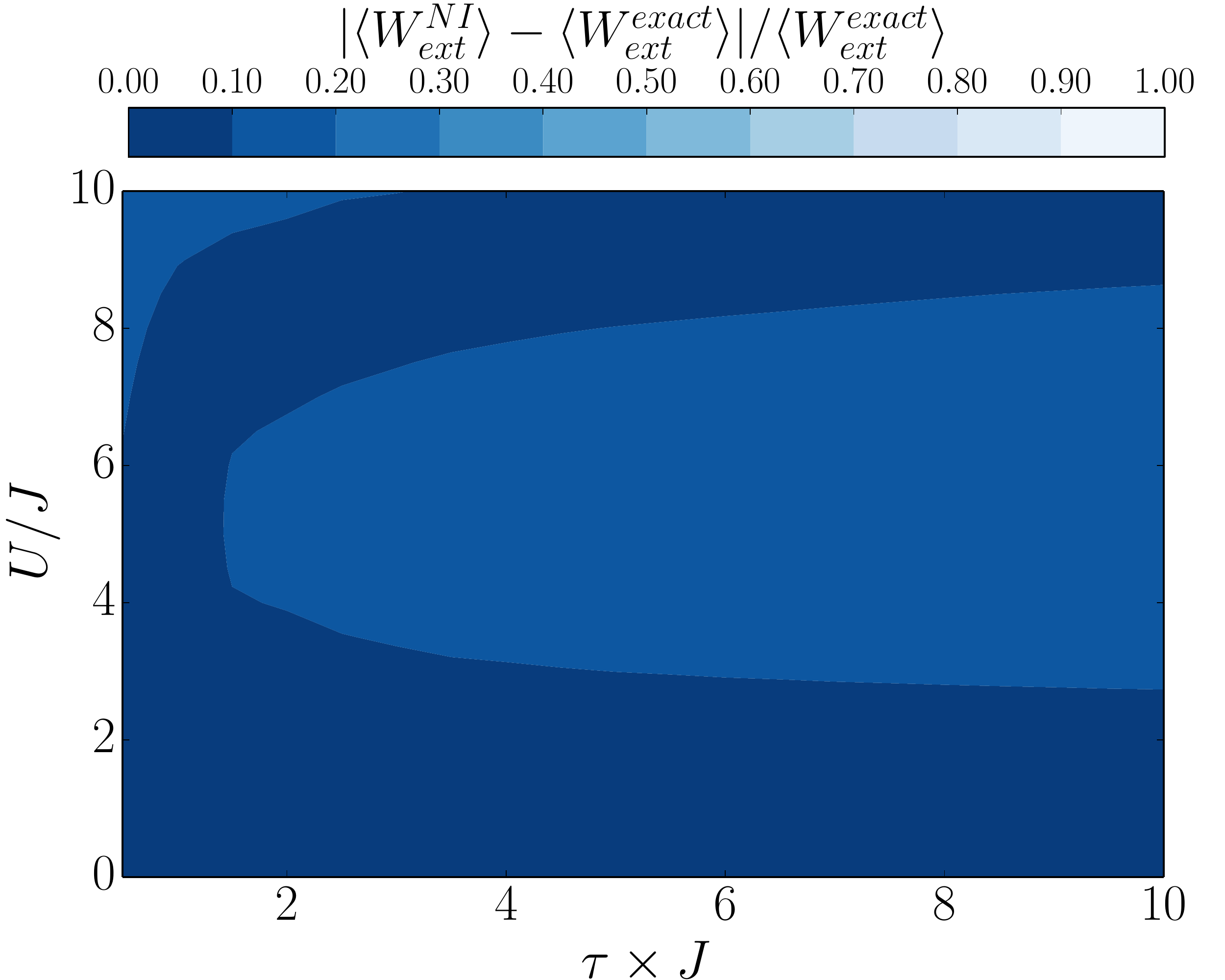}}
\caption{Panels (a) to (c): Work extracted in the NI approximation for  6 site chains driven by the `comb' potential.
Considered regimes go from non-interacting to strongly coupled along the $y$-axis, and from sudden quench to nearly adiabatic along the $x$-axis. The lighter the colour shade, the more work is extracted, compatible with the respective value ranges indicated above each panel.  Temperature increases from left to right panel, as indicated. \\ Panels (d) to (f): Relative error for $\langle W^{NI}_{ext} \rangle $ with respect to the exact results for the same parameters as the upper panels.  The darker the colour, the more accurate the approximation is in that regime.}
\label{fig:NI_work}
\end{figure*}

For non-interacting (NI) systems, $U=0$, the average work extracted $\langle W^{NI}_{ext} \rangle $ has no dependence on $U$.
This is shown in the upper panels of figure~\ref{fig:NI_work}, where we plot $\langle W^{NI}_{ext} \rangle $ for `comb' dynamics, with $0.5\le \tau \times J \le 10$ ($x$-axis) and  $0\le U/J \le 10$ ($y$-axis), and low (left), intermediate (middle), and high (right) temperatures.
The range of variation in $\langle W^{NI}_{ext} \rangle $  decreases with temperature (see colour bar above each panel), as the maximum driving potential becomes comparable to the thermal energy.

The lower panels of figure~\ref{fig:NI_work} show the corresponding relative error of the NI approximation, where the darker the blue, the more accurate the approximation is in that regime.  This approximation accurately captures the work extraction only in the parameters regions where interactions are weak compared to the other energy scales. These regions include higher values of $U$ as the temperature increases, see related discussion in section \ref{exact:res}.
However it is worth noting that there is very little work extracted at high temperatures, where the thermal energy is comparable to the driving potential which is then less effective.

As the lowest values of the average extracted work are driven by the freezing of the system dynamics due to strong many-body repulsion (dark areas in figure~\ref{fig:ex_work}), the NI approximation strongly overestimates the minimum work that can be extracted by a system. To see this compare the work range indicated over each panel for the mid column of figure~\ref{fig:ex_work} to the corresponding upper panels of figure~\ref{fig:NI_work}. At the other end, the value of the maximum average work extracted is quantitatively well captured by the NI approximation, {\it but} the corresponding parameter regions are qualitatively wrong, compare the shape of the light-shade areas of the mid-column panels of figure~\ref{fig:ex_work} to the corresponding areas of the upper panels of figure~\ref{fig:NI_work}.

We find a similar accuracy pattern with increasing temperature for 2 \footnote{See \cite{Herrera2017} and \cite{Herrera2018} for 2-site chain examples, and especially \cite{Herrera2017} for 2-site non-interacting results} and 4 site chains.

The `comb' driving potential corresponds to an accuracy of the results consistently in-between those of the AEF and MI potentials.

For the MI driving potential, the NI approximation works better than for the `comb' potential at all temperatures. As discussed in section \ref{exact:res}, for these dynamics many-body interactions become comparatively dominant at higher values of $U$.
As a consequence, for 6 site chains, the MI driving potential results in 10\% accuracy (or better) for $U\lesssim3J$ at low temperatures, for $U\lesssim7J$ at intermediate temperatures, and for all regimes at high temperatures: here thermal energy dominates so that the average extracted-work range is very narrow and so weakly sensitive to parameter changes.

For the AEF driving potential, the results in the NI approximation are in general worse than with the `comb' driving potential, and comparatively worsen as the temperature increases. At difference with the MI and comb driving potentials, the maximum AEF potential difference between nearby sites becomes at most of $4J$, so that even a Coulomb repulsion of just $U{\sim}1J$ will remain relevant. The NI approximation is then bound to fail, even at high temperatures, where, for 6 site chains, we get an accuracy of 10\% for $U\lesssim3J$ only.

\subsubsection{Exact initial system with non-interacting evolution operator}

To try and improve the estimate of the work extracted, we shall consider to still implement a non-interacting evolution, but this time starting from the exact many-body initial state.  The rationale is that a many-body evolution is in general a more challenging part of the calculation (and hence here it is approximated), while an accurate estimate for the initial state would be more readily available. This approximation is referred to as $\langle W^{exact+NI} \rangle$ in table \ref{tab:approxes}.

\begin{figure*}
\centering
\subfloat[$T=0.2J/k_B$.]{\includegraphics[width=0.3\textwidth]{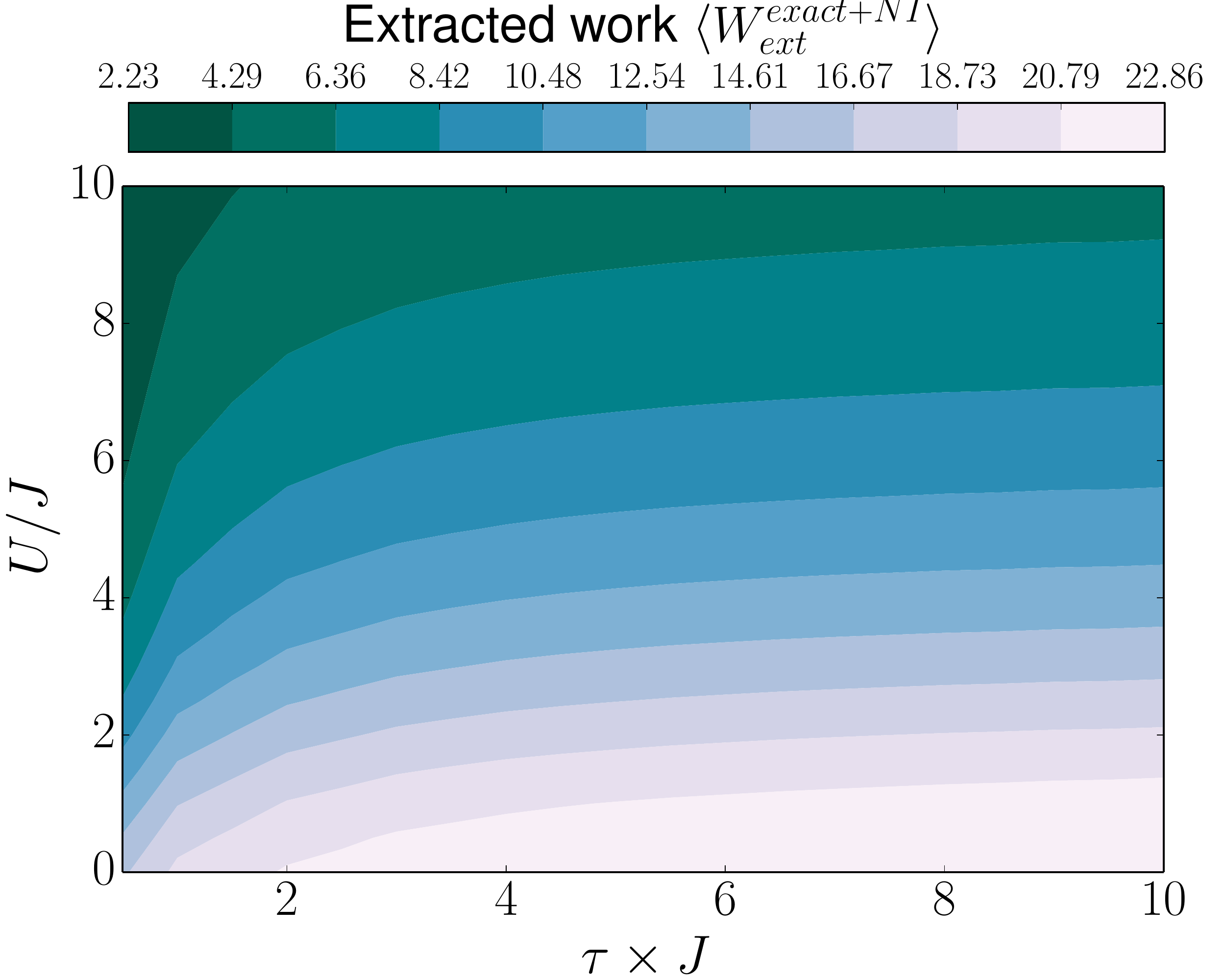}}
\subfloat[$T=2.5J/k_B$.]{\includegraphics[width=0.3\textwidth]{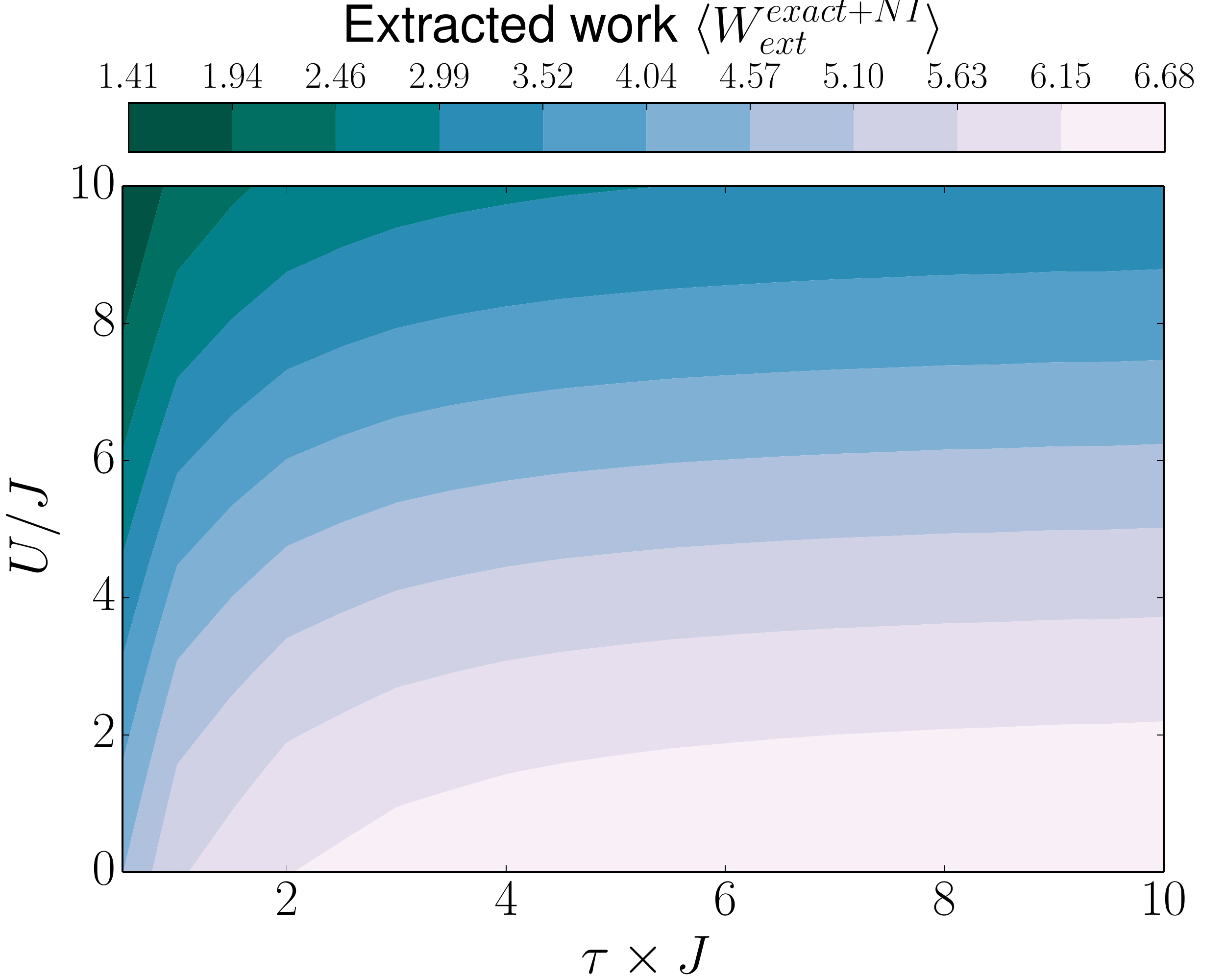}}
\subfloat[$T=20J/k_B$.]{\includegraphics[width=0.3\textwidth]{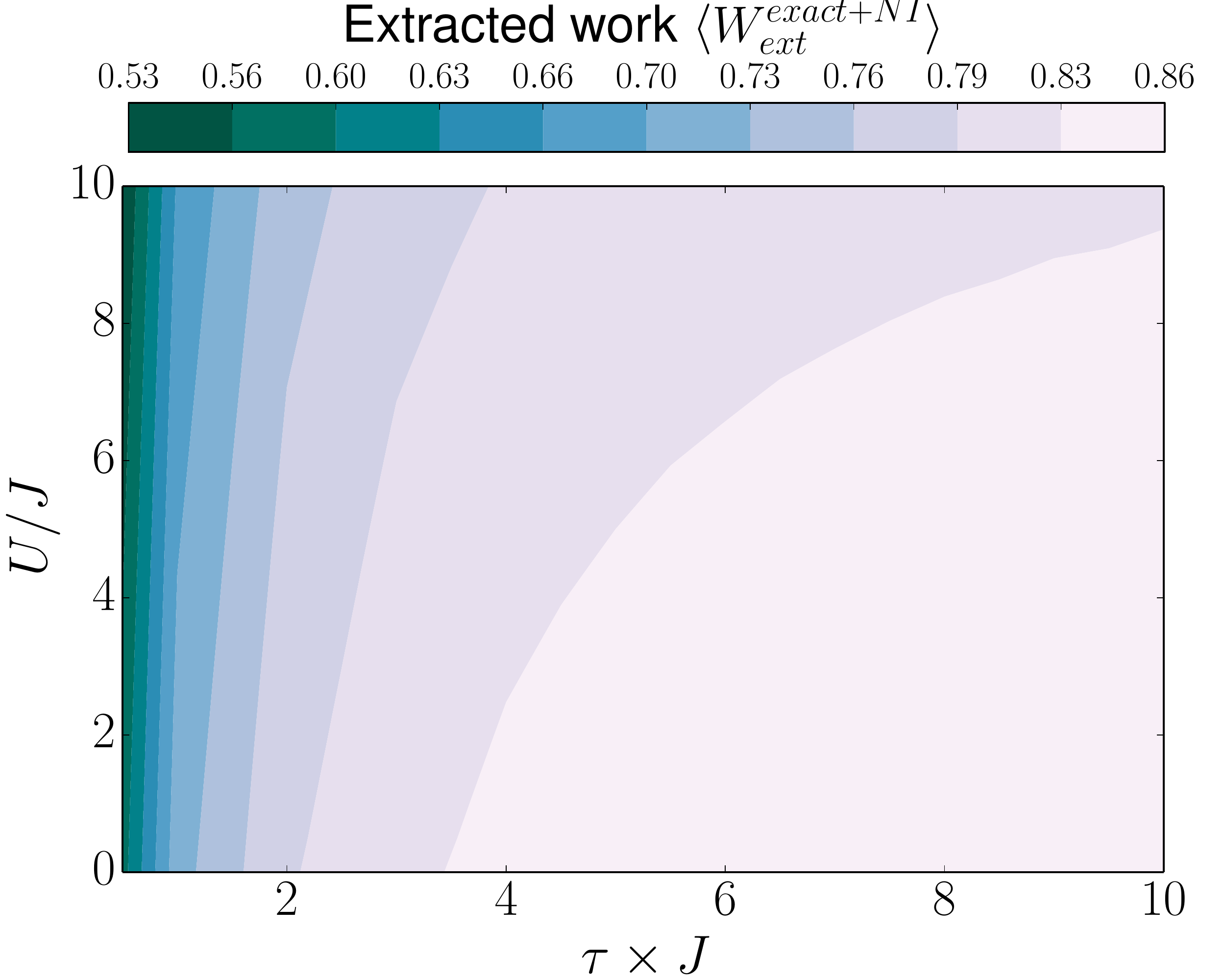}}
\label{fig:ex+NI_work_extracted}
\centering
\subfloat[$T=0.2J/k_B$.]{\includegraphics[width=0.3\textwidth]{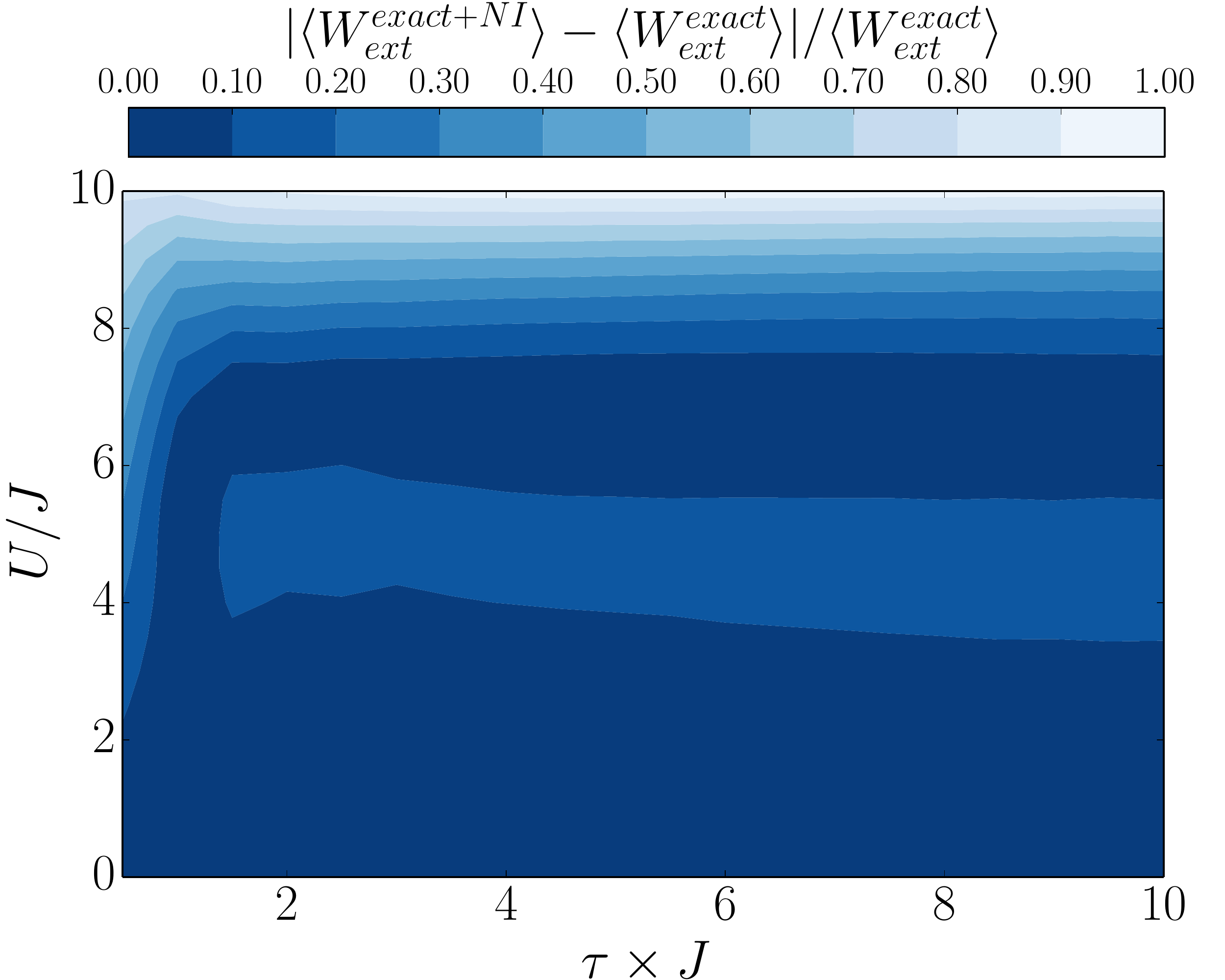}}
\subfloat[$T=2.5J/k_B$.]{\includegraphics[width=0.3\textwidth]{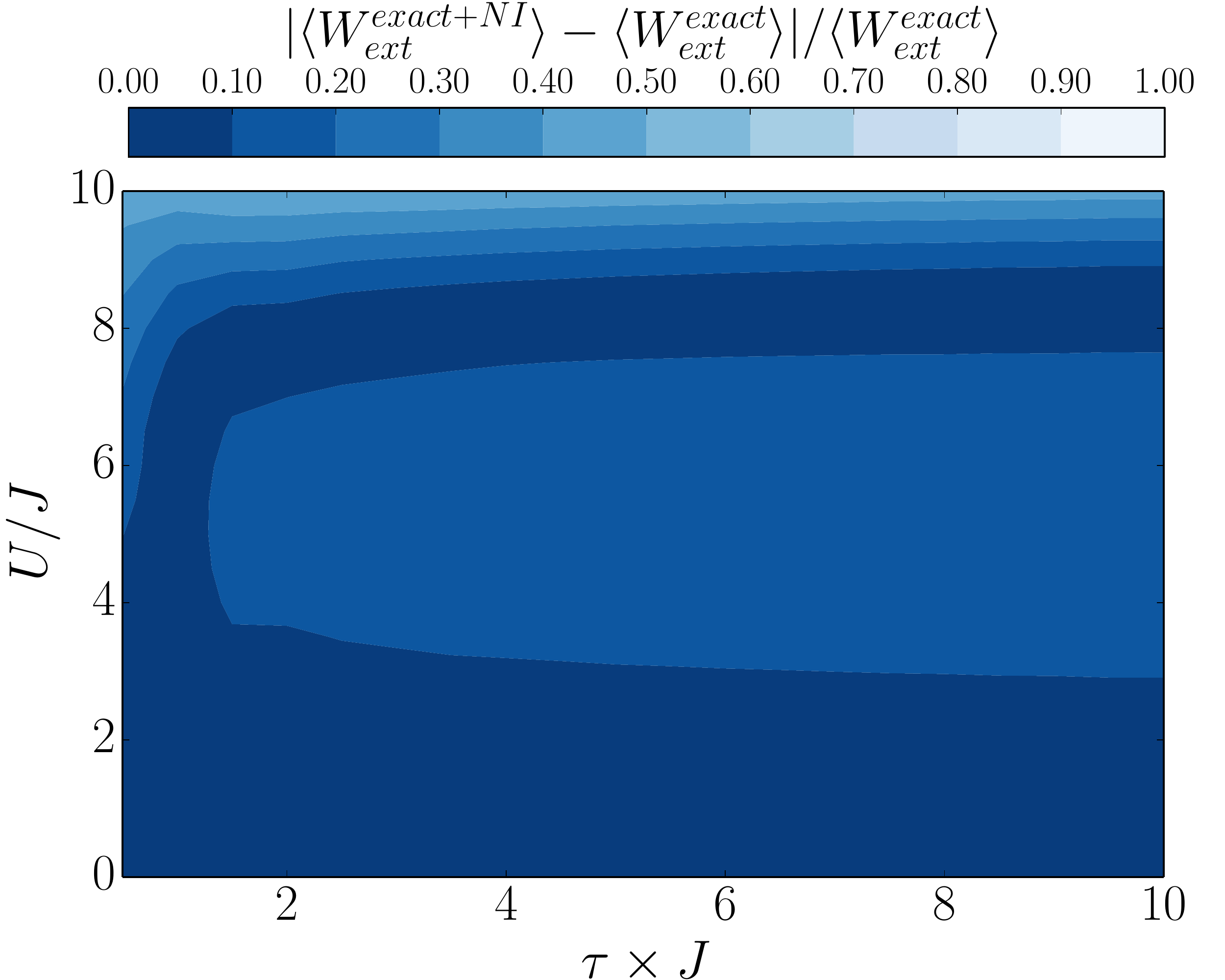}}
\subfloat[$T=20J/k_B$.]{\includegraphics[width=0.3\textwidth]{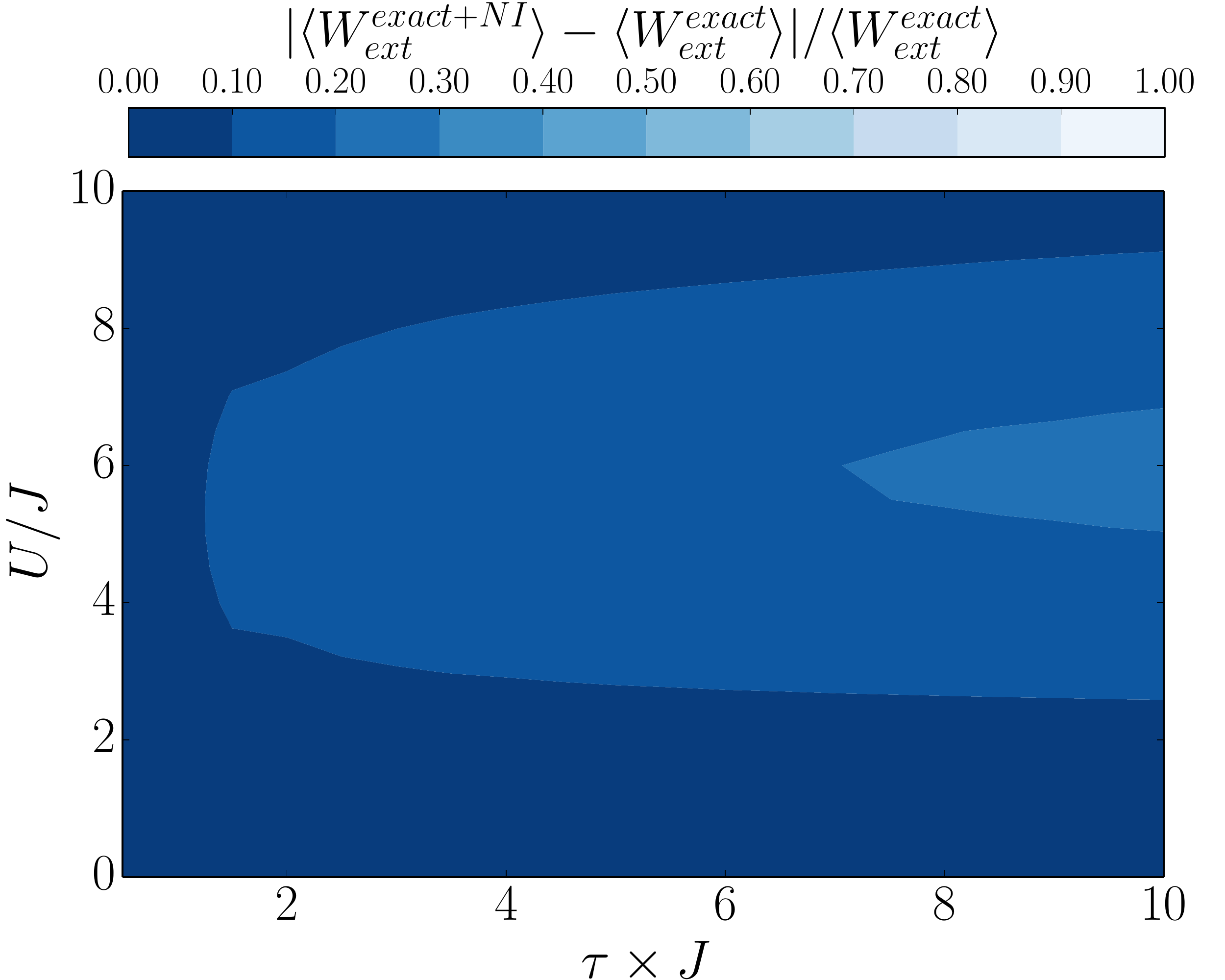}}
\caption{Panels (a) to (c): Work extracted in the `exact + NI' approximation for  6 site chains driven by the `comb' potential.
Considered regimes go from non-interacting to strongly coupled along the $y$-axis, and from sudden quench to nearly adiabatic along the $x$-axis. The lighter the colour shade, the more work is extracted, compatible with the respective value ranges indicated above each panel.  Temperature increases from left to right panel, as indicated. \\ Panels (d) to (f): Relative error for $\langle W^{exact+NI}_{ext} \rangle $ with respect to the exact results for the same parameters as the upper panels.  The darker the colour, the more accurate the approximation is in that regime.
}
\label{fig:ex+NI_work}
\end{figure*}

Indeed this simple approximation leads to strikingly improved accuracy.
Results are presented in figure~\ref{fig:ex+NI_work}: $\langle W^{exact+NI}_{ext} \rangle$ in the upper three panels, and its relative error with respect to the exact results in the lower panels. Parameters are the same as in figure~\ref{fig:NI_work}.

By comparing $\langle W^{exact+NI}_{ext}\rangle$ to the corresponding exact results in panels (b), (e), and (h) of figure~\ref{fig:ex_work} we see that including interactions just through the initial state is sufficient to recover the qualitative (and in great part quantitative) behaviour at low and intermediate temperatures. The greatest improvement is seen in the low temperature, where $\langle W^{exact+NI} \rangle$ recaptures the correct work to 10-20\% for most regimes up to $U \approx 9J$.

At high temperature the qualitative behaviour is not recovered, but, as the work extracted varies only slightly at this temperature, quantitatively the approximation remains overall good, as it reproduces rather well the overall work variation range (compare the colour bar limits of figures~\ref{fig:ex_work} and \ref{fig:ex+NI_work}). In general the `exact + NI' approximation significantly improves for the minimum extracted average work  over the NI value and at all temperatures.

A similar pattern is seen in all the other systems considered, i.e. for 2 and 4 site chains, and for the MI and AEF evolutions (see appendix), demonstrating the scalability and versatility of this approximation.
The $\langle W^{exact+NI} \rangle $ approximation handles weak to medium correlated systems well in all regimes and temperatures, from adiabatic to sudden quench, {and} from low to high temperatures.

\subsubsection{Adiabatic behaviour and the `exact + NI' approximation}
The most striking improvement of the `exact + NI' approximation with respect to the NI one, is the recovery of the qualitative behaviour of the average quantum work in the quasi-adiabatic regime for low to intermediate temperatures. In this regime the exact $\langle W_{ext} \rangle $ is becoming independent from $\tau$ -- as to be expected in the adiabatic regime -- but strongly depends on $U$, see  figure~\ref{fig:ex_work}. The recovery of this behaviour by the `exact + NI' approximation  can be explained as follows.

As discussed in section \ref{adiab_ex}, the dependency of $\langle W_{ext} \rangle $ over $\tau$ is contained in the conditional probabilities $p_{m|n}^\tau=|\langle\Psi_m^f| \mathcal{U}_{evo}(\tau)|\Psi_n^0\rangle|^2$. In the NI approximation these would read $p_{m|n}^{\tau,NI}=|\langle\Psi_m^{f,NI}| \mathcal{U}_{evo}^{NI}(\tau)|\Psi_n^{0,NI}\rangle|^2$
which, would lead, as expected, to a result potentially dependent on $\tau$ but completely independent from $U$.
However in the `exact + NI' approximation we have $p_{m|n}^{\tau,exact+NI}=|\langle\Psi_m^{f,NI}| \mathcal{U}_{evo}^{NI}(\tau)|\Psi_n^{0}\rangle|^2$, which could lead to dependency on $U$ (through $|\Psi_n^{0}\rangle$)  as well as on $\tau$ \footnote{For the `exact + NI' approximation, the results of calculating $<W>$ using (\ref{eq:q_work_trace}) or (\ref{W_PW}) differ for the $t=0$ terms.  The $t=0$ term via (\ref{eq:q_work_trace}) can be written as $\Sigma_{i,n}E_i^{NI,0} p_n^0 \left| a_{n,i} \right|^2 $, while the corresponding term via (\ref{W_PW}) is $\Sigma_n p_n^0 E_n^{NI,0}$.  The $t = \tau$ terms give instead the same results.}.

Let us name $|\Psi_m^{NI}(t)\rangle$ the $m$-th eigenstate of $H^{NI}(t)$: then $|\Psi_m^{NI}(\tau)\rangle= |\Psi_m^{f,NI}\rangle$, independent of $\tau$, and $|\Psi_n^{0}\rangle = \sum_j a_{j,n} |\Psi_j^{NI}(0)\rangle$, with $a_{j,n}=a_{j,n}(U)$. Consider
\begin{eqnarray}
 && \left.\sum_{n,m} p_{n,m}^{\tau}E_m^f \right|_{exact+ NI} = \sum_{n,m}E_m^f p_n^0 |\langle\Psi_m^{f,NI}| \mathcal{U}_{evo}^{NI}(\tau)|\Psi_n^{0}\rangle|^2 \nonumber \\
   &&= \sum_{n,m}E_m^f p_n^0 |\langle\Psi_m^{f,NI}| \sum_j a_{j,n} \mathcal{U}_{evo}^{NI}(\tau)|\Psi_j^{NI}(0)\rangle|^2. \label{p_nm}
\end{eqnarray}
In the adiabatic regime $\mathcal{U}_{evo}^{NI}(\tau)|\Psi_j^{NI}(0)\rangle= |\Psi_j^{f,NI}\rangle$, so that (\ref{p_nm}) becomes
%
\begin{align}
    \left.\sum_{n,m} p_{n,m}^{\tau}E_m^f \right|_{exact+ NI} \nonumber \\ 
     \stackrel{\mbox{adiab.}}{=} \sum_{n,m}&E_m^f p_n^0 |\langle\Psi_m^{f,NI}| \sum_j a_{j,n}|\Psi_j^{f,NI}\rangle |^2 \nonumber \\
    =  \sum_{n,m} E_m^f & p_n^0 |a_{m,n}(U)|^2,
 \label{p_nm_ad}
\end{align}
which indeed depends on $U$ but not on $\tau$, as observed.

At high temperatures, inclusion of many-body interactions just via the initial state is a too-weak correction to counter the high thermal energy, which becomes even more dominant than in the exact case: at high temperatures for the `exact+NI' approximation a behaviour closer to the fully NI approximation is recovered (compare  figure~\ref{fig:ex+NI_work} panel (c) to  figure~\ref{fig:NI_work} panel (c)). We observe this for all driving potentials (see appendix, panel (c) of figures~\ref{fig:ex+NI_work_teeth} and \ref{fig:ex+NI_work_slope}).

\subsubsection{Sudden quench and the `exact + NI' approximation}
With respect to the NI approximation, the `exact + NI' approximation recovers the qualitative and, in great part, quantitative exact behaviour for the average quantum work also in the sudden quench limit. This can be seen by comparing results for the small-$\tau$ parameter region of the upper panels of figure~\ref{fig:ex+NI_work} with the corresponding panels in the central column of figure~\ref{fig:ex_work}. We give an explanation for this below.

In the quasi-sudden-quench regime $\tau<<1/J$, $\rho_{f}\approx \rho_0 +\delta\rho(\tau)$, with the second term a small correction.
By using this and the fact that $\Delta H=\hat{H}_{f}-\hat{H}_0$ is a constant, we can approximate (\ref{eq:q_work_trace}) as
\begin{eqnarray}
& &\langle W \rangle \approx \mathrm{Tr}\left[ (\rho_0 +\delta\rho(\tau)) \hat{H}_{f} - \rho_0 \hat{H}_0 \right]\nonumber\\
& & =  \mathrm{Tr}\left[ \rho_0\Delta H  +\delta\rho(\tau) \hat{H}_{f}\right] \label{eq:q_work_trace_approx_qsq}\\
& & \stackrel{\tau=0}{=}  \mathrm{Tr}\left[ \rho_0\Delta H \right].
\label{eq:q_work_trace_approx_sq}
\end{eqnarray}

In the `exact + NI' approximation, (\ref{eq:q_work_trace_approx_qsq}) and (\ref{eq:q_work_trace_approx_sq}) become 
\begin{align}
&\langle W^{exact+NI} \rangle \approx  \mathrm{Tr}\left[ \rho_0^{exact}\Delta H^{NI}  +\delta\rho(\tau)^{exact+NI} \hat{H}_{f}^{NI}\right] \label{eq:q_work_trace_approx_qsq_eNI}\\
& \stackrel{\tau\to 0}{=}  \mathrm{Tr}\left[\rho_0^{exact}\Delta H^{NI}
\right],
\label{eq:q_work_trace_approx_sq_eNI}
\end{align}
where we have explicitly stressed which terms on the r.h.s. are to be computed exactly and which ones in a specific approximation.
In contrast to a fully NI approximation, (\ref{eq:q_work_trace_approx_sq_eNI}) is now $U$-dependent through $\rho_0^{exact}$, which means that this dependence for $\tau=0$ is recovered. Similarly, the correction for small (but finite) $\tau$ values in (\ref{eq:q_work_trace_approx_qsq_eNI}) contains a $U$-dependence through $\delta\rho(\tau)^{exact+NI}$, and will then improve both qualitatively and quantitatively over NI results, as comparison of  figures \ref{fig:ex_work}, \ref{fig:NI_work}, and \ref{fig:ex+NI_work} demonstrates.

\section{Entropy variation }
\subsection{Exact results}

Let us now examine the exact results for the entropy, $\Delta S$ from (\ref{eq:entropy}). This quantity corresponds to the heat which the system would have to disperse in the environment to return to thermodynamic equilibrium at the end of the dynamics.
Apart from the average quantum work, the other key ingredient for $\Delta S$ is the free-energy variation, shown by (\ref{eq:free_energy}). Since our final Hamiltonians are independent of $\tau$,  the free energy only depends on $U$ and $\beta$.

\begin{figure*}
\subfloat[MI driving potential.]{\includegraphics[width=0.3\textwidth]{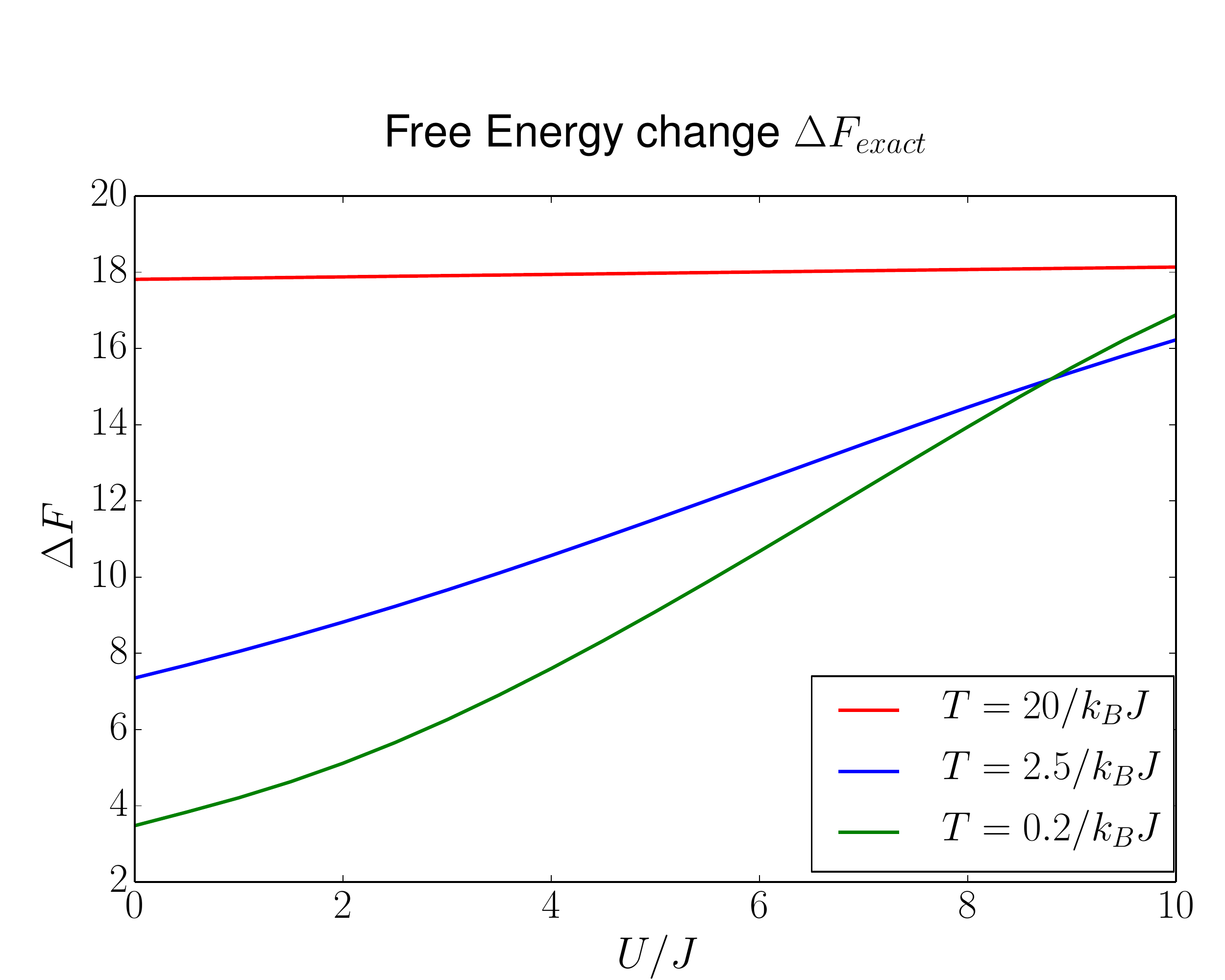}}
\subfloat[Comb driving potential.]{\includegraphics[width=0.3\textwidth]{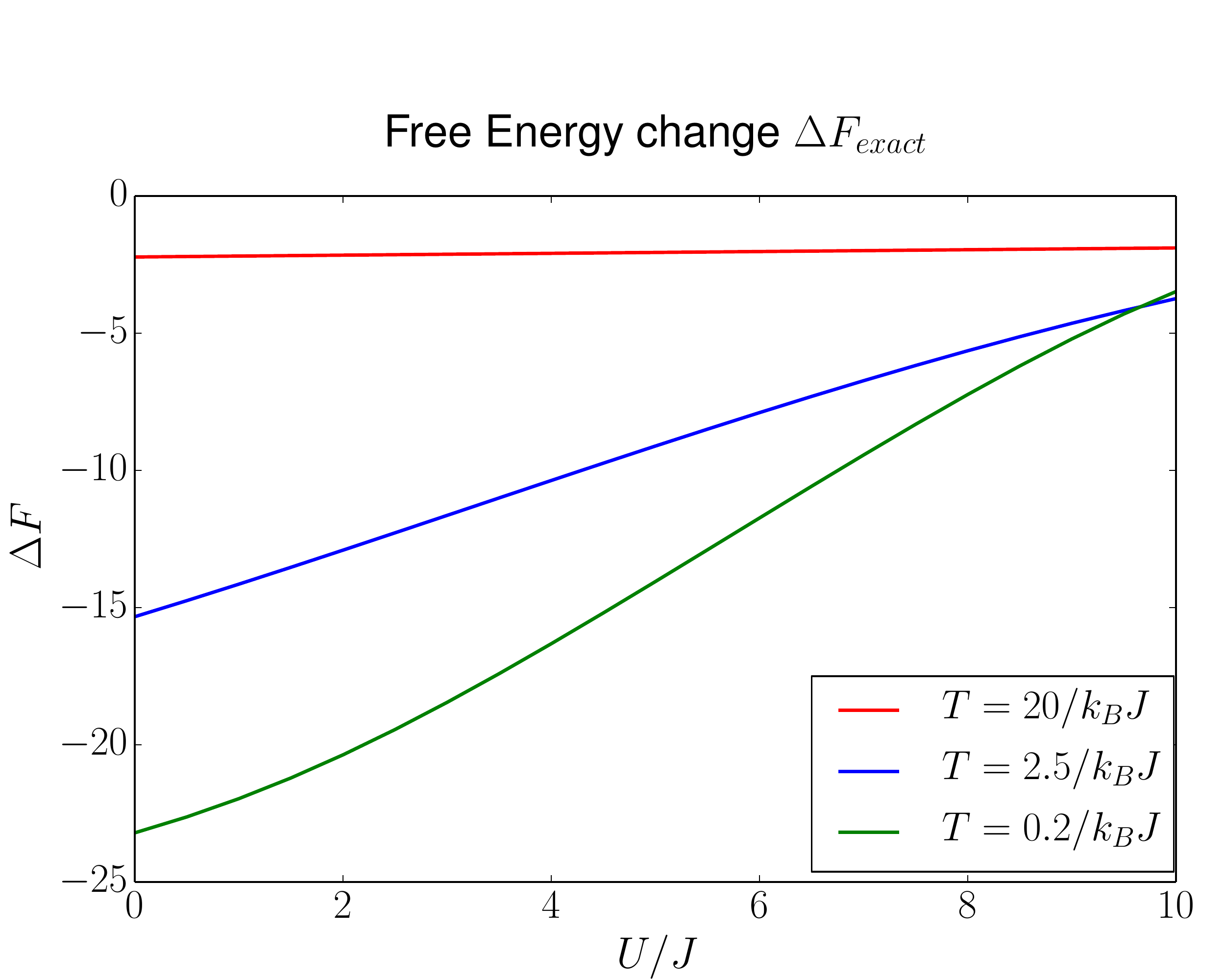}}
\subfloat[AEF driving potential.]{\includegraphics[width=0.3\textwidth]{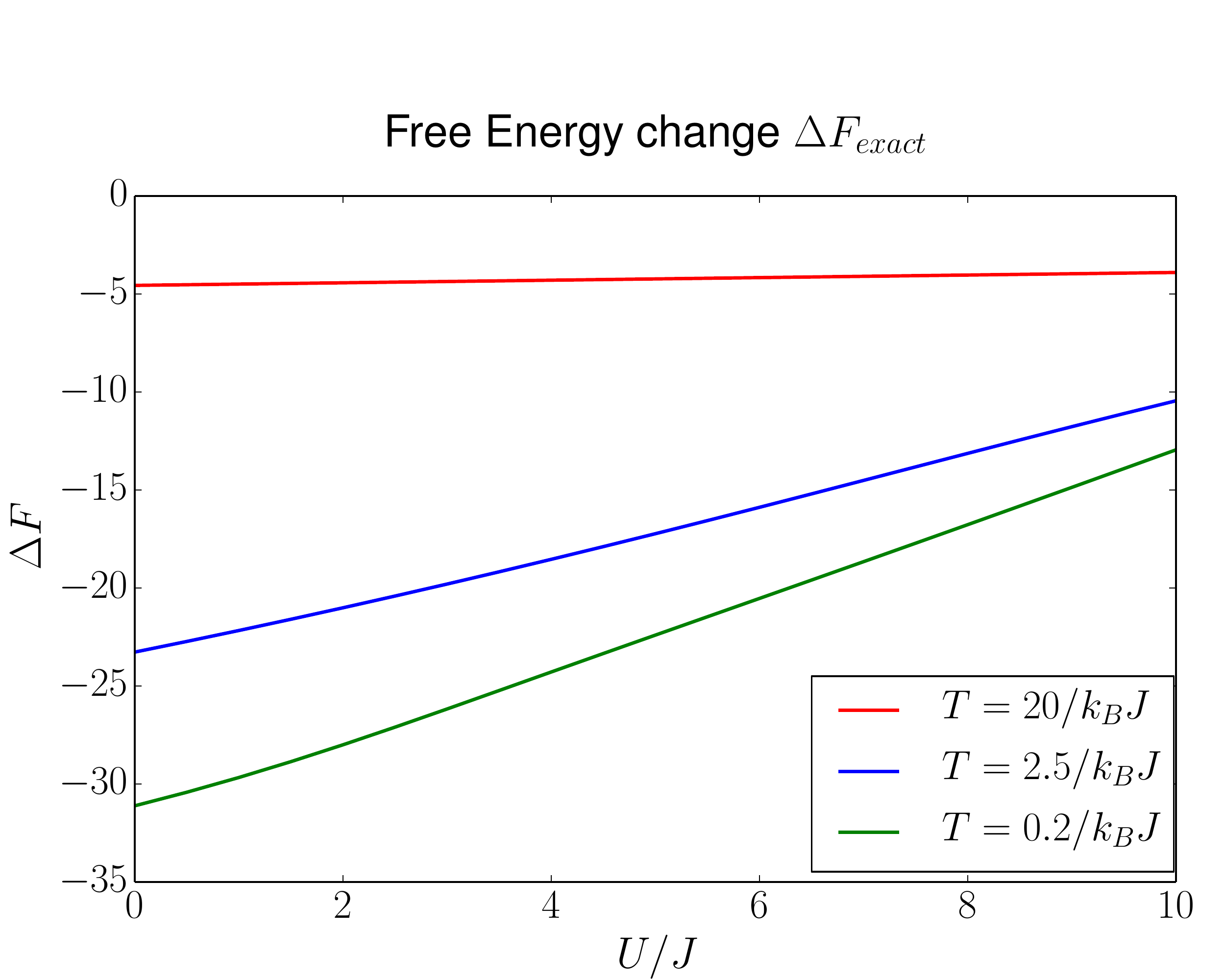}}
\par
\centering
\subfloat[$T=0.2J/k_B$ with MI driving potential.]{\includegraphics[width=0.3\textwidth]{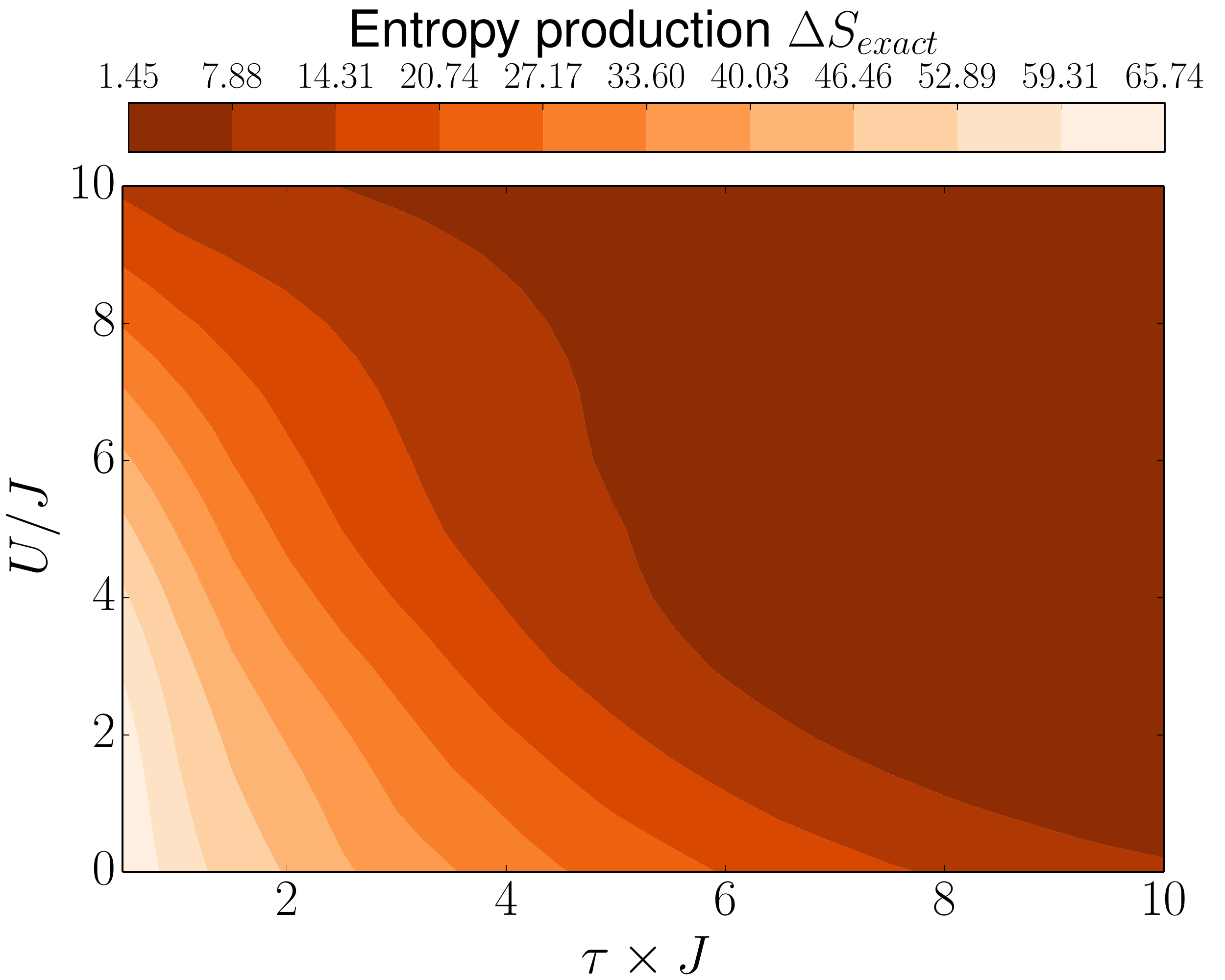}}
\subfloat[$T=0.2J/k_B$ with comb driving potential.]{\includegraphics[width=0.3\textwidth]{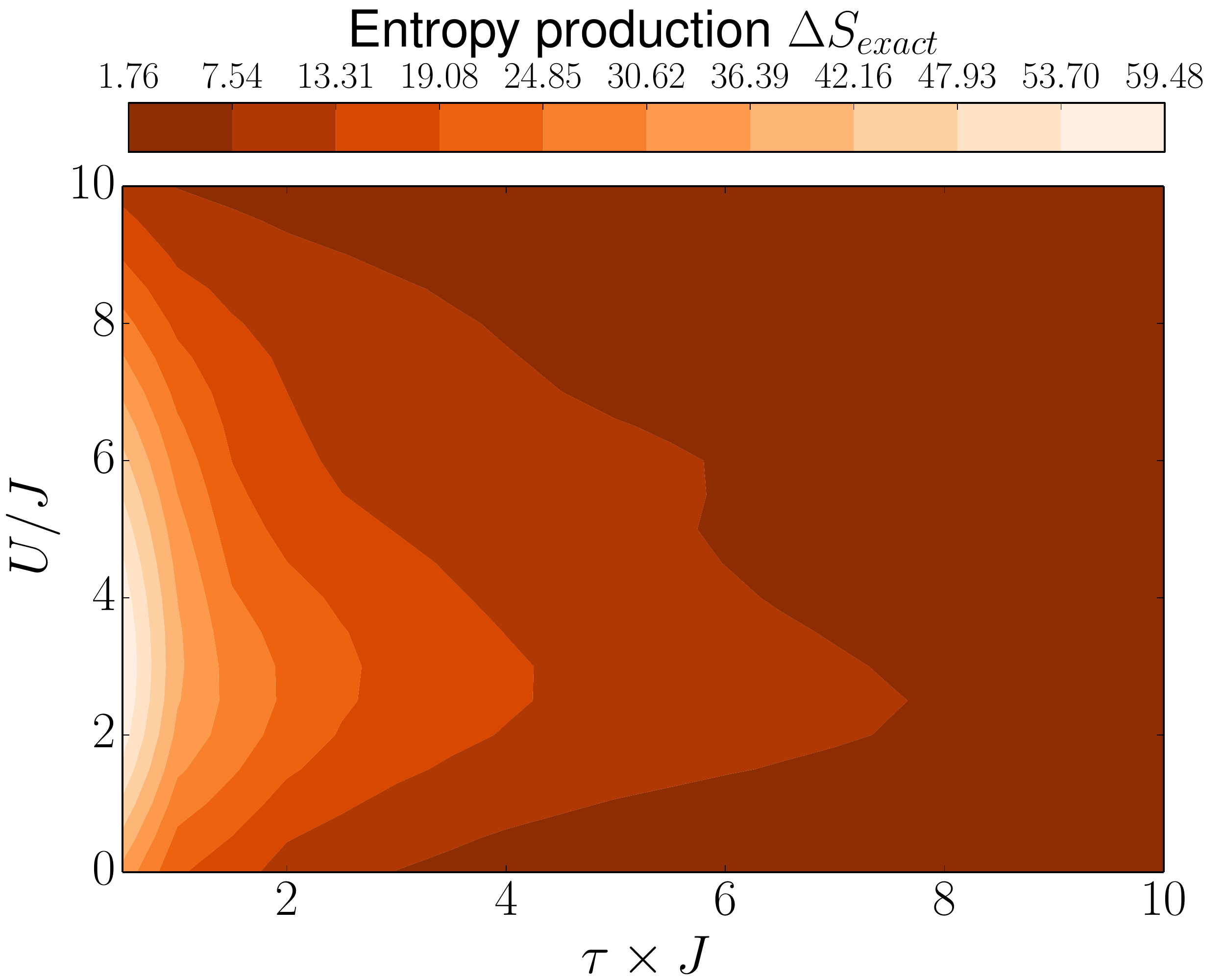}}
\subfloat[$T=0.2J/k_B$ with AEF driving potential.]{\includegraphics[width=0.3\textwidth]{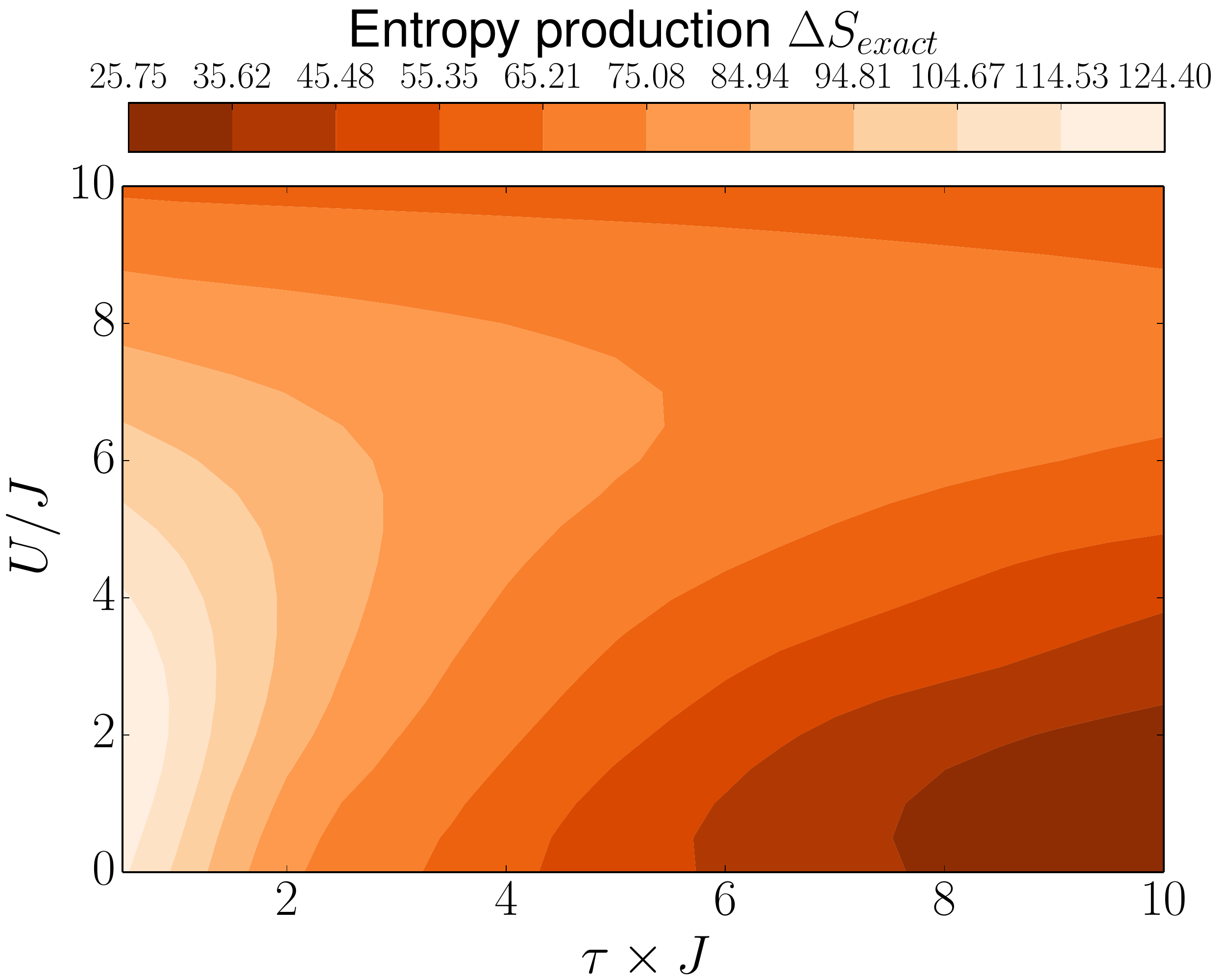}}
\par
\subfloat[$T=2.5J/k_B$ with MI driving potential.]{\includegraphics[width=0.3\textwidth]{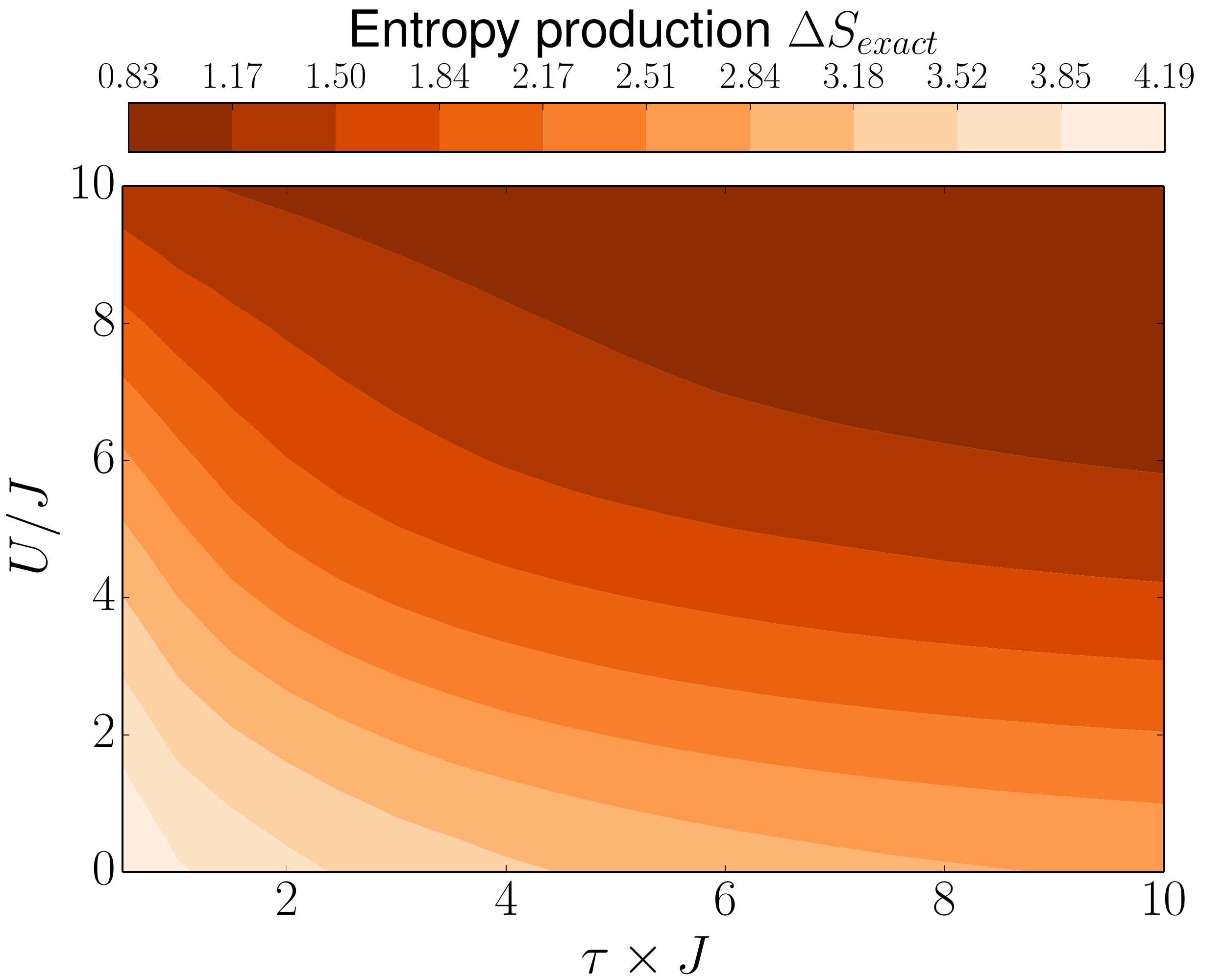}}
\subfloat[$T=2.5J/k_B$ with comb driving potential.]{\includegraphics[width=0.3\textwidth]{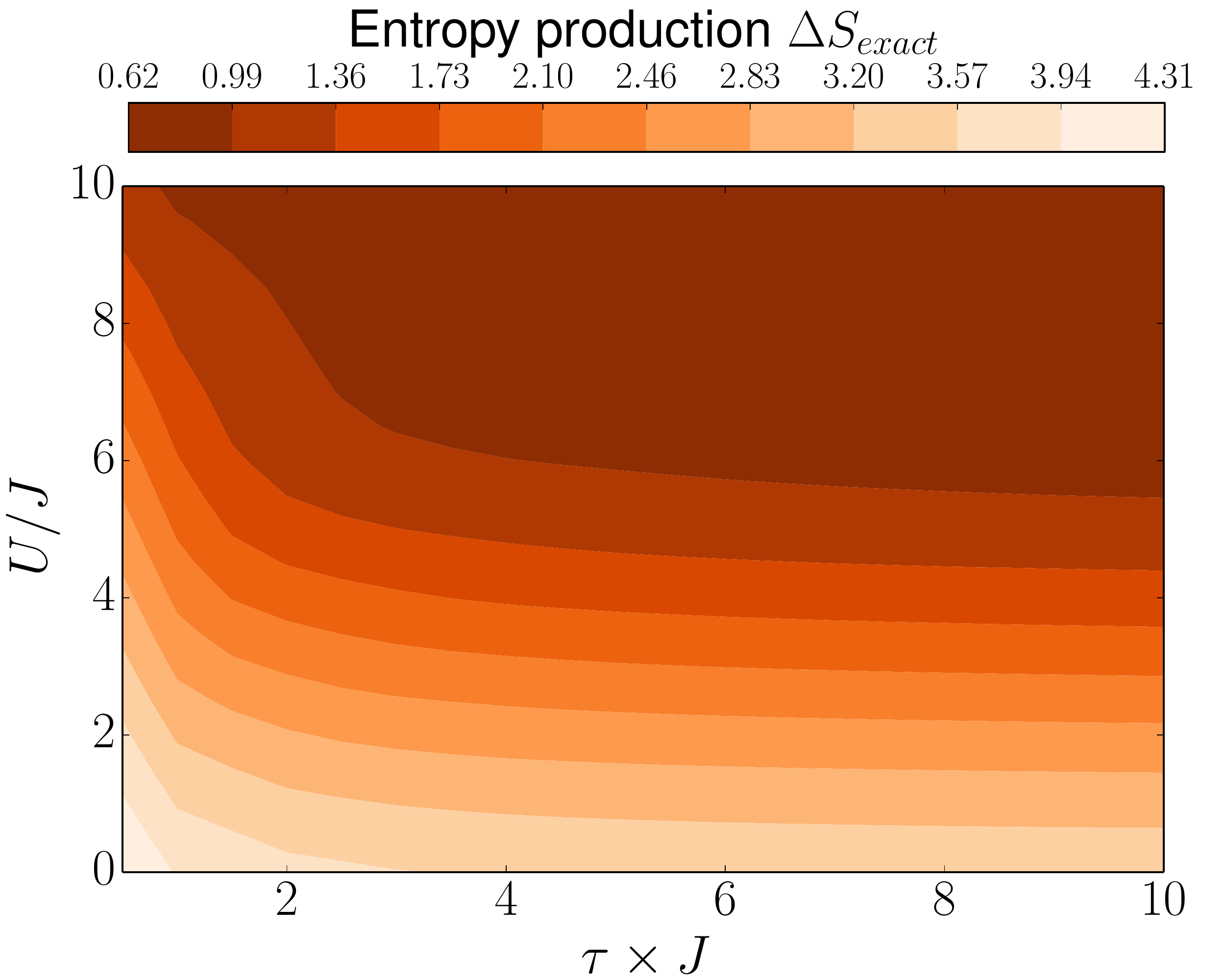}}
\subfloat[$T=2.5J/k_B$ with AEF driving potential.]{\includegraphics[width=0.3\textwidth]{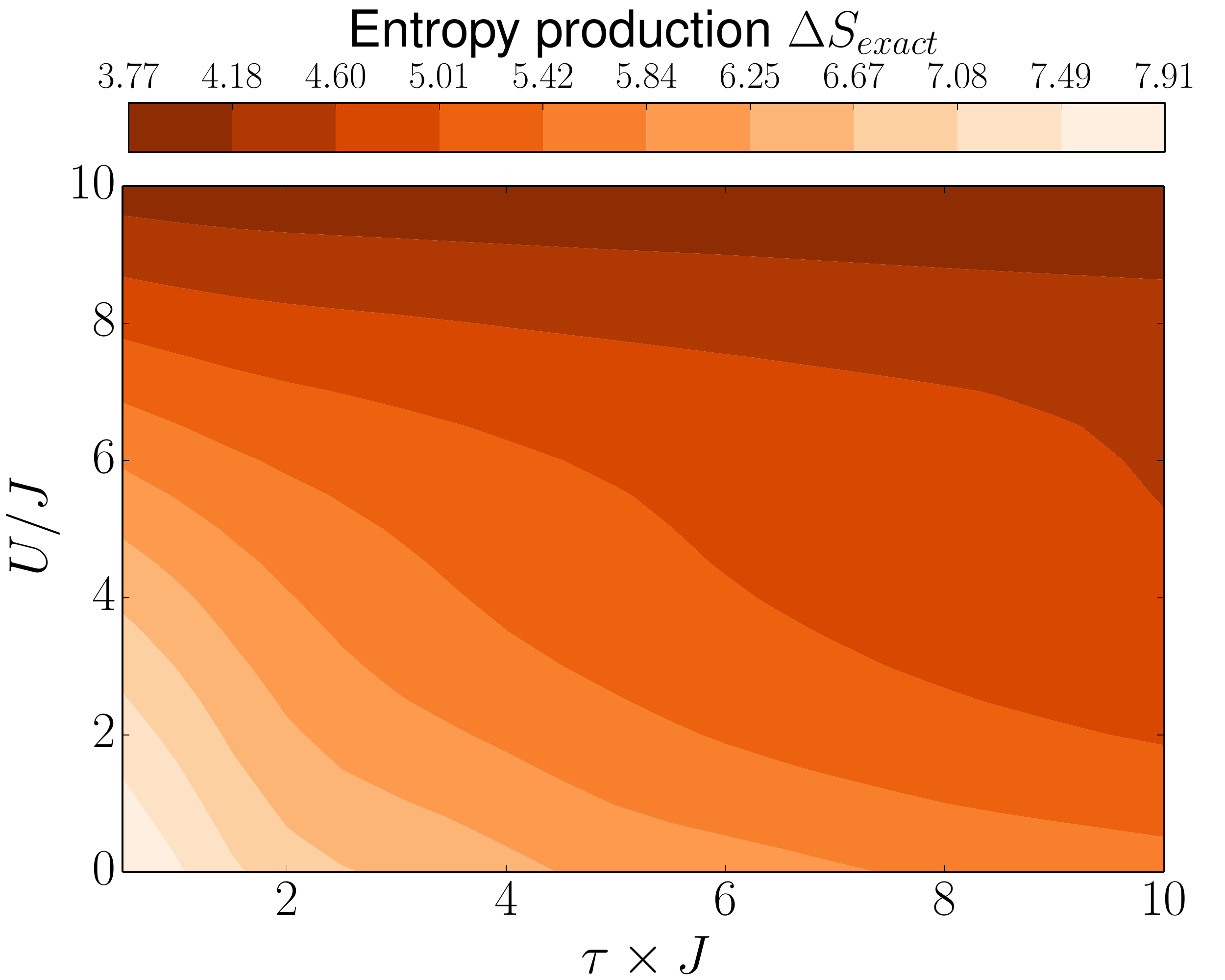}}
\par
\subfloat[$T=20J/k_B$ with MI driving potential.]{\includegraphics[width=0.3\textwidth]{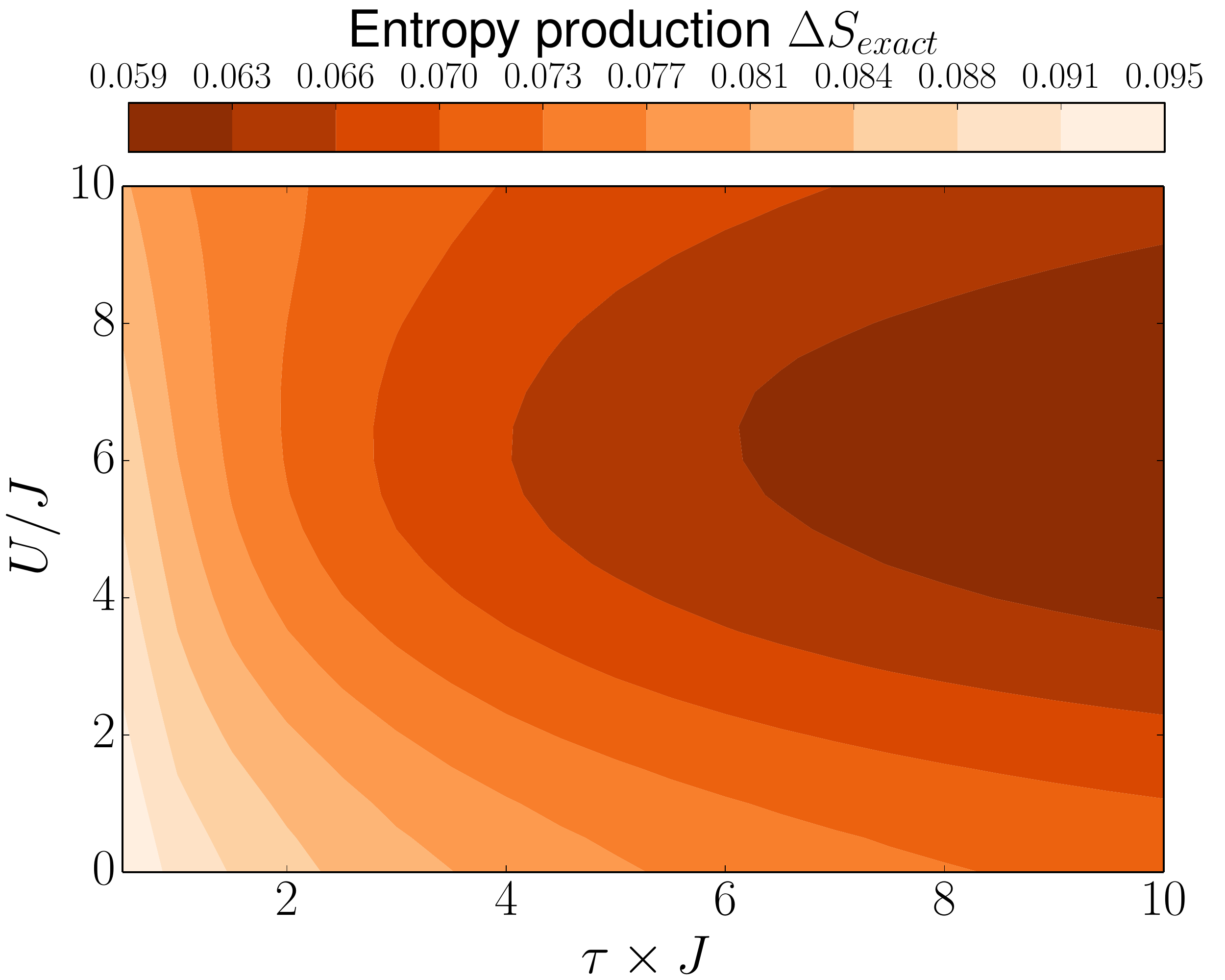}}
\subfloat[$T=20J/k_B$ with comb driving potential.]{\includegraphics[width=0.3\textwidth]{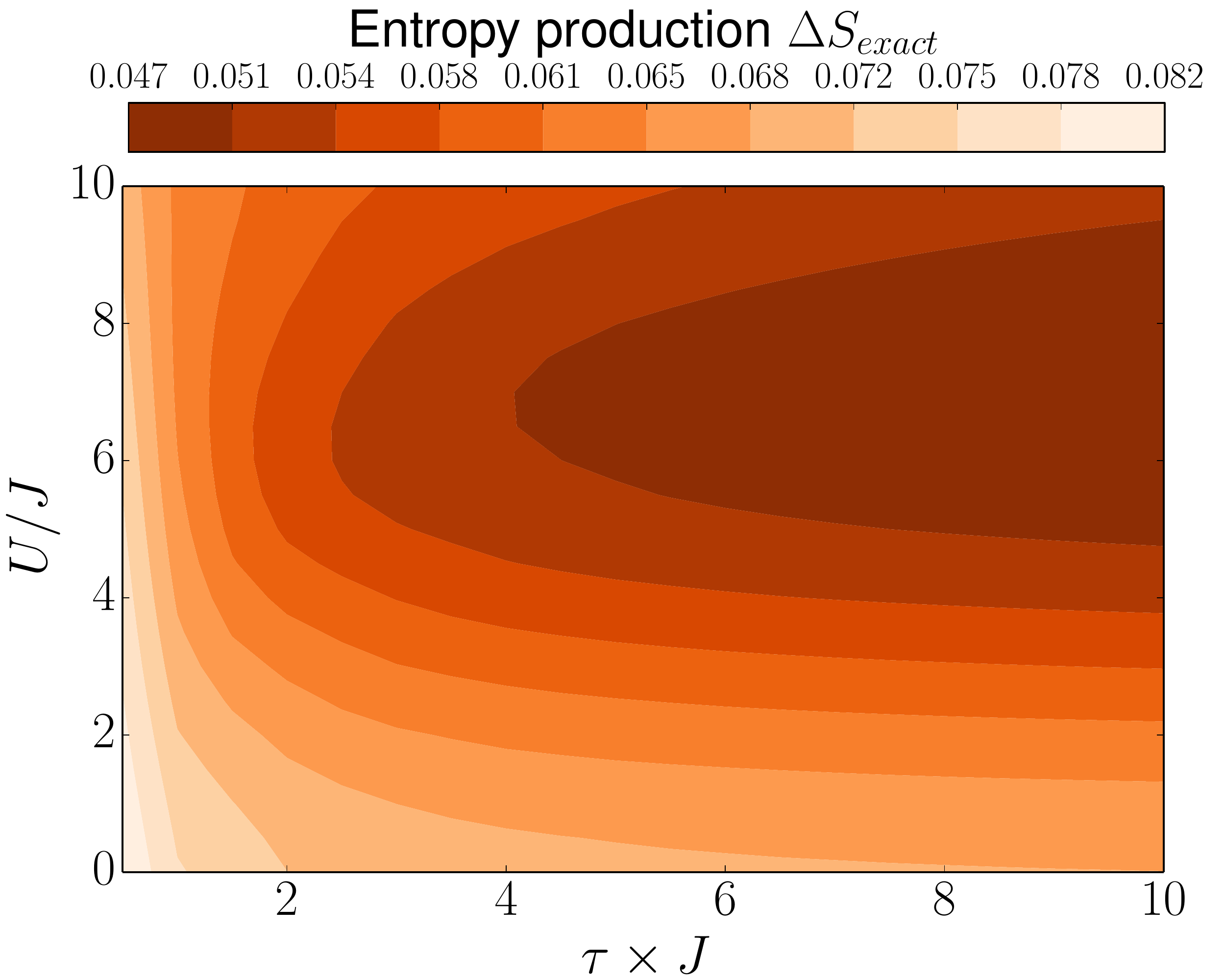}}
\subfloat[$T=20J/k_B$ with AEF driving potential.]{\includegraphics[width=0.3\textwidth]{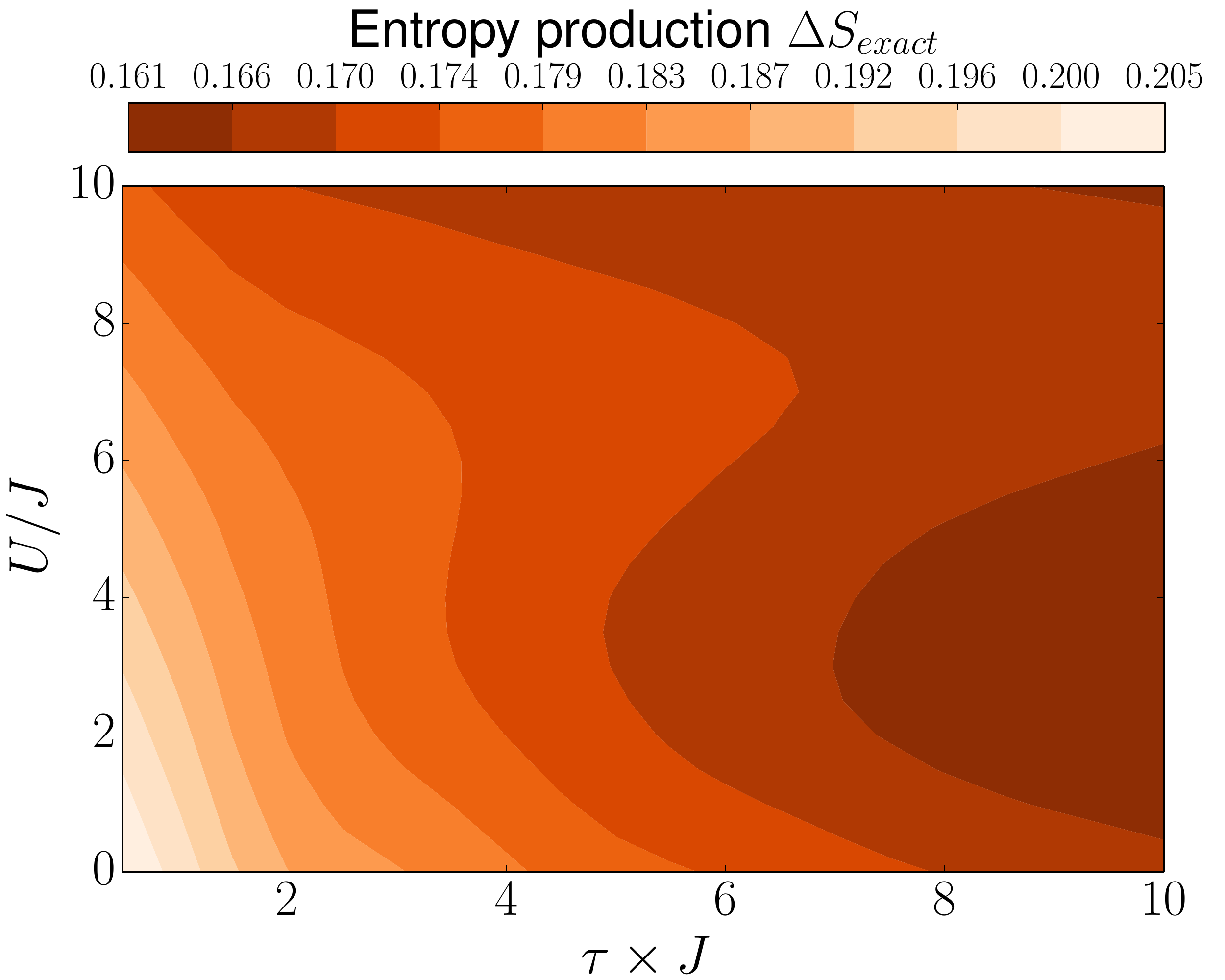}}
\caption{(a)-(c) Variation of free energy $\Delta F$ versus $U$ for 6-site chains at low (green), medium (blue), and high (red) temperatures and for the three driving potential (as indicated). (d)-(l) Exact entropy variation $\Delta S$ versus $\tau$ ($x$-axis) and $U$ ($y$-axis), for 6 site chains, with MI (left column), `comb' (middle) and AEF (right column) driving potential; temperatures as indicated.  Darker colour shades correspond to lower entropy production, whilst lighter to higher entropy production.}
\label{fig:ex_entropy}
\end{figure*}

Figure~\ref{fig:ex_entropy}, panels (a)-(c), shows the variation of free energy as $U$ varies at each of the temperatures considered (green line for low, blue for medium, and red for high temperature) and for each of the driving potentials.  $\Delta F$ is weakly dependent on $U$  for high  temperatures, while, at intermediate and, even more, at low temperatures, it significantly changes with the interaction strength. This confirms that the system behaves more and more like a non-interacting system as the temperature increases.

The exact entropy variation for all driving potentials and temperatures is shown in figure~\ref{fig:ex_entropy}, second to fourth rows: left column for MI, middle for `comb', and right for AEF driving potential.
$\Delta S$ increases as the colour shade becomes lighter; note however that the same shade corresponds to different values in different panels, as the overall entropy range significantly changes according to both temperature and type of driving potential.

The temperature affects the entropy production drastically, compare the extent of $\Delta S$ ranges between the second and last row of  figure~\ref{fig:ex_entropy}. This can be understood by comparing energy scales.  By the end of the dynamics, our driving potentials reach the maximum energy difference of $10J$ for comb, $10.5J$ for MI, and $21J$ for AEF potential.  For the low temperature $k_B T = 0.2J$ and the range of parameters explored, both the interaction strength $U$ and the driving potentials can be much bigger than the thermal energy, and so they have a large impact on the system evolution. The system can change quite drastically leading to the possibility of large work extraction and large entropy production.
However at high temperature, $k_B T = 20J$, the interaction strength and driving potential are, {\it at most}, comparable to the thermal energy: the system is not as receptive to being driven, it will remain closer to its thermal state, and hence the energy required to be dispersed to return to equilibrium (i.e. the entropy that we are here considering) will be much less.

In general at the lowest temperature the sudden quench with weak-medium strength coupling parameter region corresponds to very high $\Delta S$ values, while a dramatic reduction in entropy is seen moving towards the adiabatic regime.
Given the correspondence between $\Delta S$ and the heat to be dispersed at the end to recover equilibrium,  it stands to reason that a sudden quench would require a larger dissipation of energy to return to an equilibrium state compared to an adiabatic evolution, which remains closer to an equilibrium state at all times.

We note that systems subject to AEF potential show at low $T$ a relatively larger entropy production in the strongly-coupled regime and $\tau\sim 10/J$ than systems subject to `comb' and MI potentials. As discussed in section~\ref{adiab_ex}, the level of adiabaticity reached for the same value of $\tau$ differs with driving potential. Indeed the dynamic induced by AEF at $U \gtrsim 6J$ and $\tau\sim 10/J$ is less adiabatic than the ones by MI or `comb', as can be observed by comparing panel (c) to panels (a) and (b) of  figure~\ref{fig:ex_work}.
This leads to a larger amount of entropy production occurring with AEF even in this strongly-coupled large-$\tau$ region.

\subsection{Non-interacting approximation}

We extend the approximations for the extracted work to the entropy production. $\Delta S^{NI}$ will be calculated from $\langle W^{NI} \rangle= - \langle W_{ext}^{NI}\rangle$ and by setting $U=0$ in the calculation of the free energy
($U=0$ values in figure~\ref{fig:ex_entropy}, upper panels).

\begin{figure*}
\centering
\subfloat[$T = 0.2J/k_B$]{\includegraphics[width=0.3\textwidth]{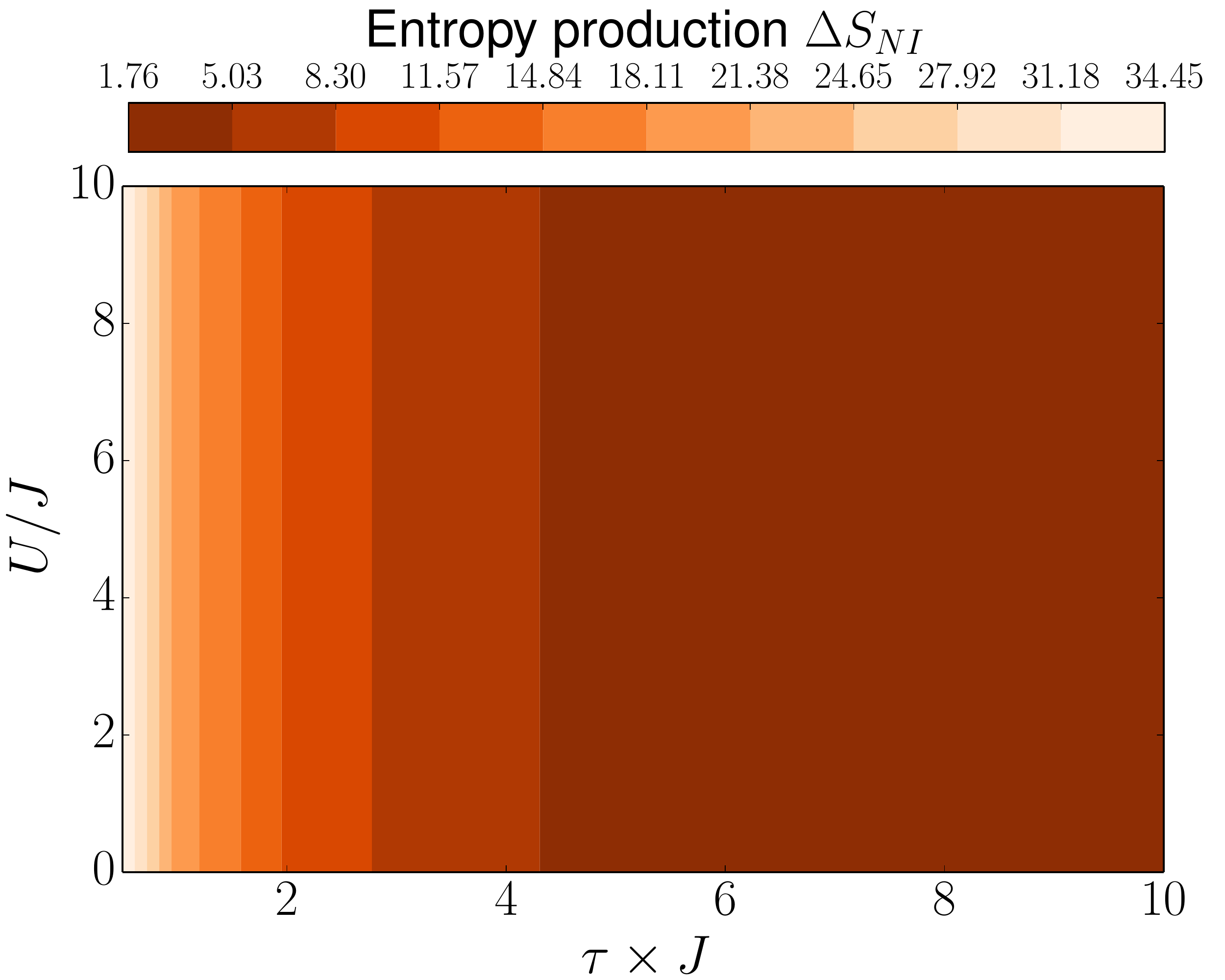}}
\subfloat[$T = 2.5J/k_B$]{\includegraphics[width=0.3\textwidth]{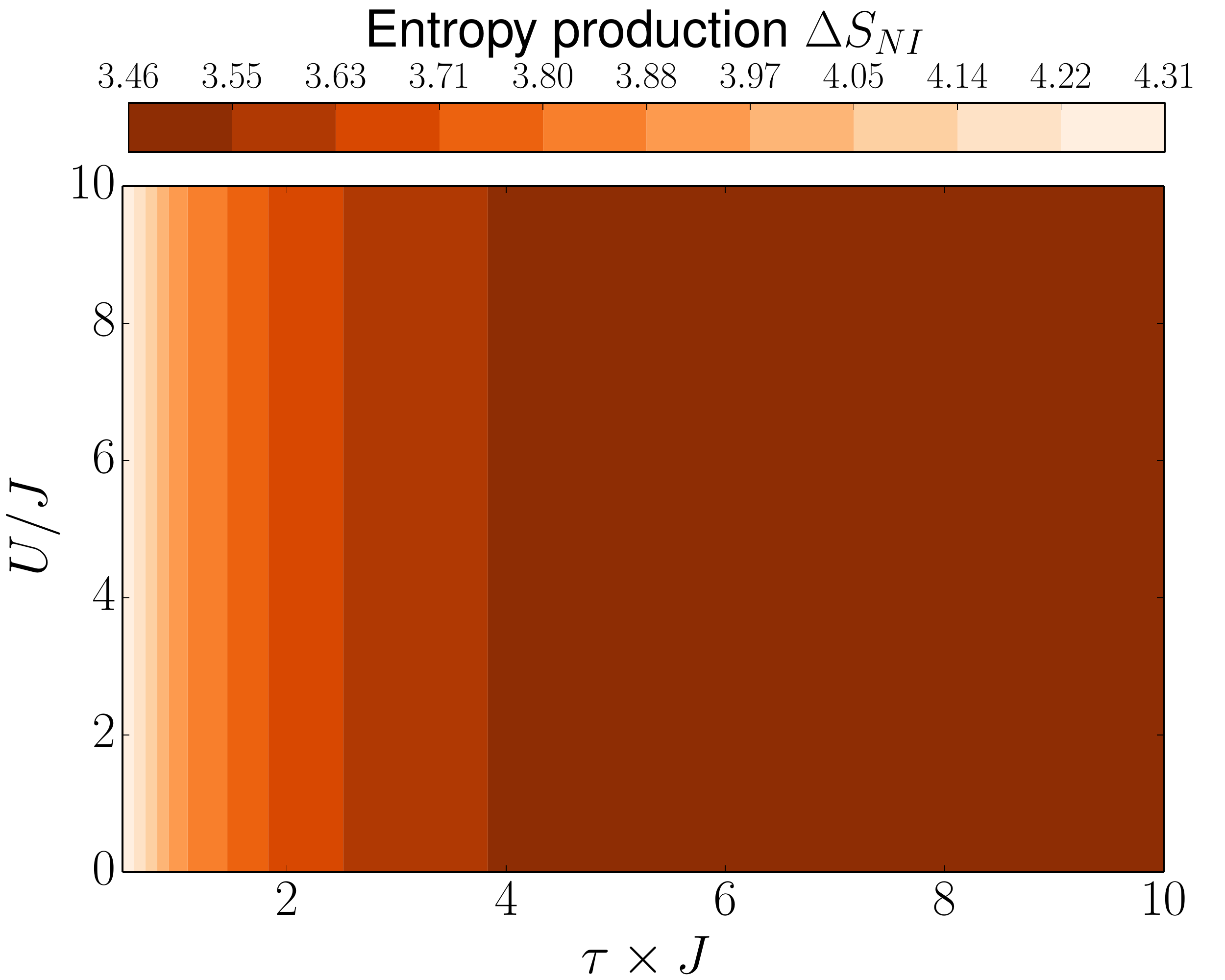}}
\subfloat[$T = 20J/k_B$]{\includegraphics[width=0.3\textwidth]{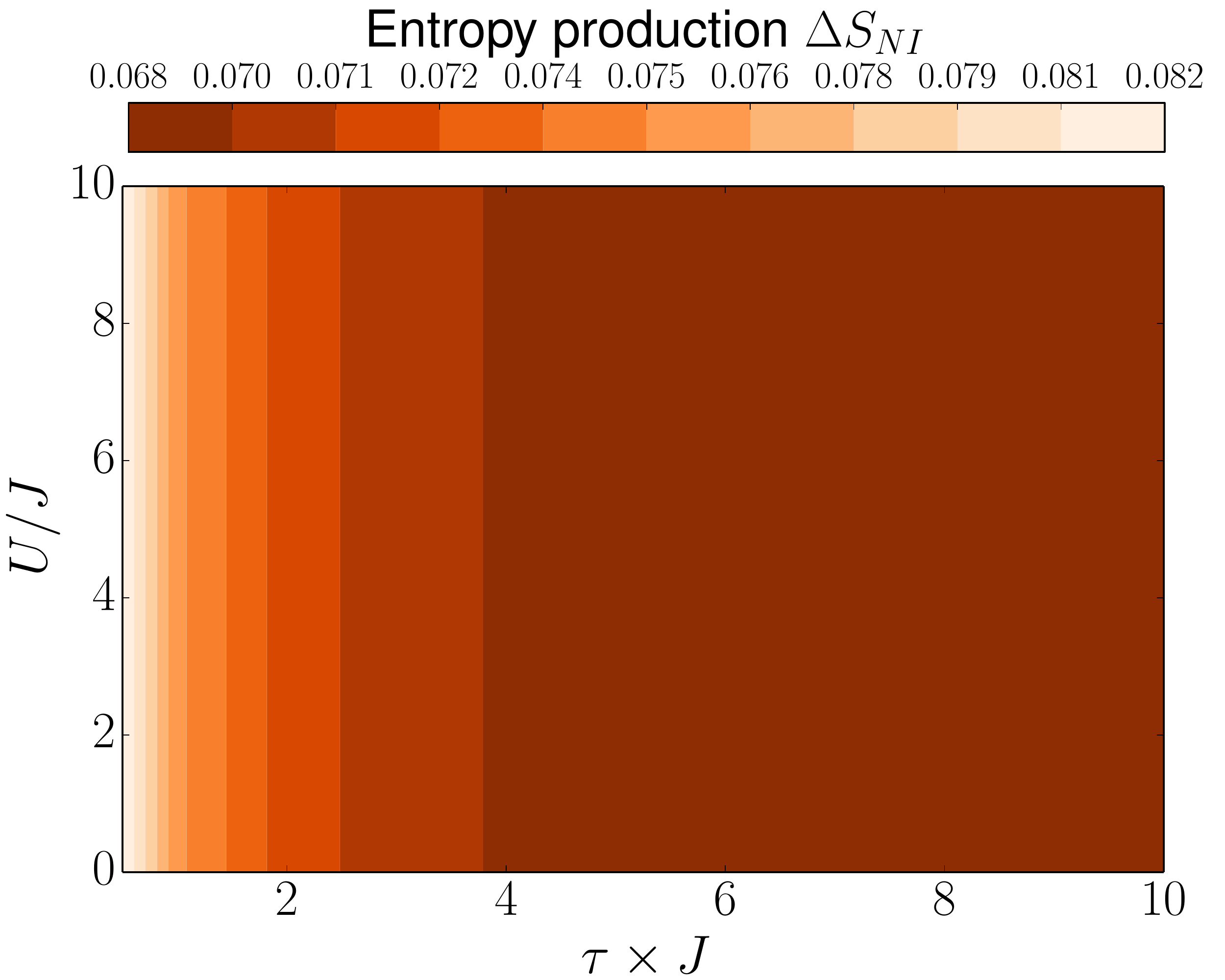}}

\centering
\subfloat[$T = 0.2J/k_B$]{\includegraphics[width=0.3\textwidth]{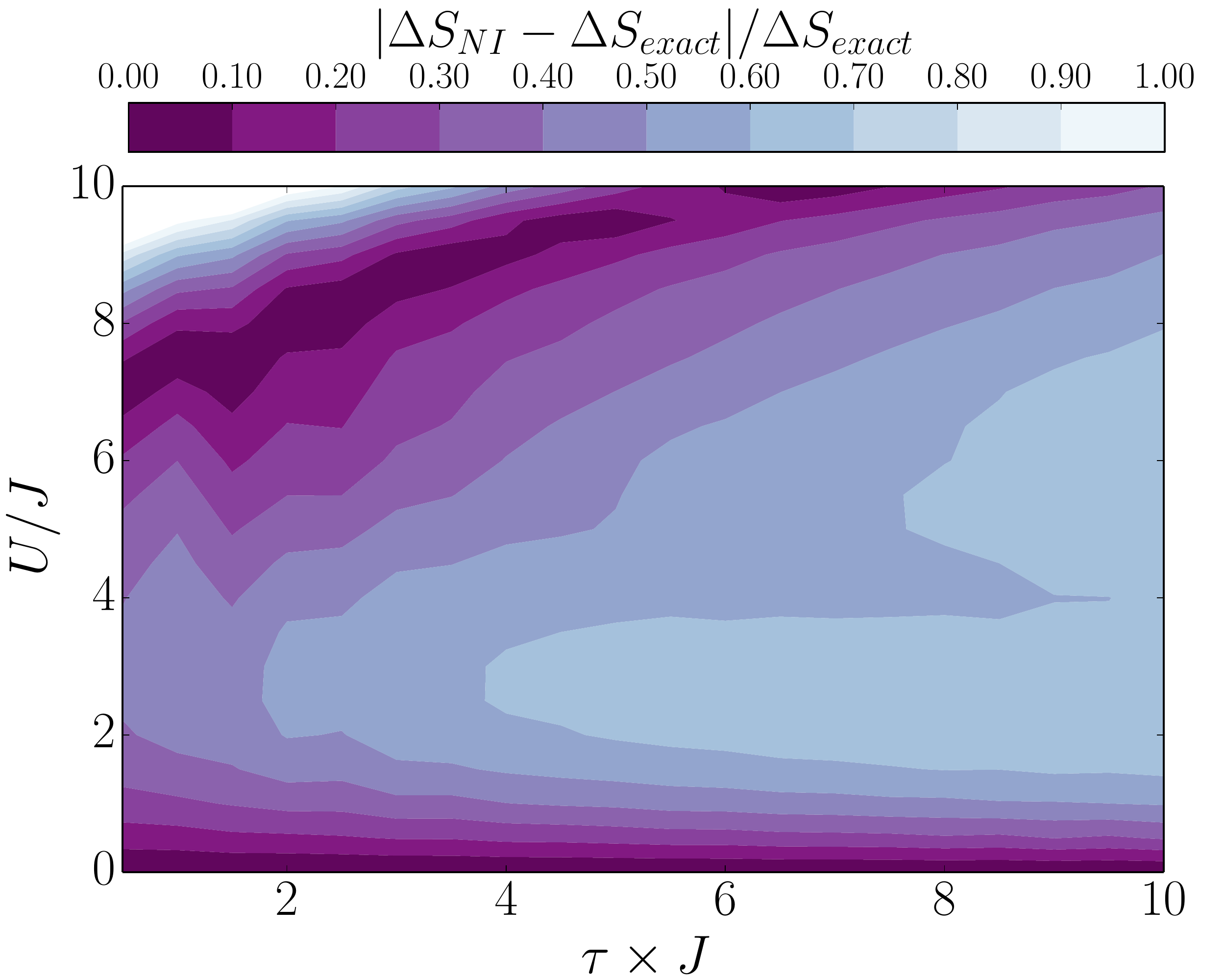}}
\subfloat[$T = 2.5J/k_B$]{\includegraphics[width=0.3\textwidth]{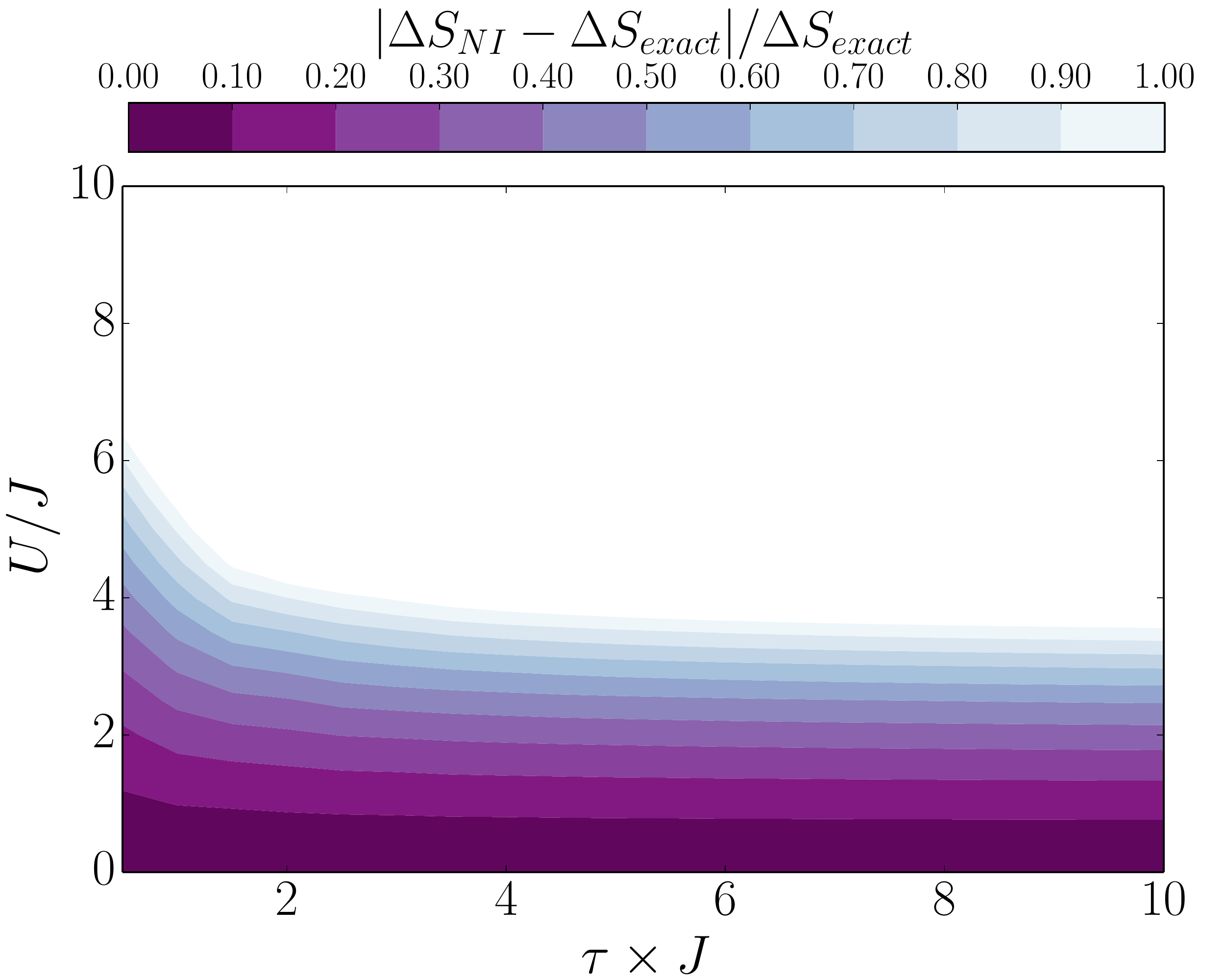}}
\subfloat[$T = 20J/k_B$]{\includegraphics[width=0.3\textwidth]{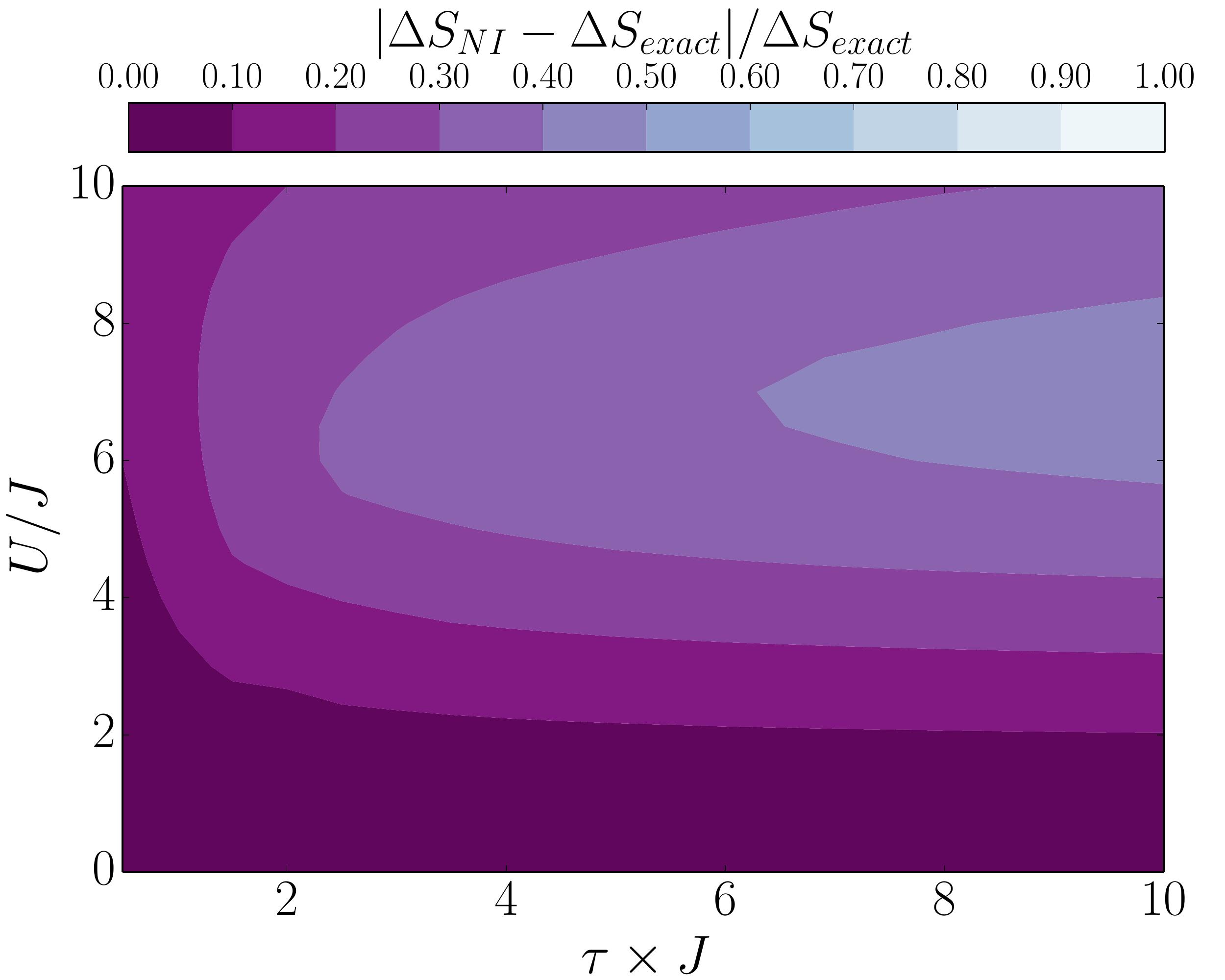}}
\caption{Upper panels: Non-interacting entropy production versus $\tau$ ($x$-axis) and $U$ ($y$-axis) for 6-site chains with comb driving potential.
Lower panels: Non-interacting entropy production relative difference for the same parameters as the upper panels.}
\label{fig:NI_entropy}
\end{figure*}

Figure~\ref{fig:NI_entropy} presents the $\Delta S^{NI}$ results for 6 sites, `comb' driving potential, and increasing temperature (left to right). The upper panels show the non-interacting entropy variation, and the lower ones the relative difference with the exact entropy variation (the darker the purple, the more accurate the approximation is in that region).

For each given temperature, the entropy $\Delta S^{NI}$ is just $\langle W_{ext}^{NI}\rangle$ with an added constant, so, much like the non-interacting work, for all driving potentials this approximation is unable to {\it qualitatively} describe the exact entropy produced.

Comparing with the NI-work approximation accuracy, the overall {\it quantitative} accuracy of the entropy is in general reduced for all three temperatures, which is to be expected since, on top of the work, we are severely approximating the free energy as well. As $U$ increases, we are exploring highly correlated systems, including systems subject to a precursor to the Mott-insulator transition, but these Coulomb correlations are completely neglected by the NI approximation.

For the AEF driving potential, the NI approximation works better then for `comb' drive at all temperatures, giving a 10\% accuracy (or better) for $U\lesssim1.5J$ at low and intermediate temperatures and for almost all regimes at high temperatures. For the MI driving potential, the accuracy of the results is comparable to the `comb' driving potential.

\subsection{Exact initial state with non-interacting evolution}

Let us now see how considering the exact initial state affects the estimate of the entropy production. Initially the entropy $\Delta \tilde{S}^{exact+NI}$ is calculated from (\ref{eq:entropy}) using
$\langle W^{exact+NI} \rangle= - \langle W_{ext}^{exact+NI}\rangle$ and the {\it exact} free energy variation: in this approximation, we are already assuming that we can diagonalise the initial Hamiltonian to get the exact initial state, we then make the same assumption for $H_f$, as this operation would have the same calculation costs/difficulties. This leads to the exact free energy variation.
However we note that, with the implementation described above, this approximation could lead to the nonphysical occurrence of negative entropy: in fact the two contributions to the entropy have opposite sign, and one of them (the work) has been approximated, so the occurrence of a negative sign cannot apriori be excluded.
We then further  impose that $\Delta S^{exact+NI}=\max\{\Delta \tilde{S}^{exact+NI},0\}$ to correct for it. 

\begin{figure*}
\centering
\subfloat[$T = 0.2J/k_B$]{\includegraphics[width=0.3\textwidth]{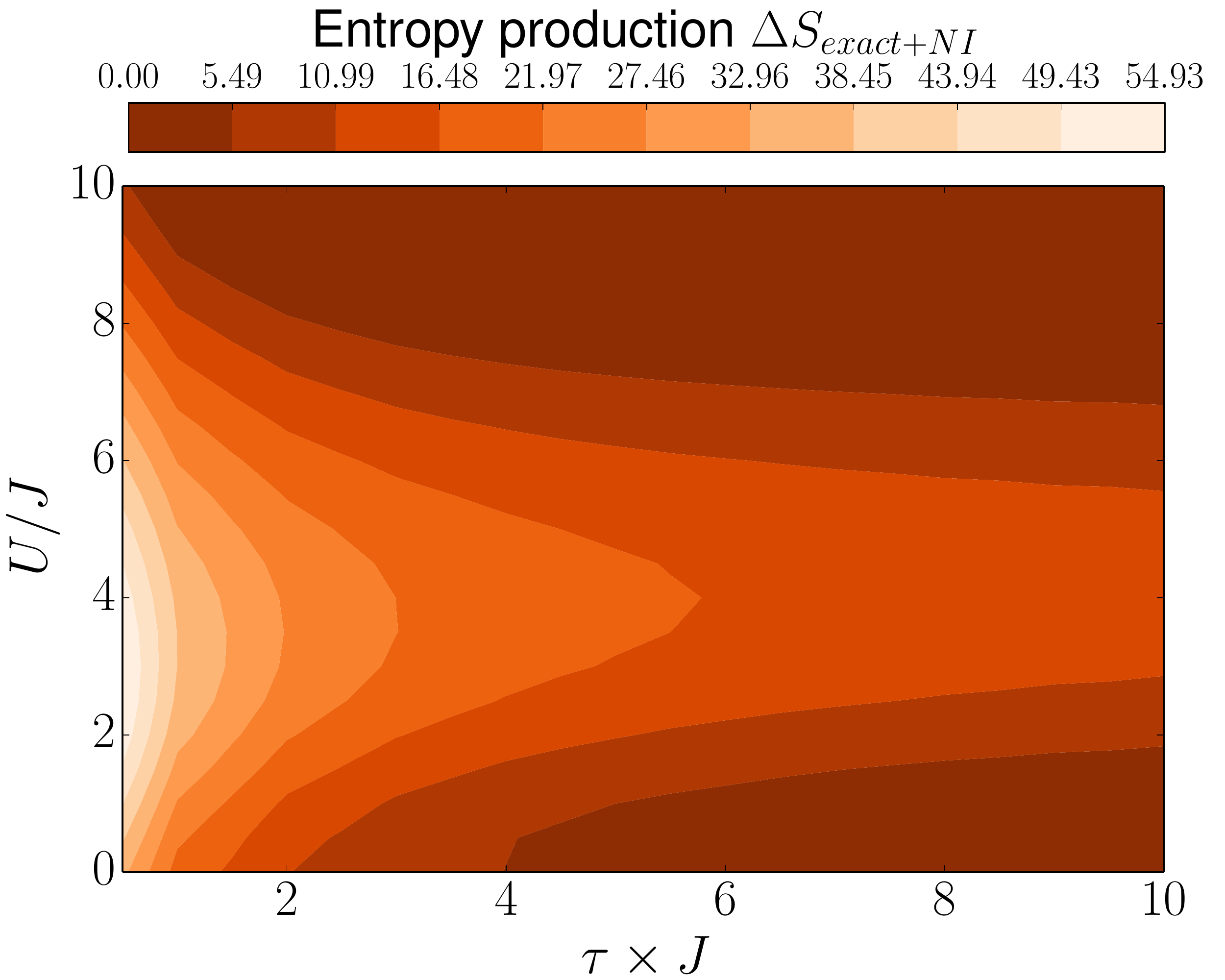}}
\subfloat[$T = 2.5J/k_B$]{\includegraphics[width=0.3\textwidth]{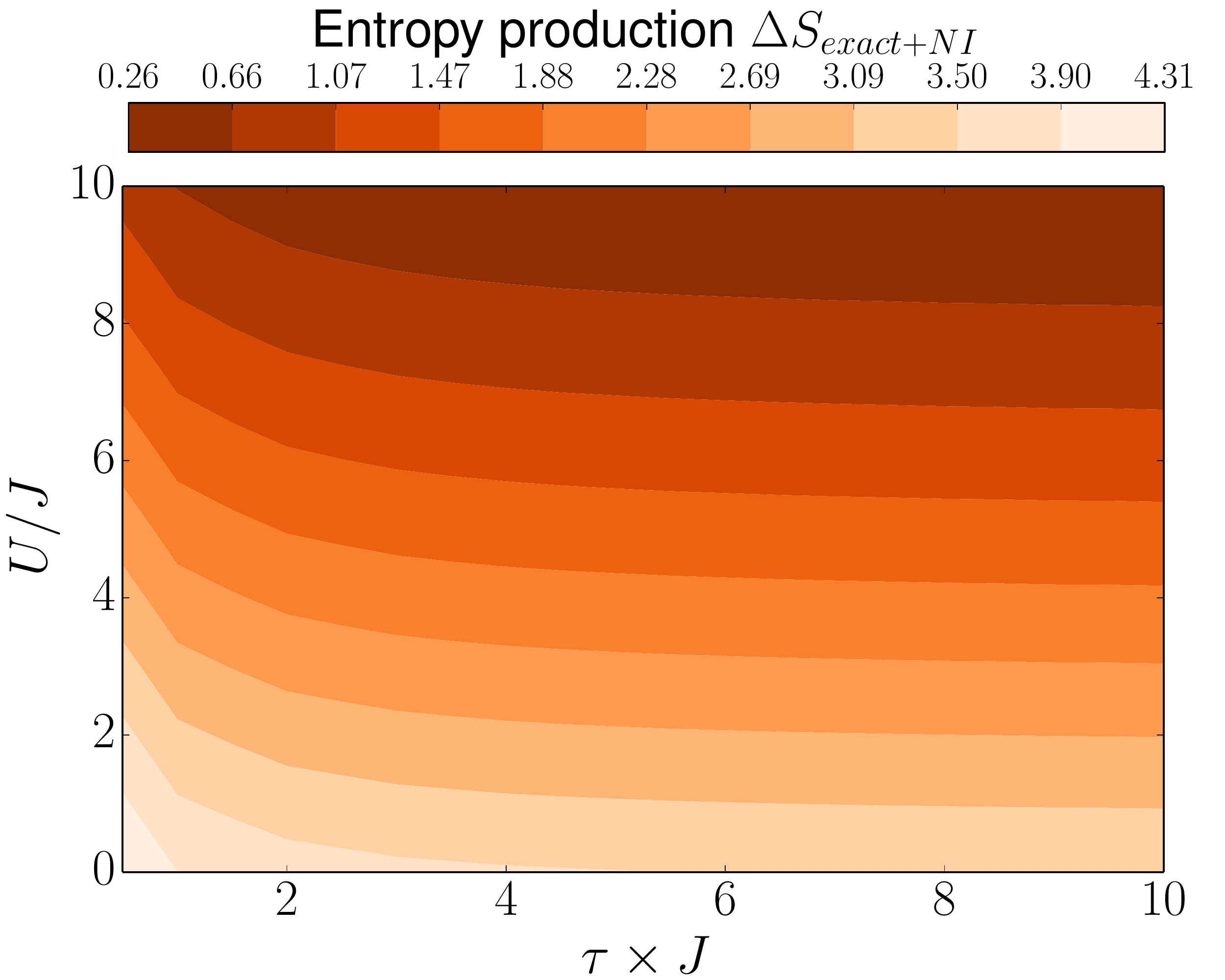}}
\subfloat[$T = 20J/k_B$]{\includegraphics[width=0.3\textwidth]{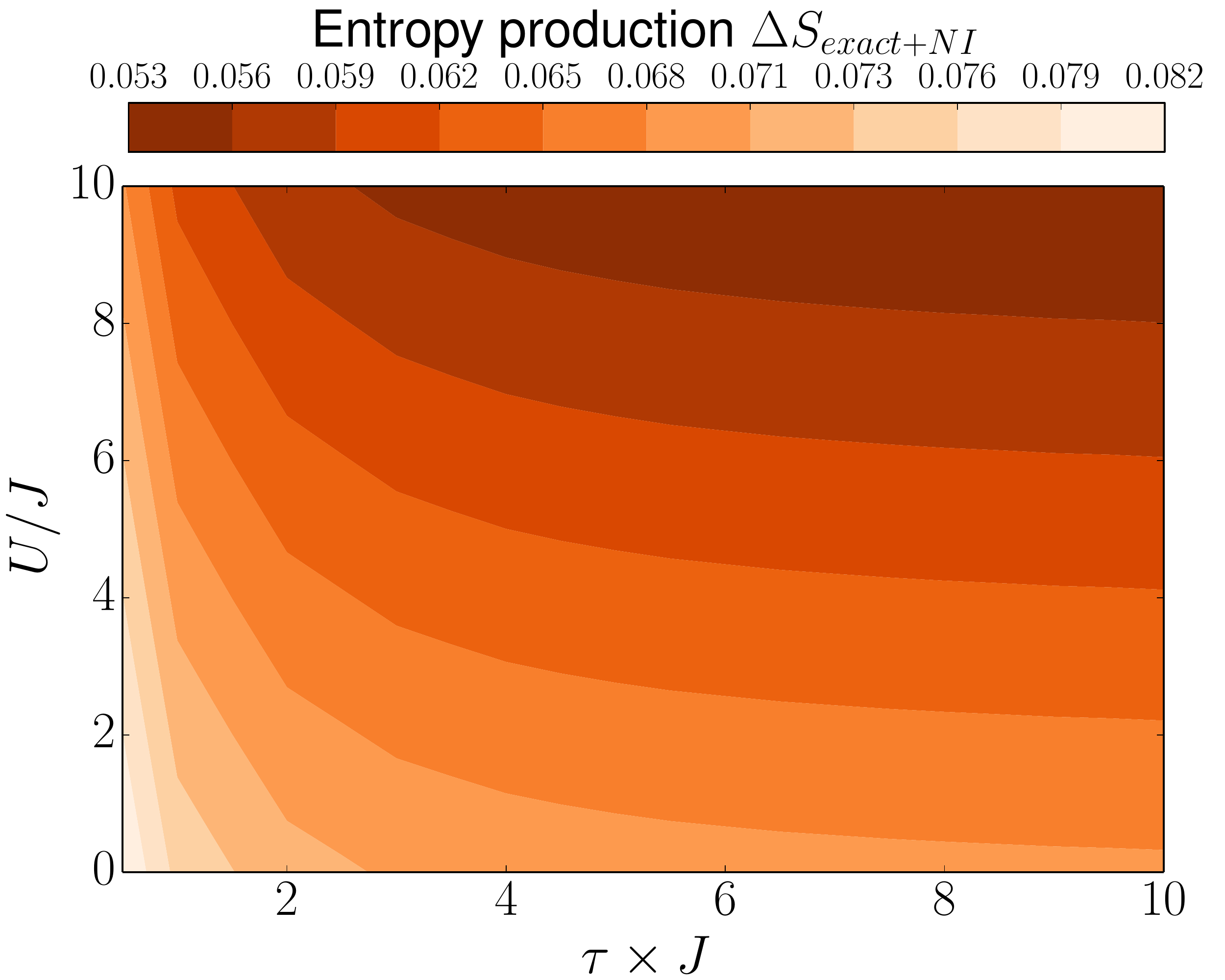}}

\centering
\subfloat[$T = 0.2J/k_B$]{\includegraphics[width=0.3\textwidth]{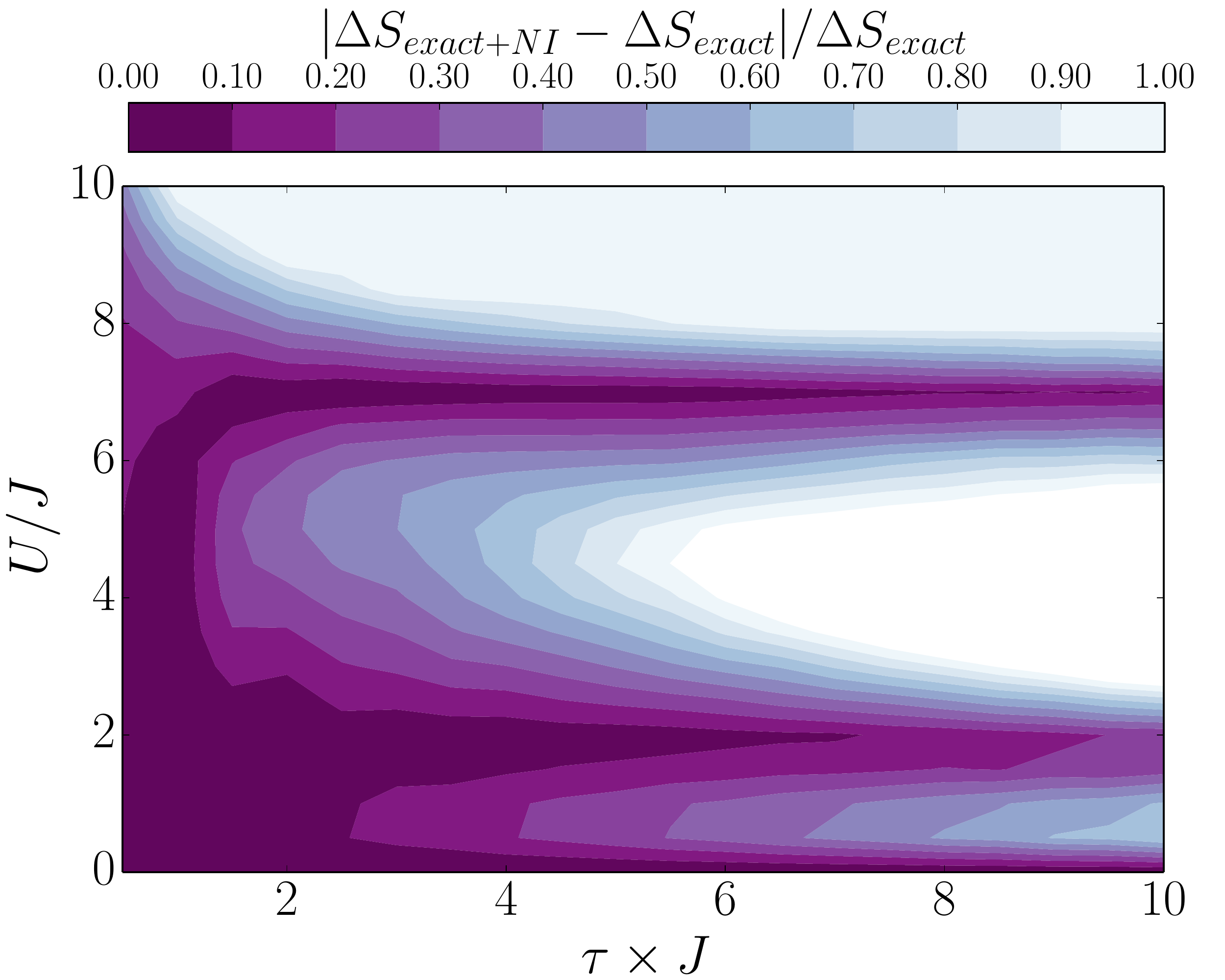}}
\subfloat[$T = 2.5J/k_B$]{\includegraphics[width=0.3\textwidth]{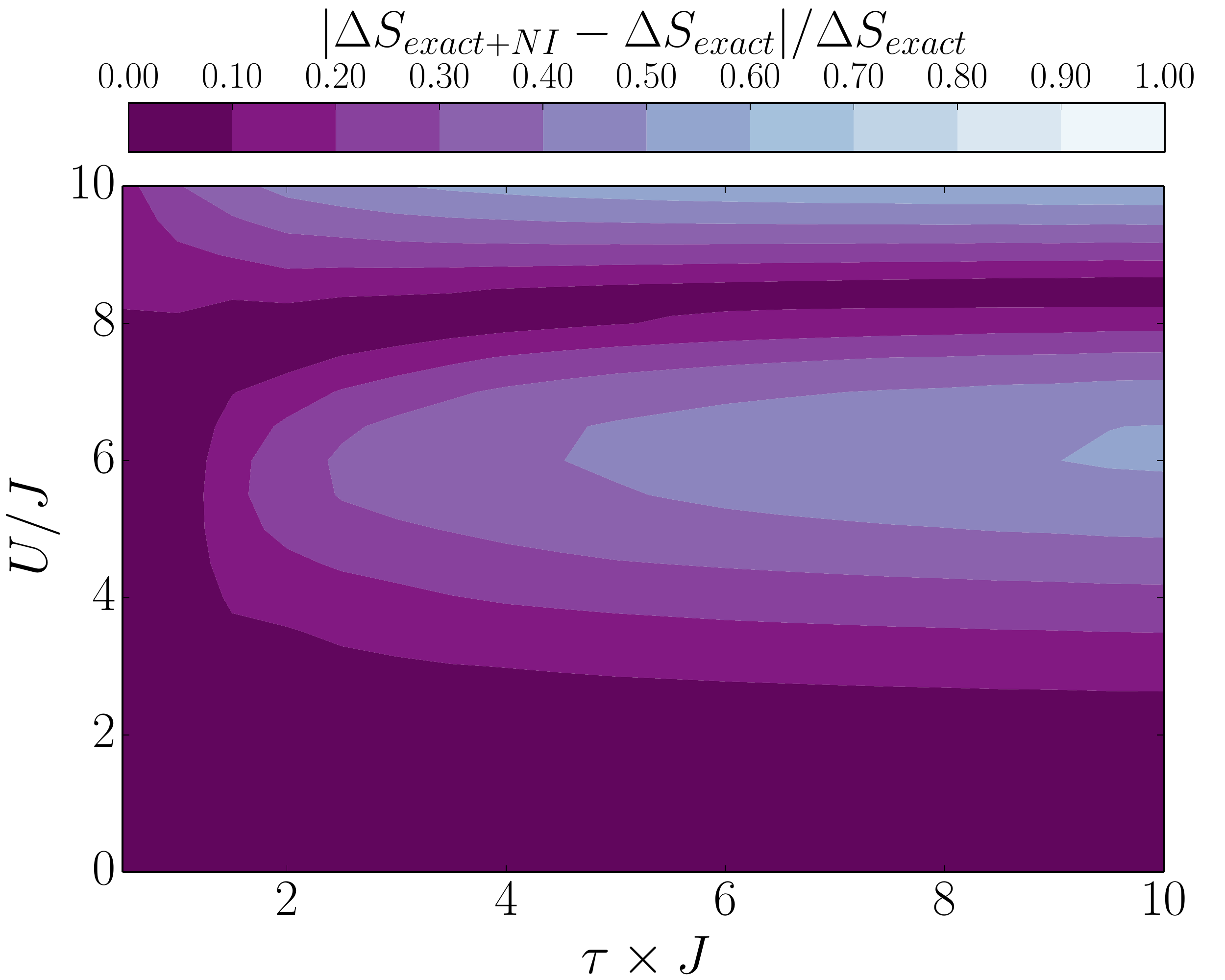}}
\subfloat[$T = 20J/k_B$]{\includegraphics[width=0.3\textwidth]{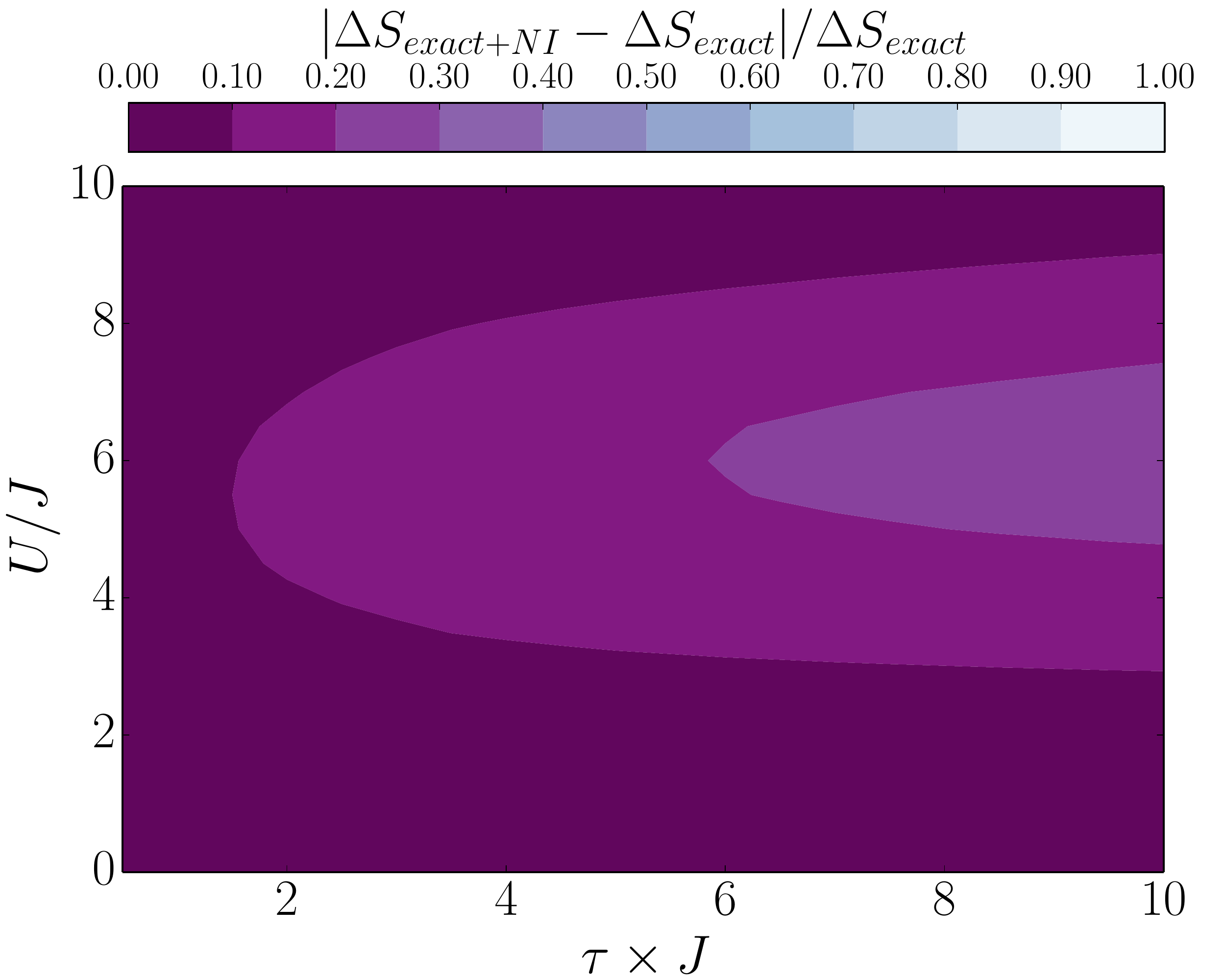}}
\caption{Upper panels: Exact + non-interacting entropy production versus $\tau$ ($x$-axis) and $U$ ($y$-axis) for 6-site chains with comb driving potential.
Lower panels: Exact + non-interacting entropy production relative difference for the same parameters as the upper panels.}
\label{fig:ex+NI_entropy}
\end{figure*}

Related results are plotted in figure~\ref{fig:ex+NI_entropy} for 6 sites, `comb' driving potential, and increasing temperature (left to right).
By comparing the upper panels of figure~\ref{fig:ex+NI_entropy} to the upper panels of figure~\ref{fig:NI_entropy} and to the corresponding ones in the mid column of  figure~\ref{fig:ex_entropy}, we note a marked improvement in the {\it qualitative} behaviour of the approximation. As with the work, we can see an improvement  also in the quantitative results.  The high temperature accuracy is very much akin to that of the work, and is accurate within 30\% for all regimes.  As the temperature decreases, however, the quantitative accuracy also decreases: inaccuracy comes into the entropy calculations through the approximation of the average work, so the regimes of greater/lesser accuracy fairly mirror those of the work.

Results for the other two driving potentials confirm these trends and are shown in the appendix.

Similarly to the `comb' potential, the `exact+NI' approximation with the MI and AEF driving potentials recover to a good extent the qualitative  behaviour of $\Delta S^{exact}$ for low and intermediate temperatures. For high temperatures the qualitative behaviour is recovered only for $U\lesssim2$.
Quantitatively, the areas of worse performance are related to the areas of worse performance for the corresponding $\langle W_{ext}^{exact+NI}\rangle$, however the approximation performs worse for $\Delta S^{exact+NI}$ than $\langle W_{ext}^{exact+NI}\rangle$ for MI, and better for AEF. Overall the approximation improves its quantitative performance with temperature, as it reproduces well the limits of the entropy variation range, and especially so at high temperature.

\section{Conclusion}

We have presented a comprehensive study of the extracted average quantum work and entropy variation in complex many-body systems subject to a wealth of driving dynamics and dynamical regimes.
We have discussed in details the effects of the interplay of the different energy scales governing the systems -- driving potentials, many-body interactions, and thermal energy -- on $\langle W \rangle$ and $\Delta S$, and compared results as many-body correlations are turned on up to the strongly coupled regime and as the dynamical regime is continuously  changed from sudden quench to nearly adiabatic.

For all driving potentials and low to intermediate temperatures, intermediate to strong Coulomb repulsion decreases work extraction by making the system less responsive to the external drive. At weak Coulomb correlations, more work can be extracted as the system becomes adiabatic, while with stronger many-body interactions the work production becomes independent of the overall driving time $\tau$ much sooner. For the same parameter sets, at high temperatures, work extraction is greatly reduced both in absolute values and in variation range.

At low temperatures the entropy variation presents a behaviour somewhat more dependent on the applied driving potential. We observe a general tendency of lower $\Delta S$ values for weak and strong Coulomb correlations and intermediate to large $\tau$'s, but the onset of an adiabatic-like entropy dynamics varies considerably with the driving potential at intermediate coupling strengths. For a zero temperature, open-boundary, finite, homogeneous Hubbard chain, intermediate coupling strengths correspond to the transition between metallic and insulating behaviour (precursor to a Mott insulator transition). The behaviour observed at intermediate coupling strengths may be a signal of how the different driving potentials affect this transition.

Similarly to the extracted average quantum work, entropy production decreases with temperature in both absolute value and range of variation, as the system becomes less and less responsive to the applied potential.

The strong effect of the Coulomb interaction on the extracted average quantum work can be also appreciated by comparing exact results to the corresponding non-interacting approximations. Apart from the obvious independence on $U$ for all regimes, this approximation strongly overestimates the minimum work that can be extracted from a system, as it cannot reproduce the freezing of the system dynamics due to strong many-body correlations.  On the other hand, the value of the maximum average work extracted is well captured, albeit often it is attributed to the wrong parameter regions.

Including many-body interactions exactly for complex system is often an hopeless task. In this paper we have proposed a relatively simple approximation which relies on been able to provide an accurate approximation for the system's initial state, while interactions are completely neglected in the dynamics.

We found this approximation to behave surprisingly well. Including interactions just through the initial state recovers the qualitative (and in great part quantitative) behaviour at low and intermediate temperatures for all regimes, driving potentials and chain lengths considered. The greatest improvement is seen at low temperatures, where, for example, with the `comb' driving potential the `exact+NI' approximation reproduces the exact work within
10-20\% up to very strong interactions ($U \approx 9J$) for most regimes. At all temperatures, the `exact + NI' approximation reproduces fairly well the range of variation for the value of the average extracted quantum work and in particular it significantly improves the value of its minimum over the NI estimates.

We demonstrate analytically that, independent of system Hamiltonian and driving potential, including interactions in just the initial state is sufficient to recover at low to intermediate temperatures the characteristic dependency on $U$ and $\tau$ in the adiabatic region, which was completely missed by the non-interacting approximation. This is important as this is the region in which the most work can be produced, and the dependency on $U$ is very strong.

We also demonstrate analytically that this approximation recovers the characteristic $U$-dependence of the extracted work in the small-$\tau$ parameter region for all Hamiltonians and driving potentials.

Fully non-interacting estimates of the entropy variation strongly under-perform, both qualitatively and quantitatively. However at high temperatures the exact entropy variation range becomes very small and its quantitative non-interacting estimate is in the right ball-park due to the decreased influence of many-body interactions: as a result, at high temperatures even the non-interacting approximation gives reasonable quantitative (but not qualitative) results.

When we extend the `exact+NI' approximation to the entropy variation, we find qualitative improvements similar to the work extraction, with the behaviour for low and intermediate temperatures largely recovered.
Quantitatively this approximation significantly improves over the non-interacting approximation, albeit not as strikingly as for the average quantum work. Overall the `exact+NI' approximation improves its quantitative performance with temperature, as it captures well the entropy variation range, including at high temperature. As an example, at high temperature and MI driving potential, this approximation would reproduce exact results within 10\% for all parameter range, to be contrasted with the performance by the non interacting approximation which gives such an accuracy only for $U\lesssim3$.

Our results show that, even when taking a very crude approximation
for the evolution operator, starting from an accurate initial state is sufficient for greatly improving the estimate of thermodynamical quantities such as the average quantum work and the corresponding entropy variation  for the wide range of driving potentials tested, all temperatures, and the great part of interaction and dynamical regimes.

\begin{acknowledgments}
AHS thanks EPSRC for financial support. KZ thanks the Schlumberger Foundation for financial support through the Faculty for the Future program.  IDA acknowledges support from the Conselho Nacional de Desenvolvimento Cientfico e Tecnologico (CNPq, Grant: PVE Processo: 401414/2014-0) and from the Royal Society (Grant no. NA140436).  We thank M. Herrera and R. M. Serra for useful discussions in the early stages of this work.
\end{acknowledgments}

\appendix
\section{Six-site `exact+NI' approximation results for MI and AEF driving potentials: extracted average quantum work}
To demonstrate how well the the `exact+NI' approximation  fares with other driving potentials, we present here the related MI and AEF driving potential results for 6 site chains and all three temperatures.

\begin{figure*}
\subfloat[$T=0.2J/k_B$.]{\includegraphics[width=0.3\textwidth]{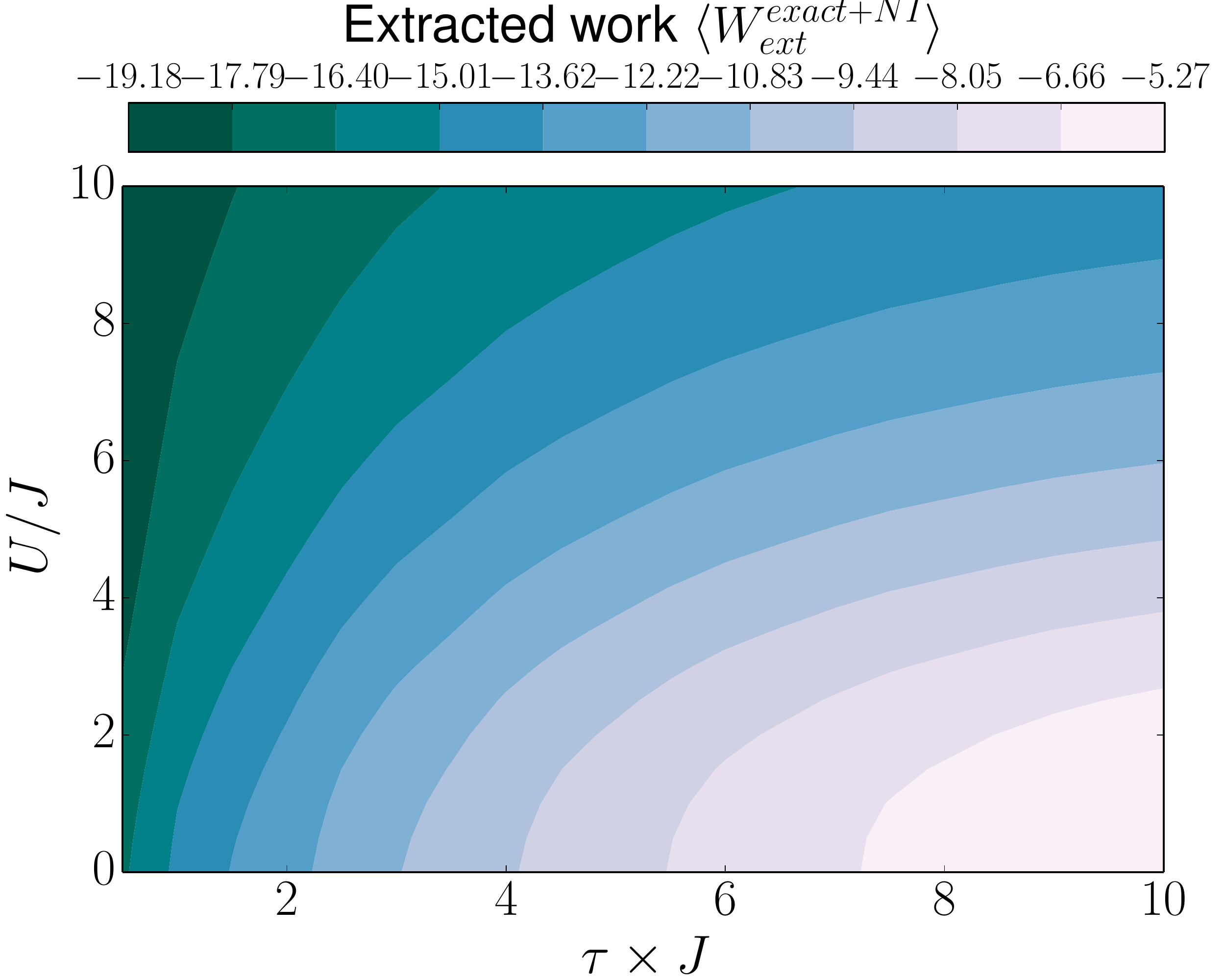}}
\subfloat[$T=2.5J/k_B$.]{\includegraphics[width=0.3\textwidth]{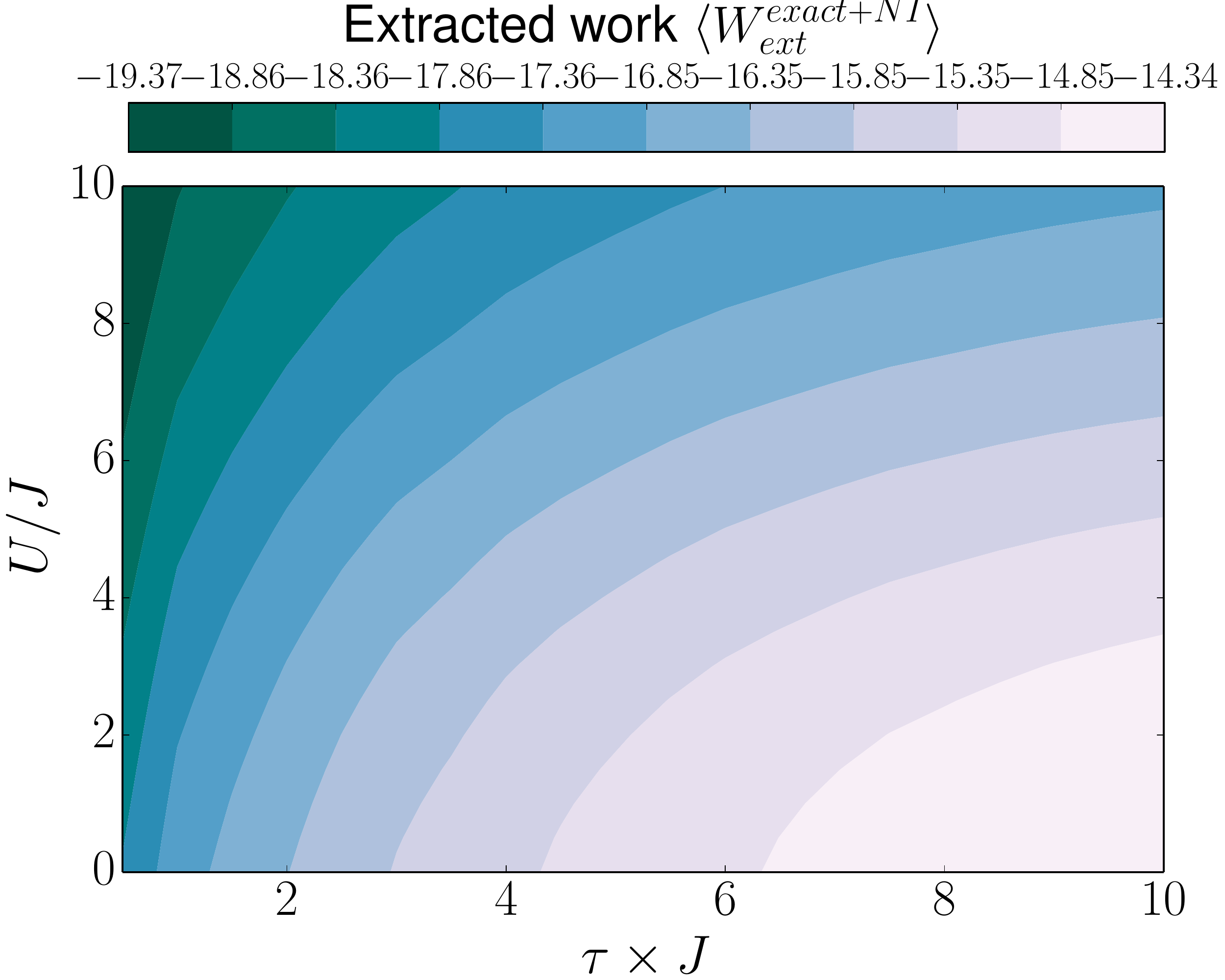}}
\subfloat[$T=20J/k_B$.]{\includegraphics[width=0.3\textwidth]{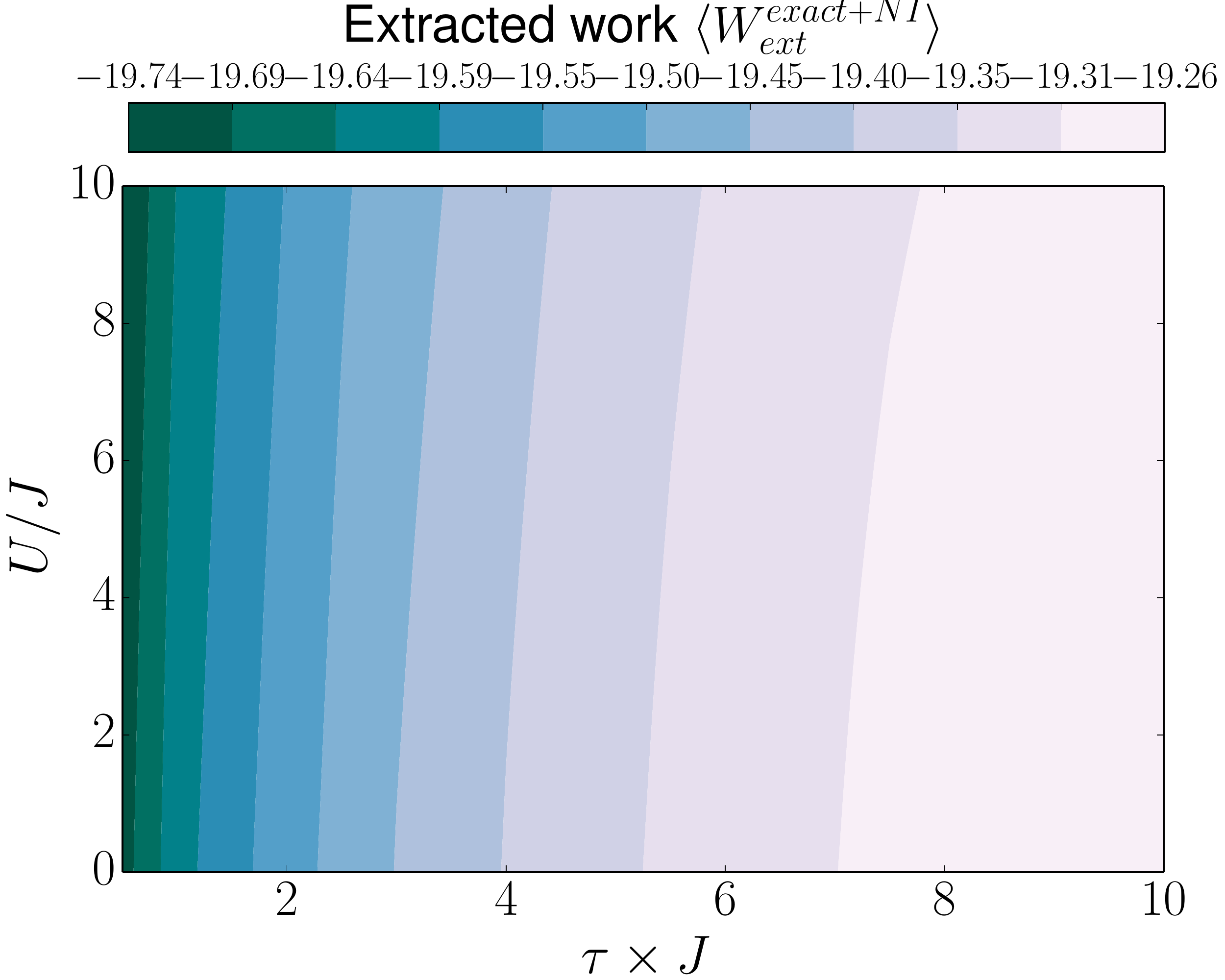}}
\label{fig:ex+NI_work_teeth_extracted}
\subfloat[$T=0.2J/k_B$.]{\includegraphics[width=0.3\textwidth]{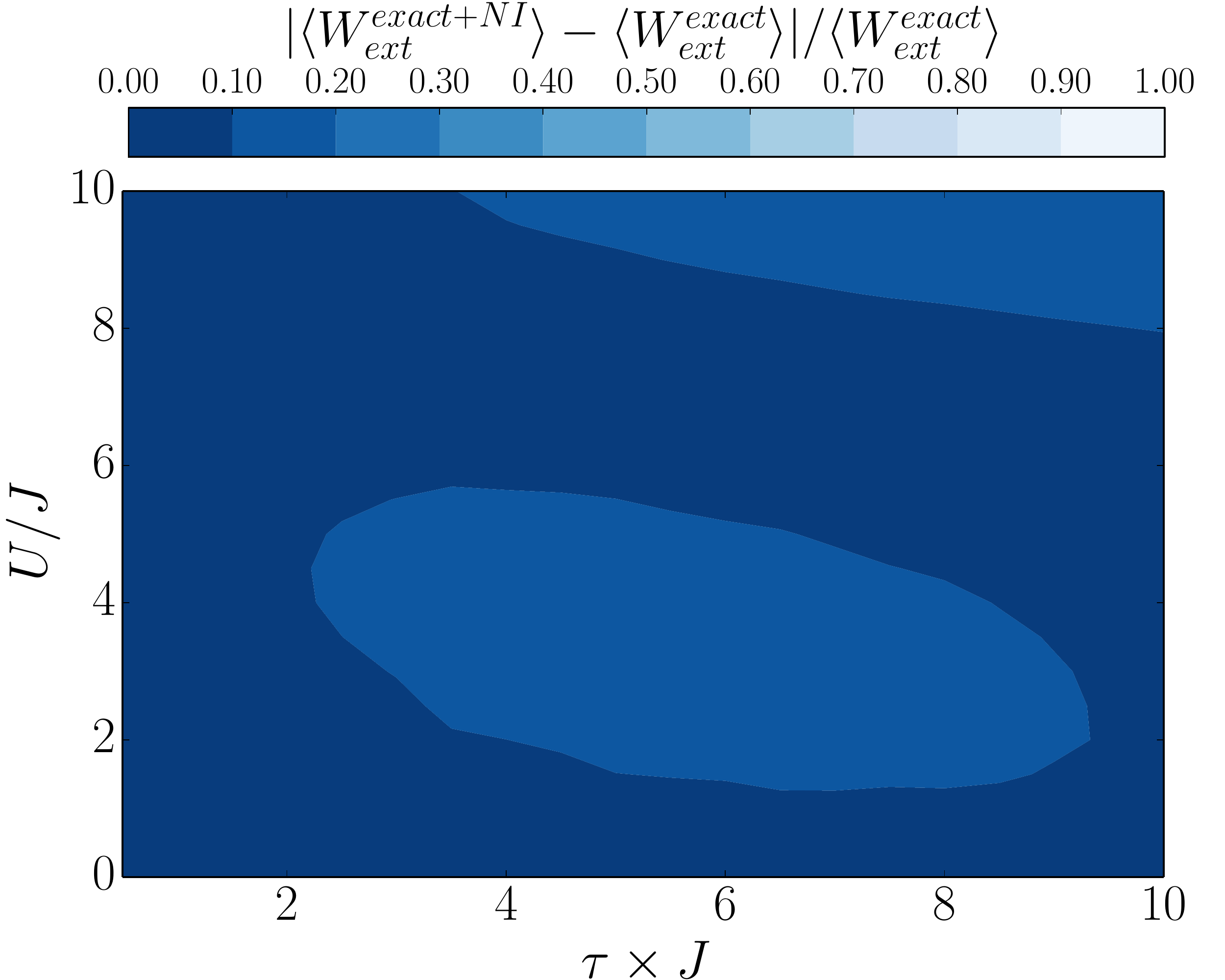}}
\subfloat[$T=2.5J/k_B$.]{\includegraphics[width=0.3\textwidth]{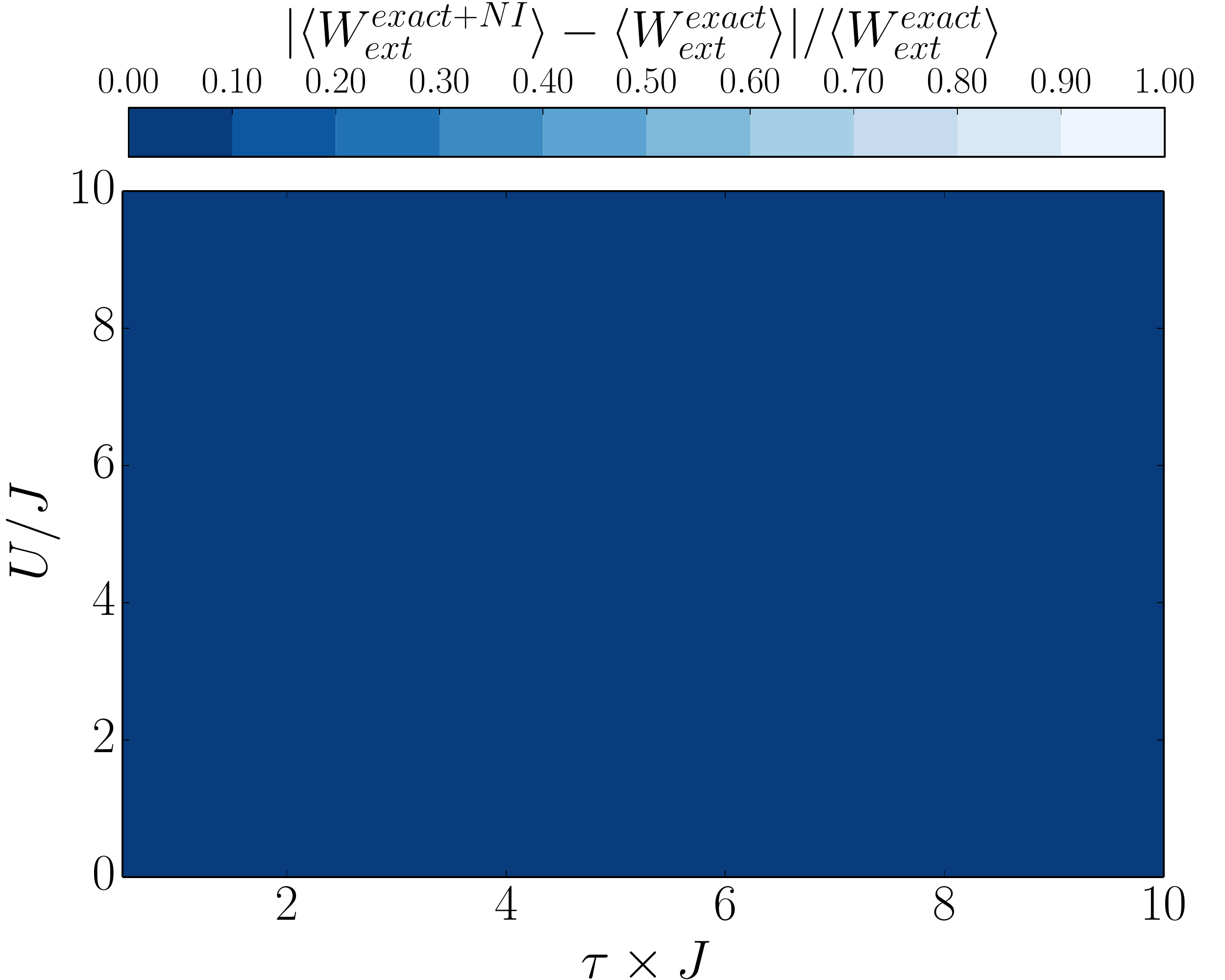}}
\subfloat[$T=20J/k_B$.]{\includegraphics[width=0.3\textwidth]{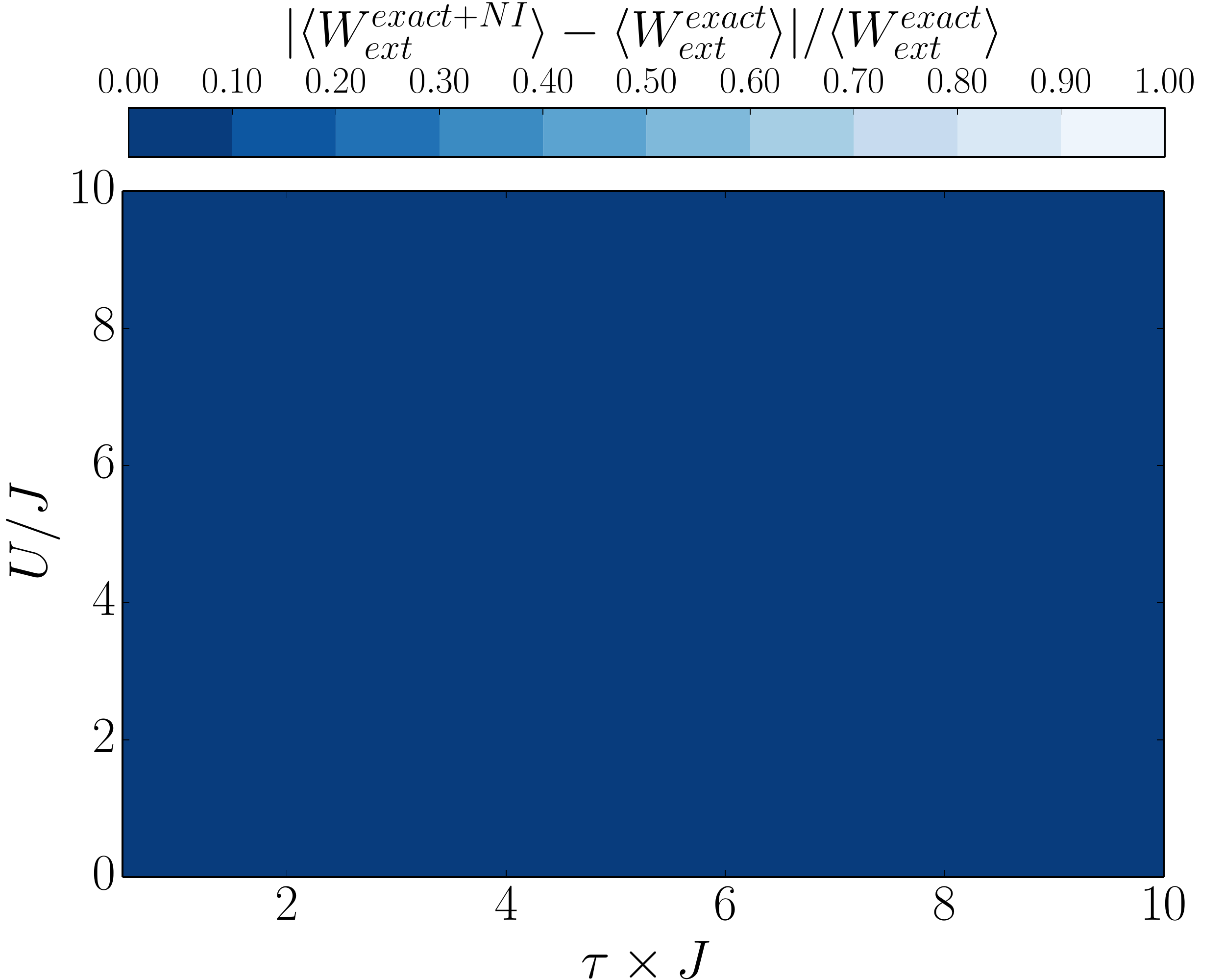}}
\caption{Panels (a) to (c): Extracted work in the `exact + NI' approximation, for 6-site chains undergoing  dynamics driven by the MI potential. The lighter the colour shade, the more work is extracted, compatible with the respective work ranges indicated above each panel.  The regimes go from non-interacting to strongly coupled along the $y$-axis, and from sudden quench closer to adiabatic along the $x$-axis. Temperature increases from left to right panel, as indicated. \\ Panels (d) to (f): Relative error for $\langle W^{exact+NI} \rangle $ with respect to the exact results for the same parameters as the upper panels.  The darker the colour, the more accurate the approximation is in that regime.
}
\label{fig:ex+NI_work_teeth}
\end{figure*}

Figure~\ref{fig:ex+NI_work_teeth} shows results for the MI driving potential: this particular dynamic is captured very well by the `exact+NI' approximation for all temperatures and regimes.  As with all other driving potentials considered, the lowest temperature is where {\it quantitatively} the approximation struggles the most, though its accuracy is, in most regions, within 10\% of the exact work, otherwise it is 20\%.  At higher temperatures the accuracy remains always within 10\%. We note that this approximation recovers to high accuracy and {\it at all temperatures} the overall range of extractable work (compare colour-scale range above panels (a) to (c) in figure~\ref{fig:ex+NI_work_teeth} with the corresponding ranges in panels (a), (d), and (g) of figure~\ref{fig:ex_work}).
 This approximation also recovers most of the work {\it qualitative} behaviour, at least at low and intermediate temperature.
At high temperature, although the trends of the exact and approximate results  are qualitatively different for $\tau\gtrsim2/J$ and $U\gtrsim J$, the exact amount of work extracted  changes very little with $\tau$ and $U$ (panel (g) of figure~\ref{fig:ex_work}), so, as the quantitative range of the approximated work is close to the exact one, the results in  figure~\ref{fig:ex+NI_work_teeth}(c) show high quantitative accuracy.

\begin{figure*}
\centering
\subfloat[$T=0.2J/k_B$.]{\includegraphics[width=0.3\textwidth]{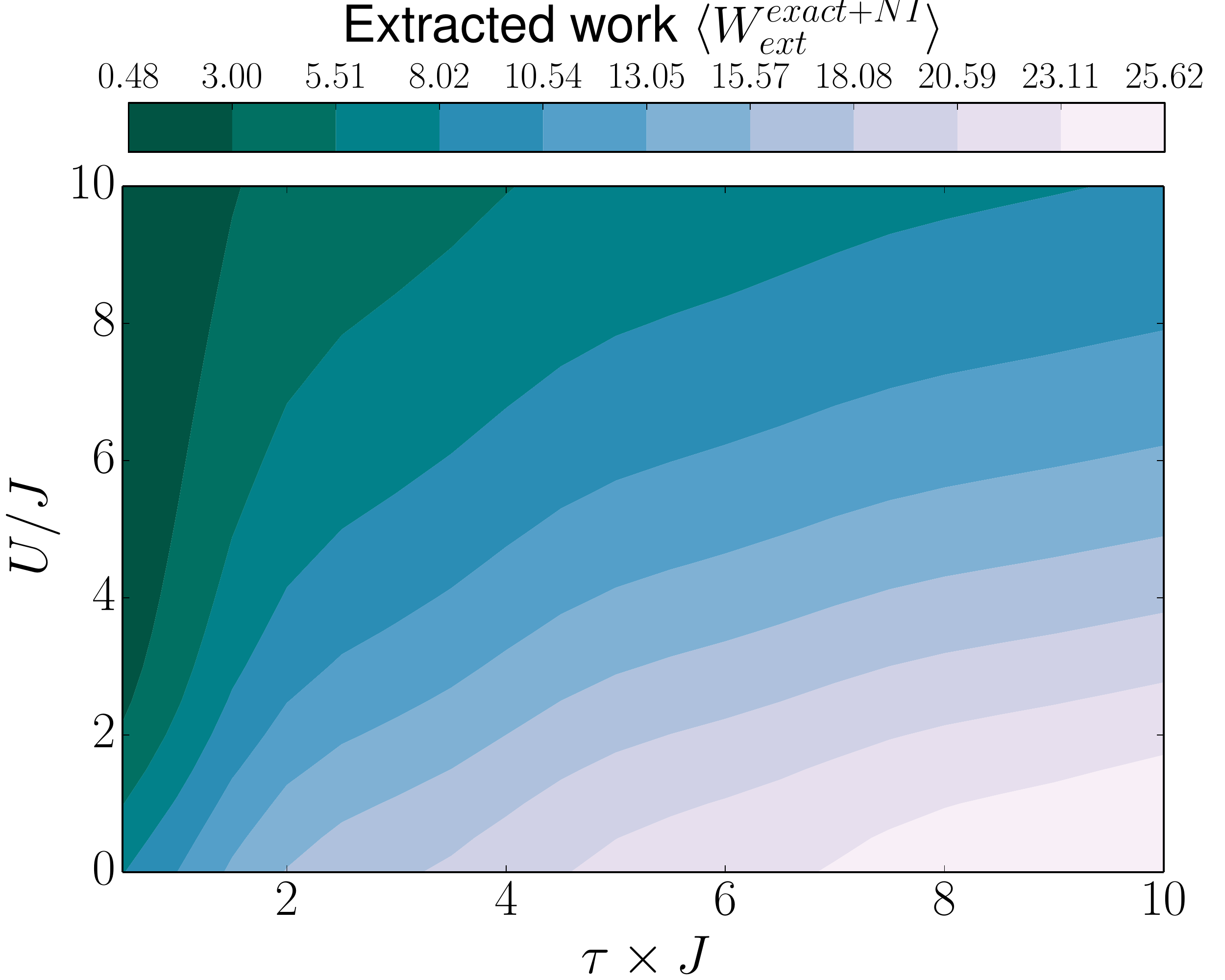}}
\subfloat[$T=2.5J/k_B$.]{\includegraphics[width=0.3\textwidth]{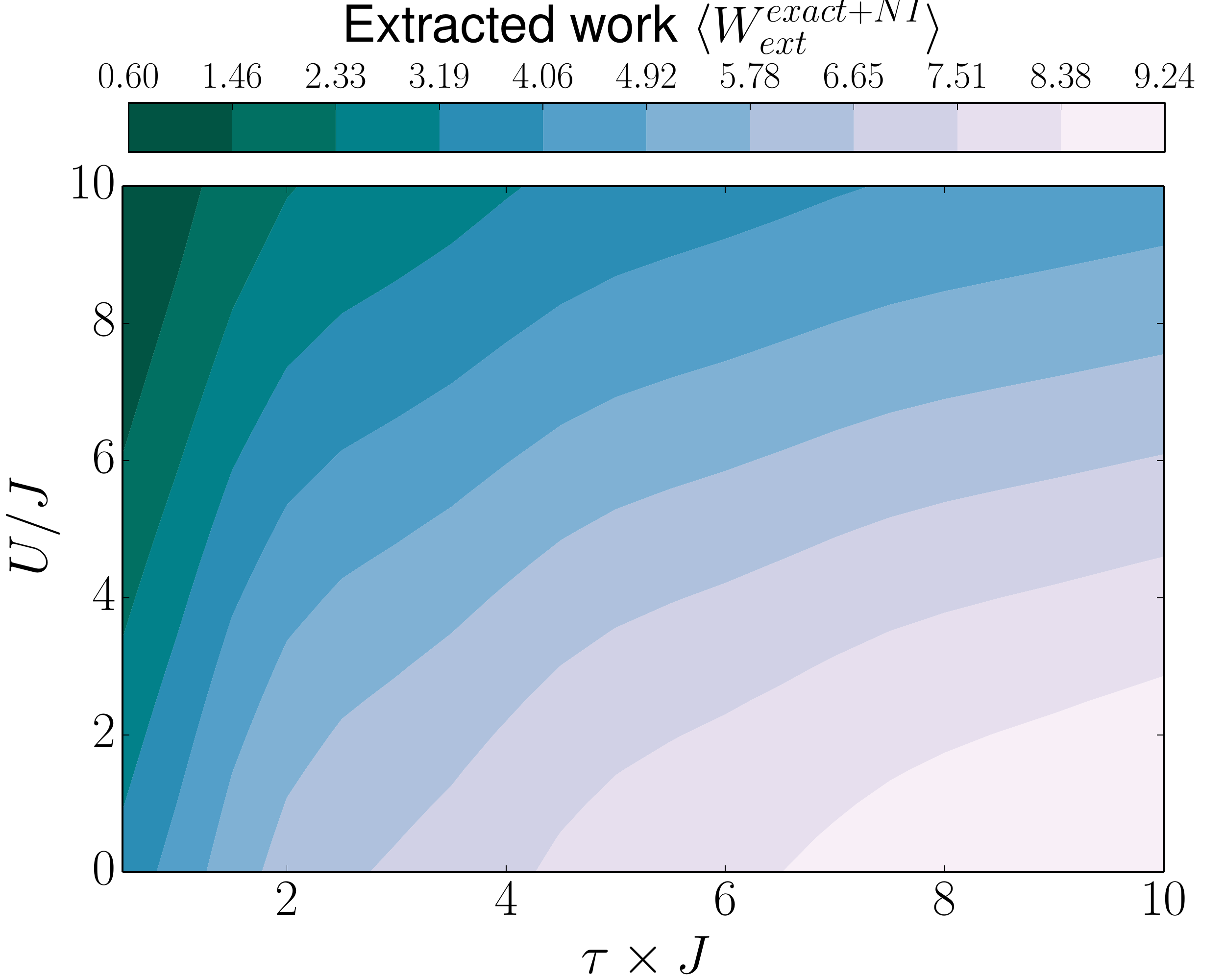}}
\subfloat[$T=20J/k_B$.]{\includegraphics[width=0.3\textwidth]{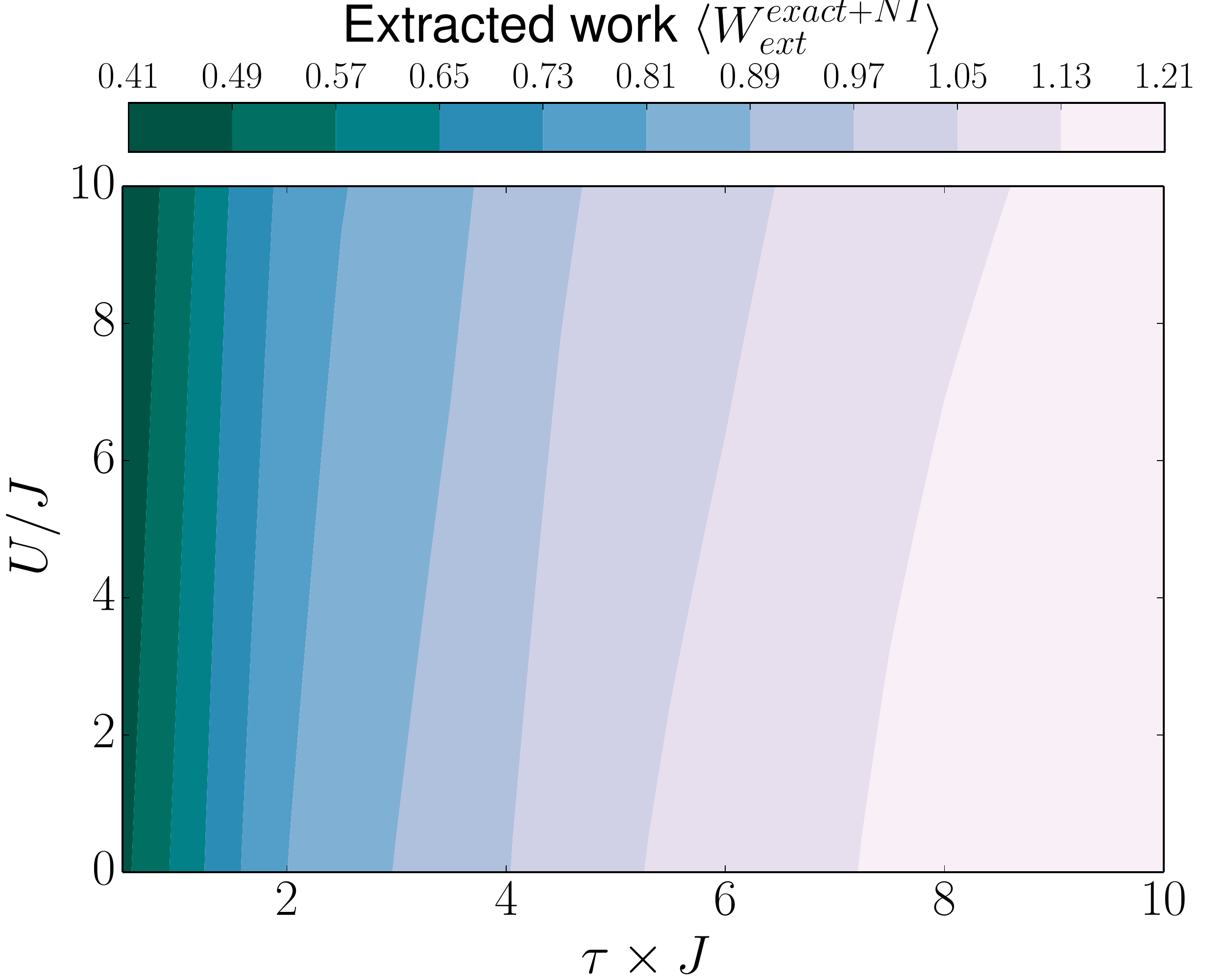}}
\label{fig:ex+NI_work_slope_extracted}
\centering
\subfloat[$T=0.2J/k_B$.]{\includegraphics[width=0.3\textwidth]{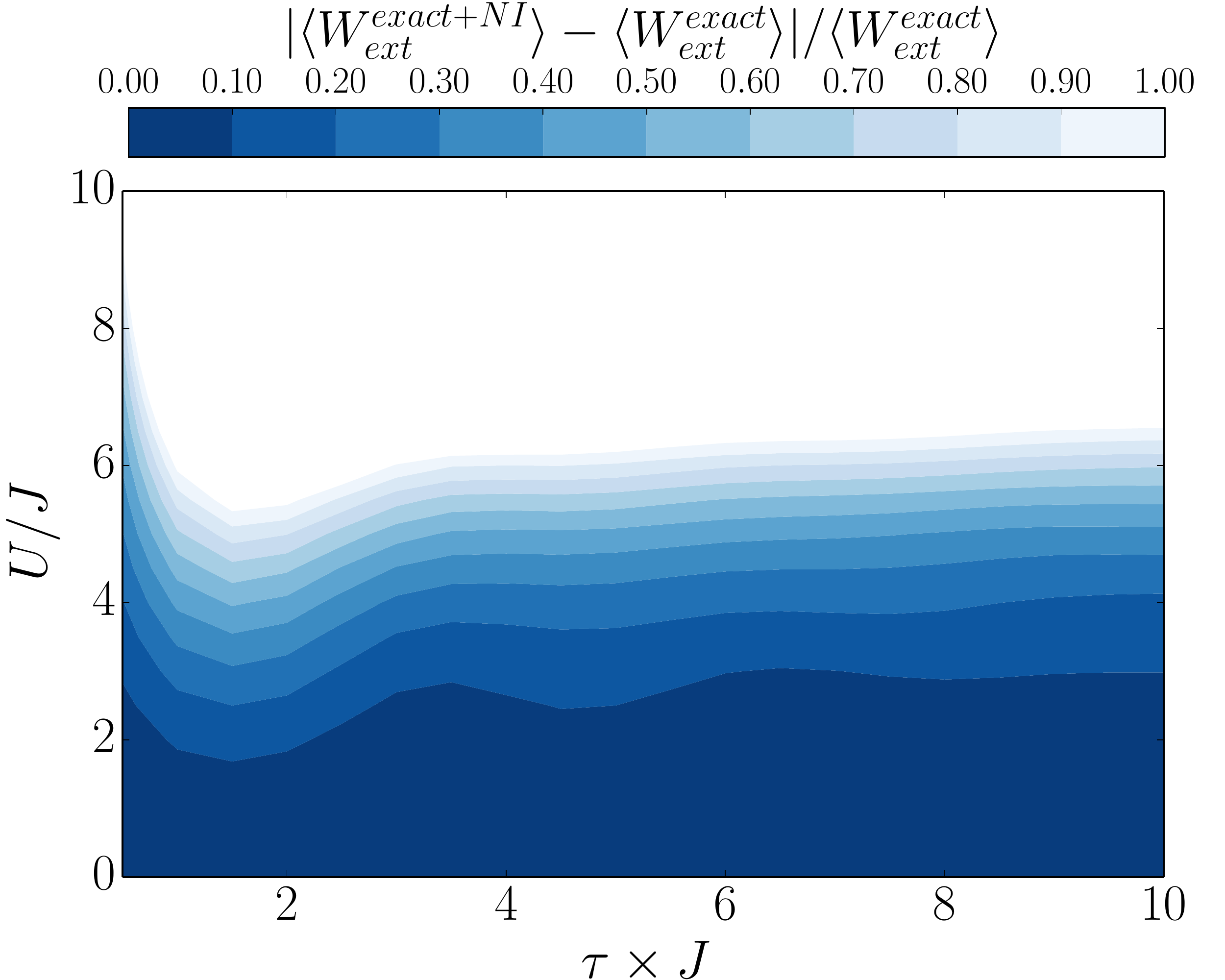}}
\subfloat[$T=2.5J/k_B$.]{\includegraphics[width=0.3\textwidth]{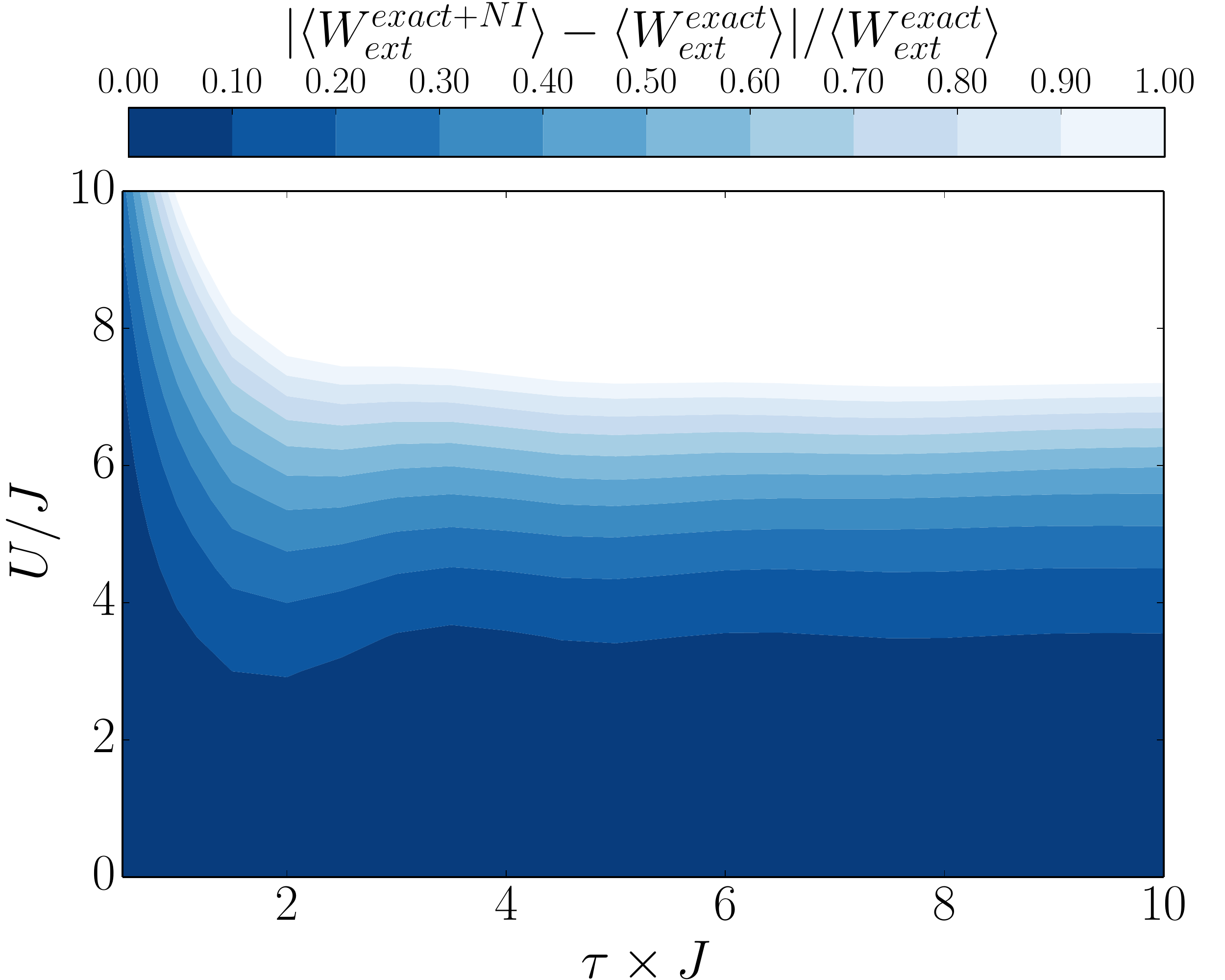}}
\subfloat[$T=20J/k_B$.]{\includegraphics[width=0.3\textwidth]{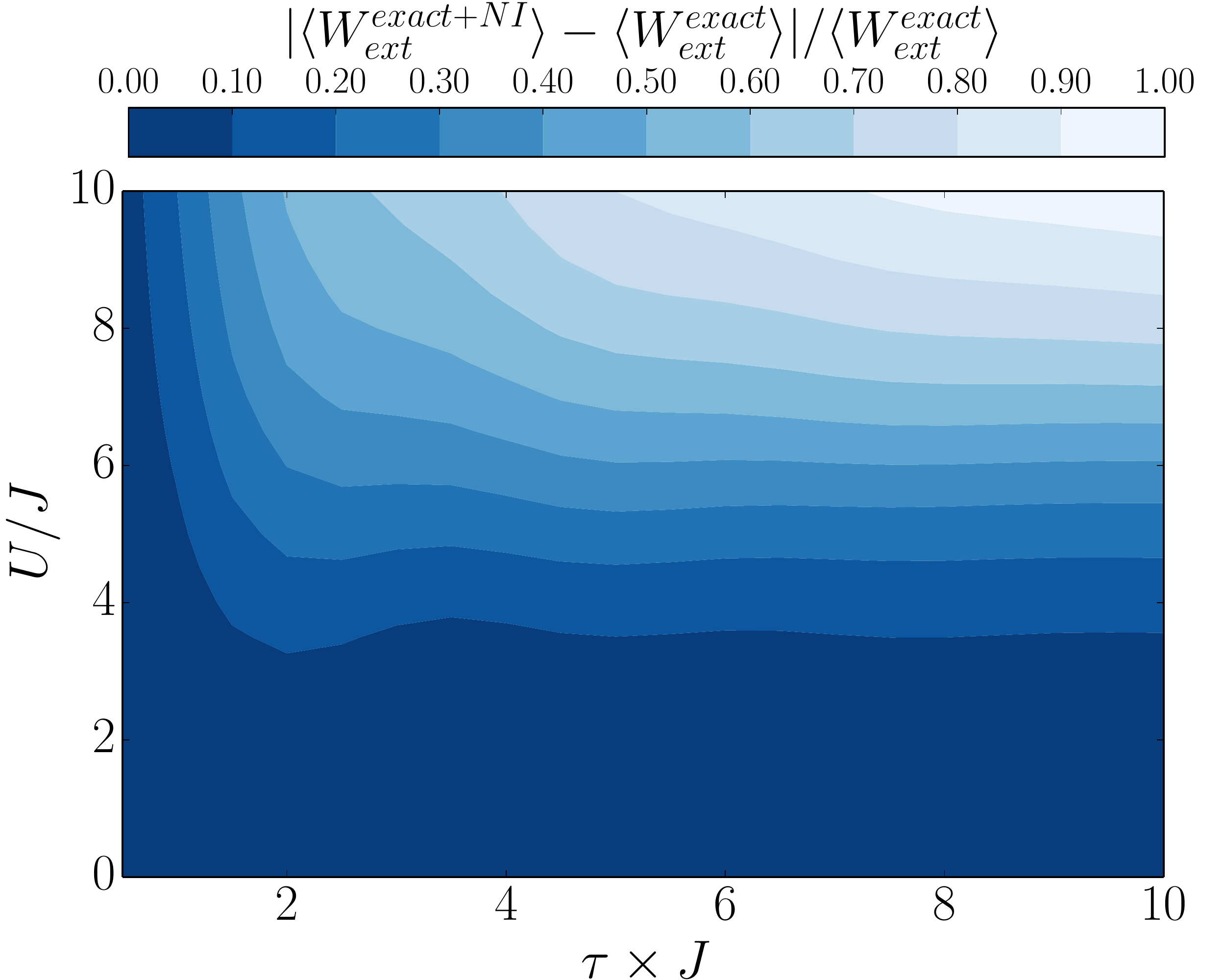}}
\caption{Panels (a) to (c): Extracted work in the `exact + NI' approximation, for 6-site chains undergoing  dynamics driven by the AEF potential. The lighter the colour shade, the more work is extracted, compatible with the respective work ranges indicated above each panel.  The regimes go from non-interacting to strongly coupled along the $y$-axis, and from sudden quench closer to adiabatic along the $x$-axis. Temperature increases from left to right panel, as indicated. \\ Panels (d) to (f): Relative error for $\langle W^{exact+NI} \rangle $ with respect to the exact results for the same parameters as the upper panels.  The darker the colour, the more accurate the approximation is in that regime.}
\label{fig:ex+NI_work_slope}
\end{figure*}

Results for $\langle W^{exact+NI} \rangle $ and the AEF driving potential are shown in figure~\ref{fig:ex+NI_work_slope}.  Once again the approximation accuracy improves with increasing temperature (panels (d) to (f)), while the {\it qualitative} trend is well reproduced for low and intermediate temperatures, but missed at high temperatures for intermediate-to-large $\tau$ and $U$ values  (compare panels (a) to (c) of figure~\ref{fig:ex+NI_work_slope} to panels (c), (f), and (i) of figure~\ref{fig:ex_work}, respectively).
For AEF -- and at difference with `comb' and MI driving potentials -- the Coulomb repulsion $U$ for $U\gtrsim4$ remains a dominant energy scale at any temperature: in fact even at $t=\tau$ the driving potential difference between nearby sites remains at most of $4J$ (see figure \ref{fig:potentials}). This results in a {\it quantitatively} poor approximation for $U\gtrsim4$ even at high temperature.

\section{Six-site `exact+NI' approximation results for MI and AEF driving potentials: entropy variation}

\begin{figure*}
\centering
\subfloat[$T = 0.2J/k_B$]{\includegraphics[width=0.3\textwidth]{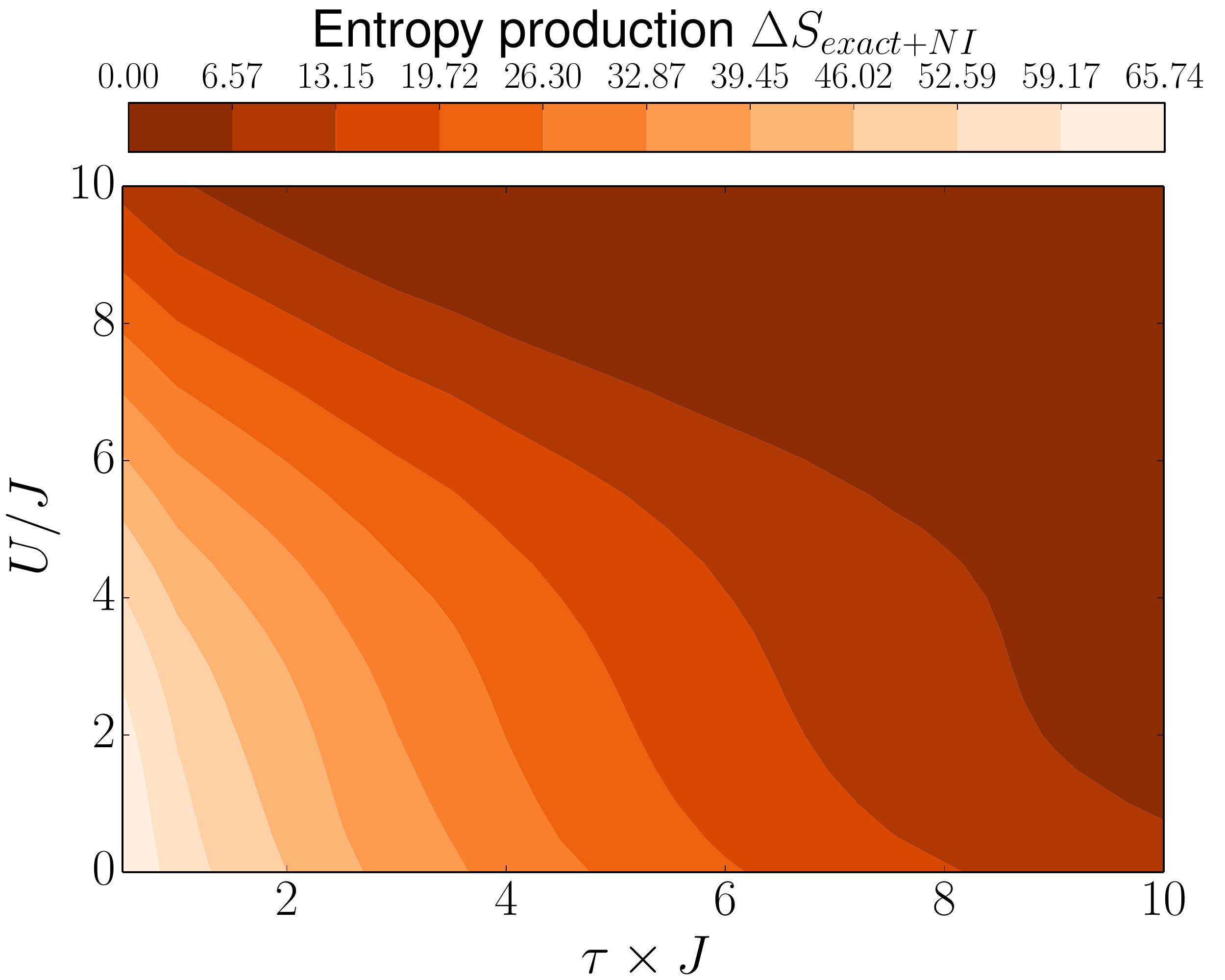}}
\subfloat[$T = 2.5J/k_B$]{\includegraphics[width=0.3\textwidth]{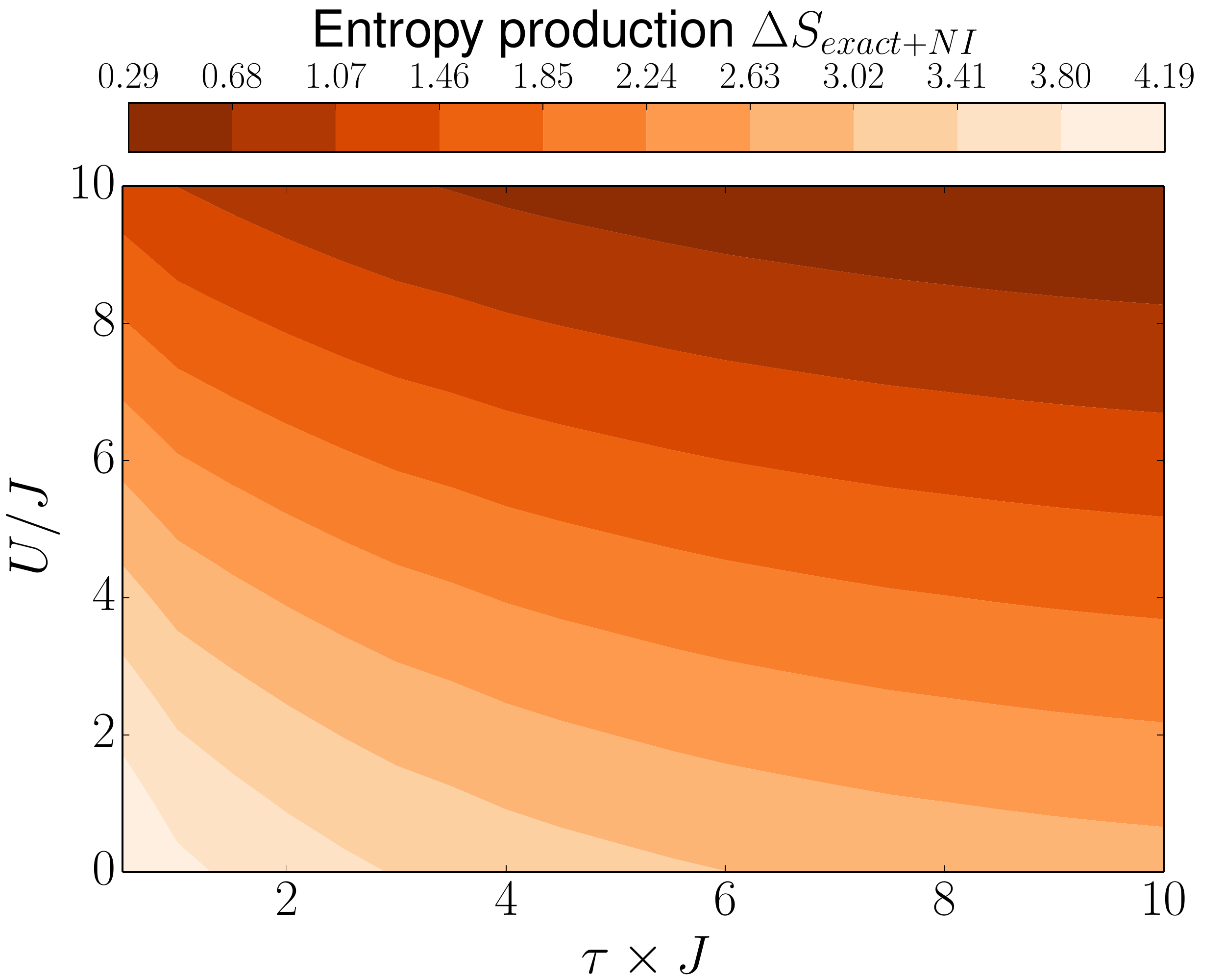}}
\subfloat[$T = 20J/k_B$]{\includegraphics[width=0.3\textwidth]{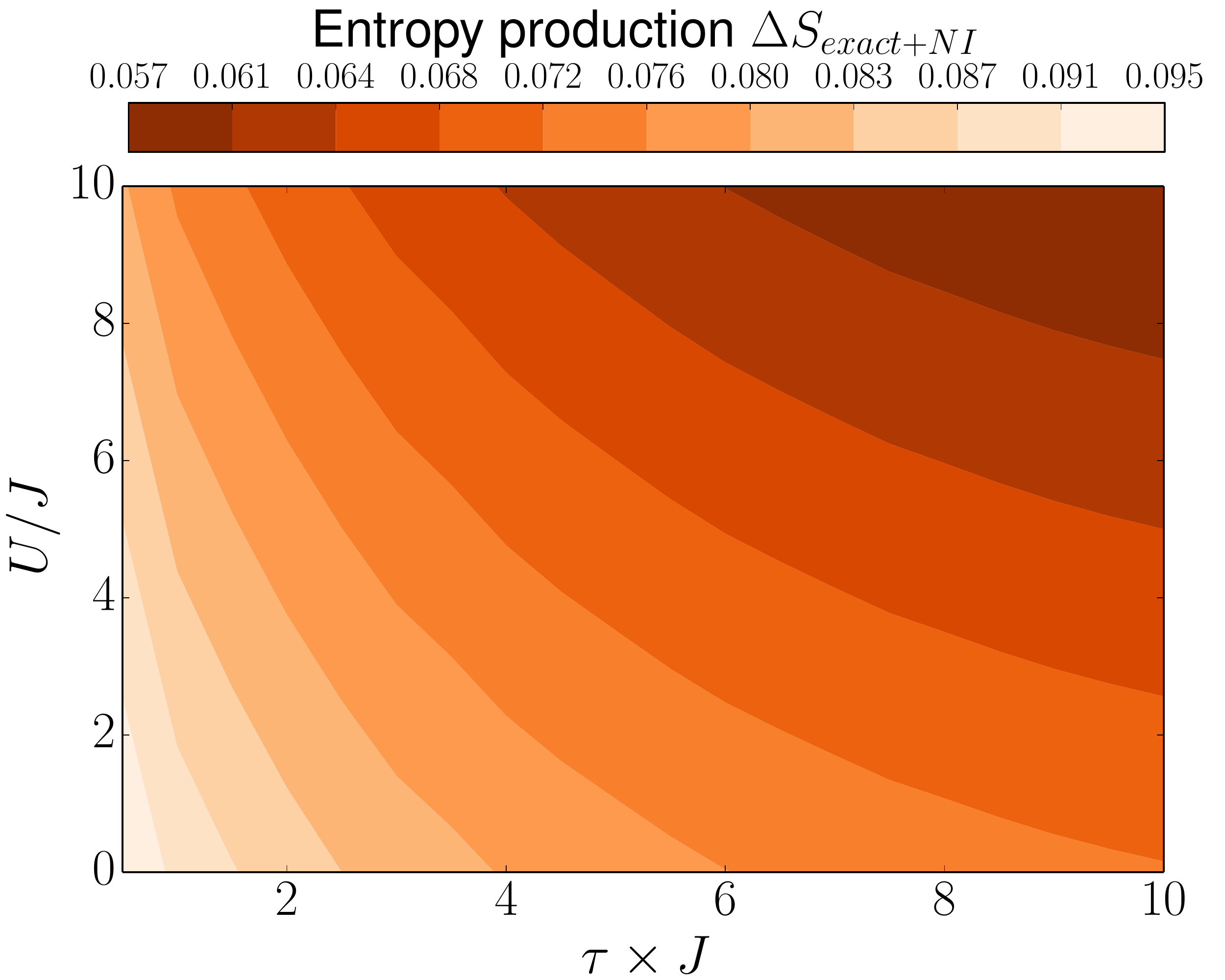}}
\label{fig:ex+NI_entropy_production_MI}
\centering
\subfloat[$T = 0.2J/k_B$]{\includegraphics[width=0.3\textwidth]{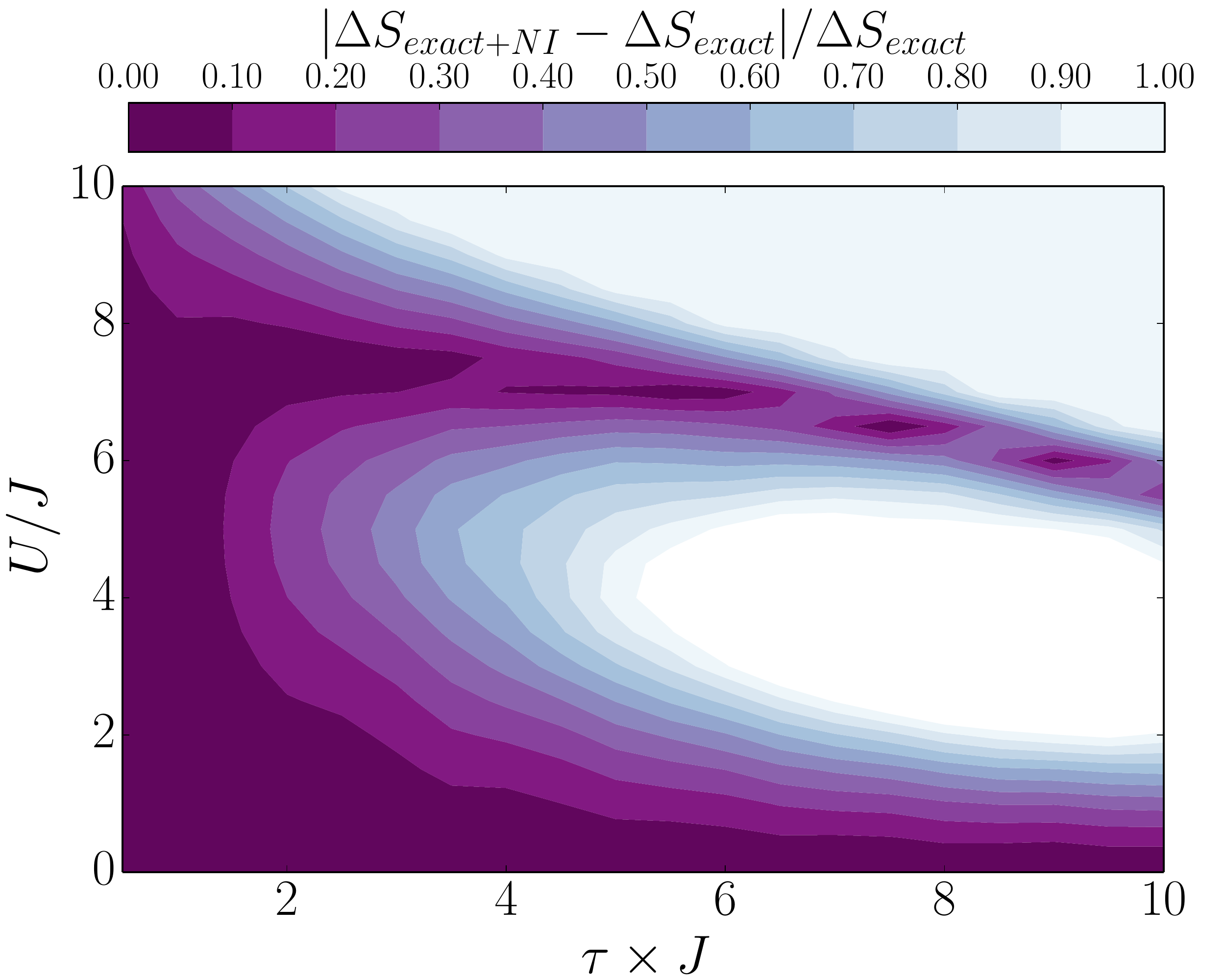}}
\subfloat[$T = 2.5J/k_B$]{\includegraphics[width=0.3\textwidth]{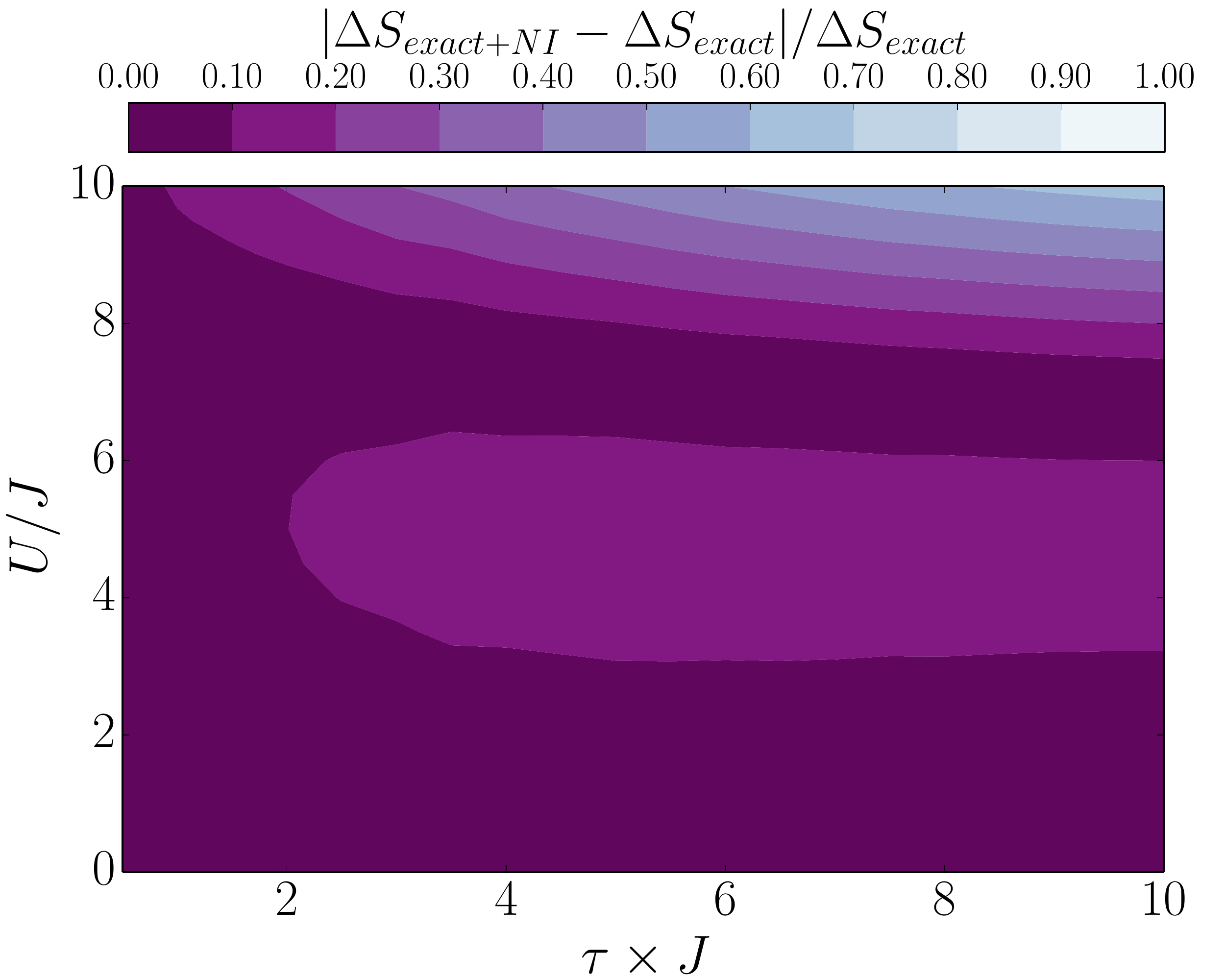}}
\subfloat[$T = 20J/k_B$]{\includegraphics[width=0.3\textwidth]{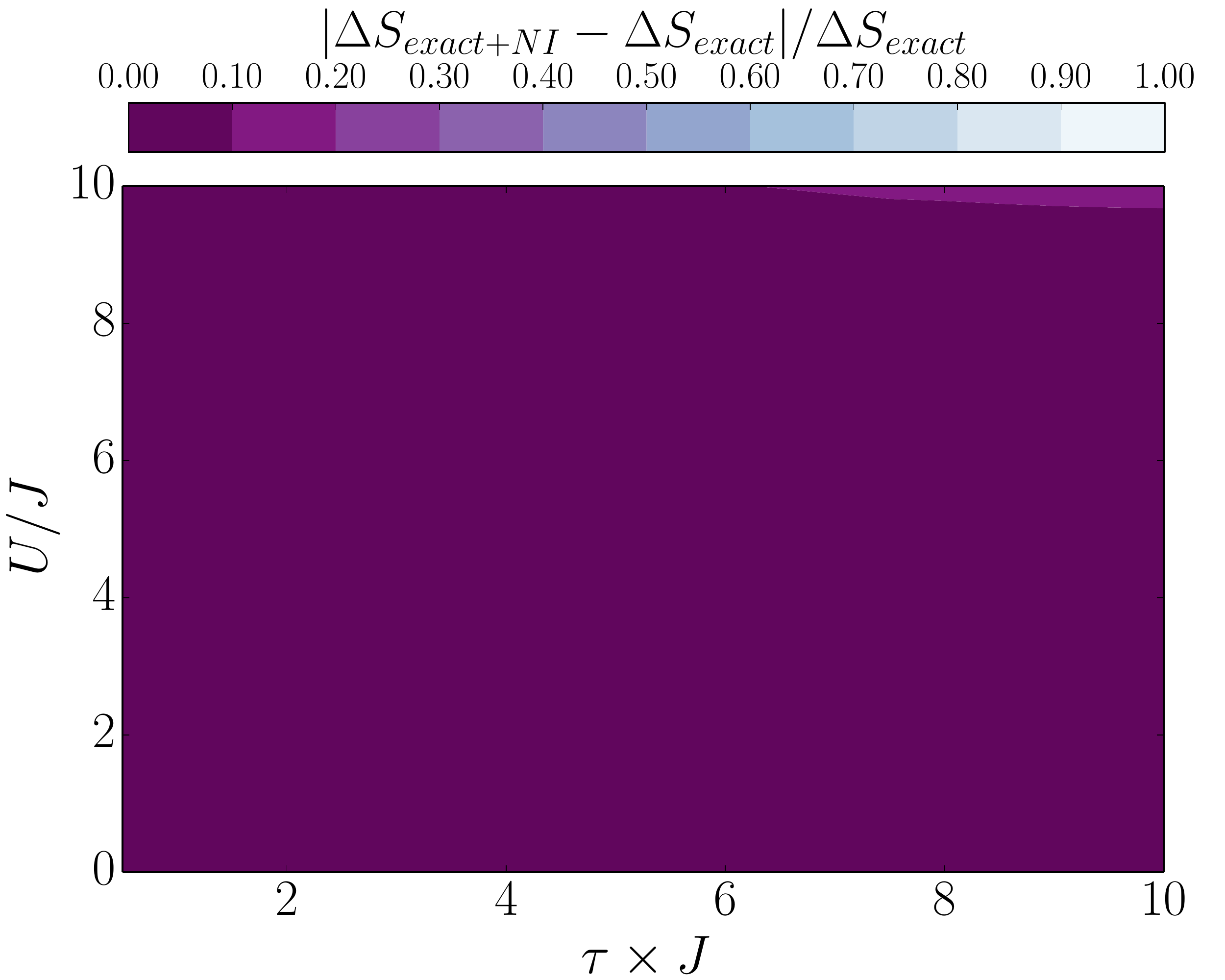}}
\caption{Upper panels: Estimate of the entropy production in the exact + NI approximation versus $\tau$ (x-axis) and $U$ (y-axis), for 6 site chains and MI driving potential. Temperatures as indicated. Lower panels: relative difference between exact and exact + NI entropy production for MI driving potential. Parameters as for the upper panels.}
\label{fig:ex+NI_entropy_MI}
\end{figure*}

\begin{figure*}
\centering
\subfloat[$T = 0.2J/k_B$]{\includegraphics[width=0.3\textwidth]{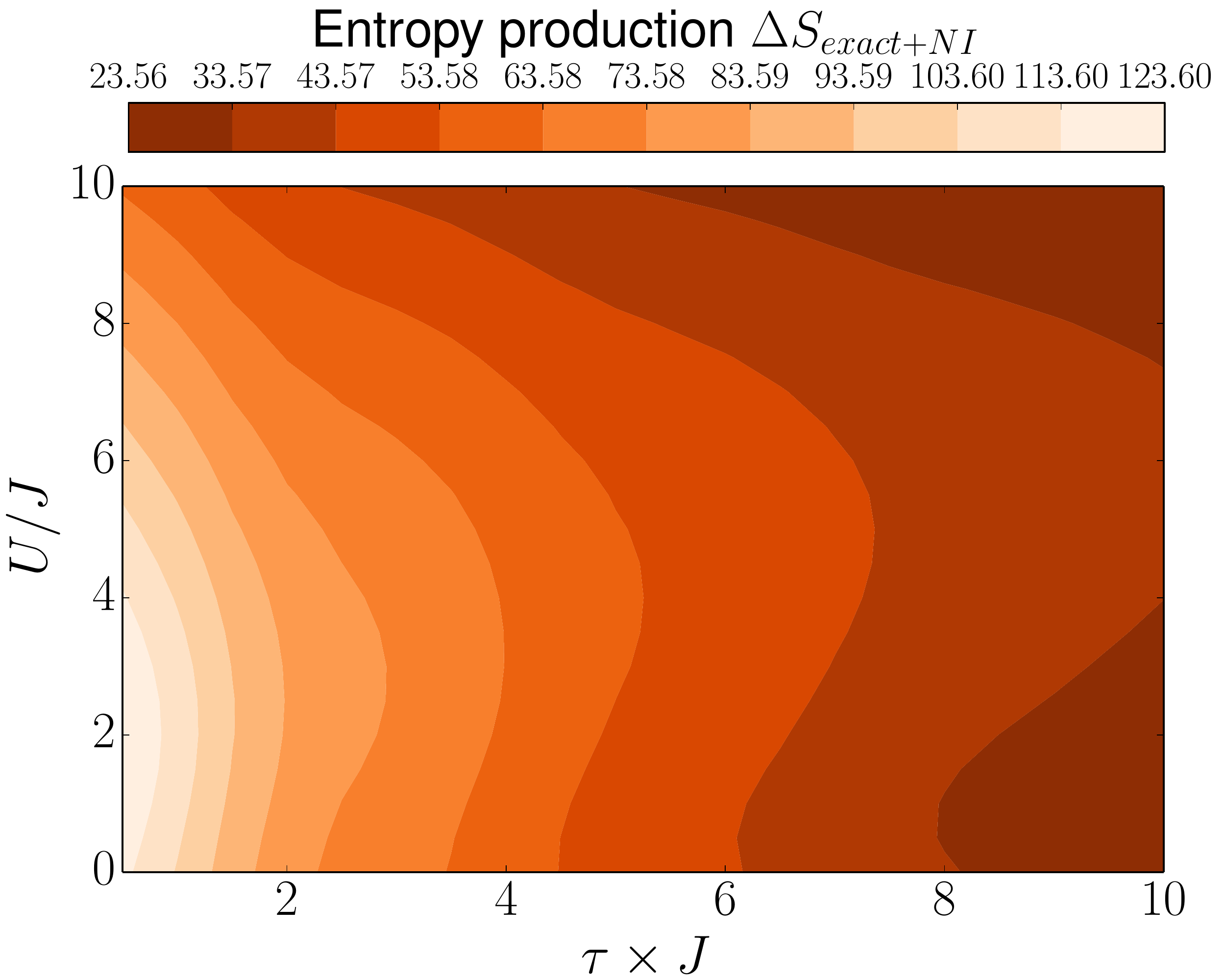}}
\subfloat[$T = 2.5J/k_B$]{\includegraphics[width=0.3\textwidth]{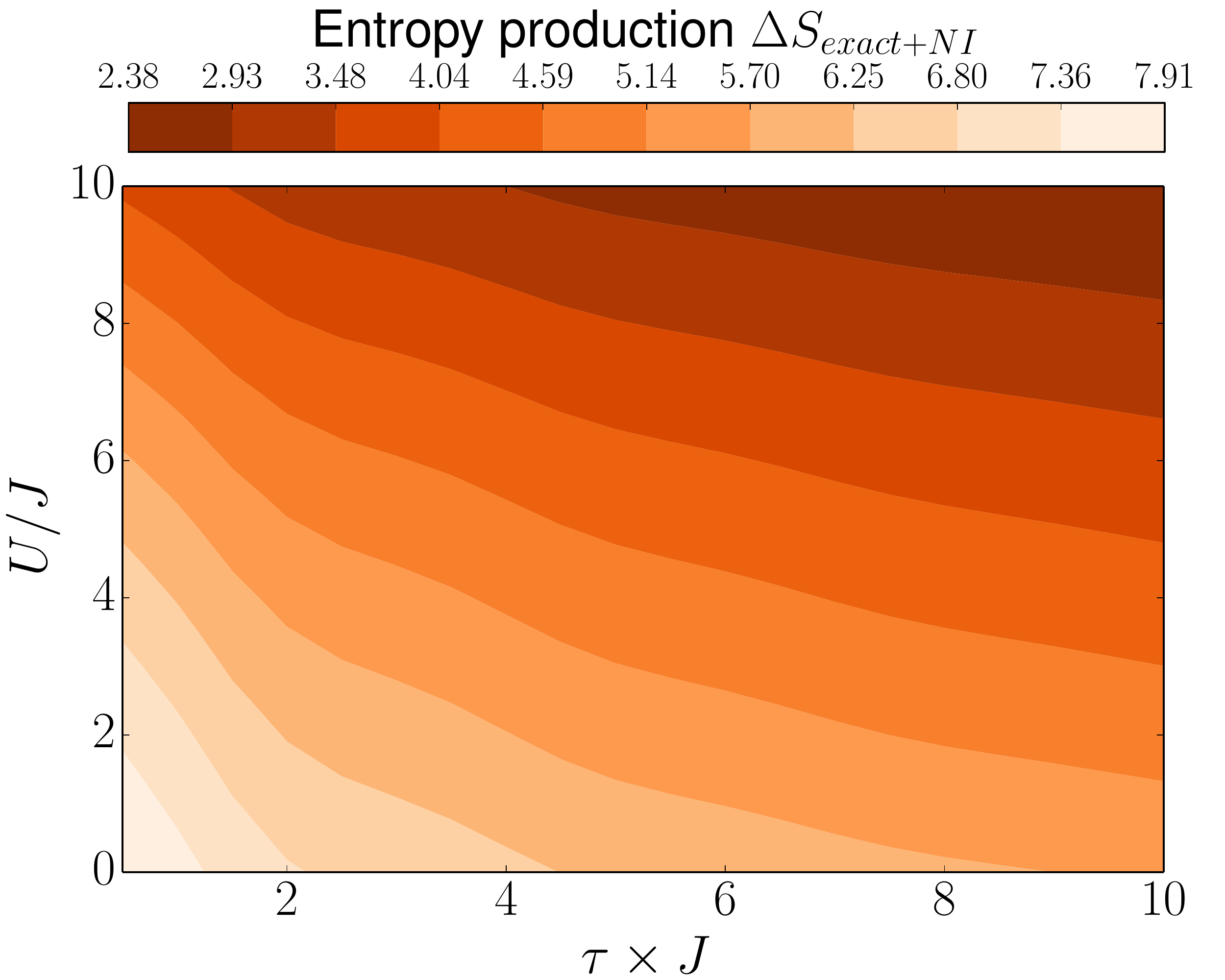}}
\subfloat[$T = 20J/k_B$]{\includegraphics[width=0.3\textwidth]{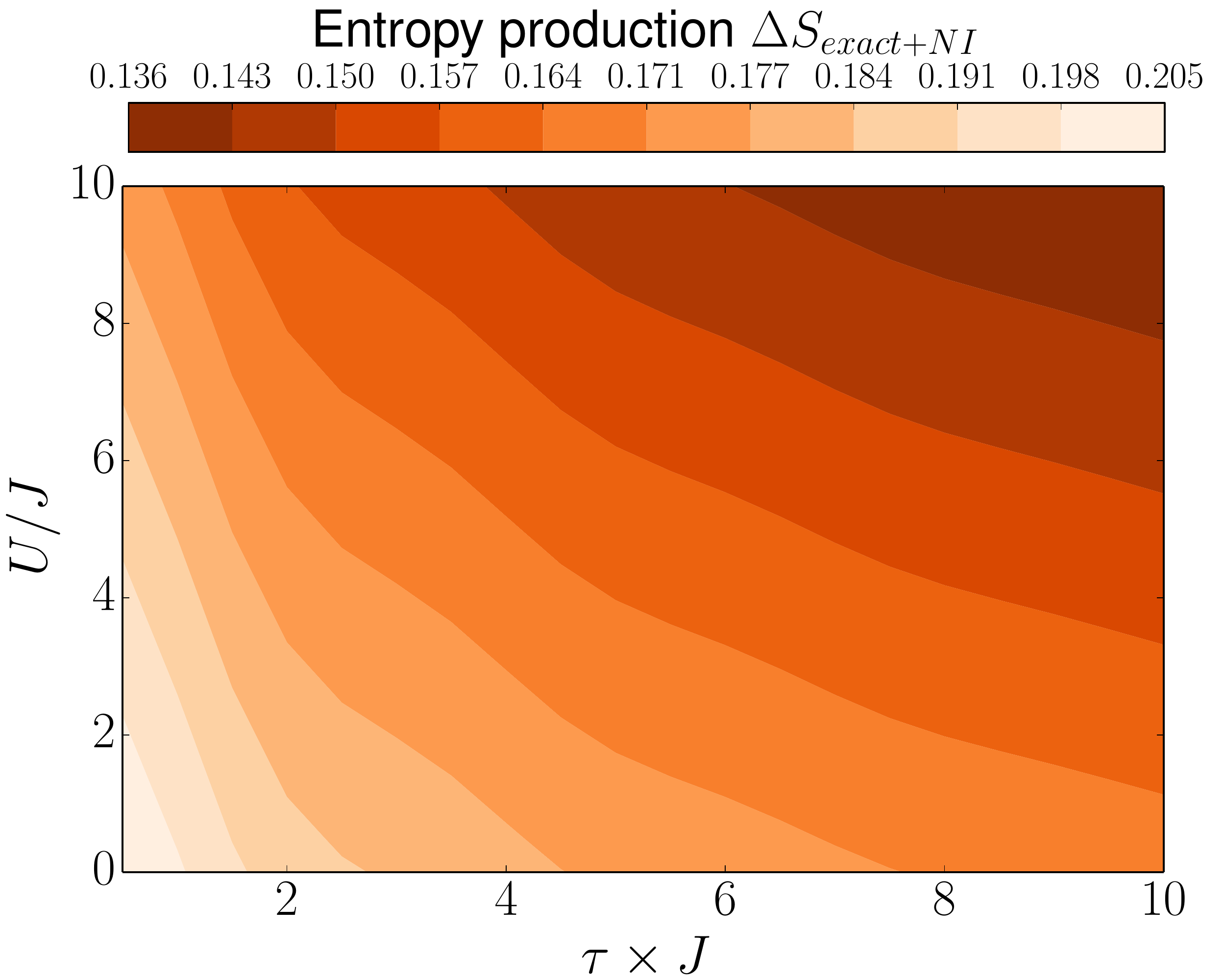}}
\label{fig:ex+NI_entropy_production_AEF}
\centering
\subfloat[$T = 0.2J/k_B$]{\includegraphics[width=0.3\textwidth]{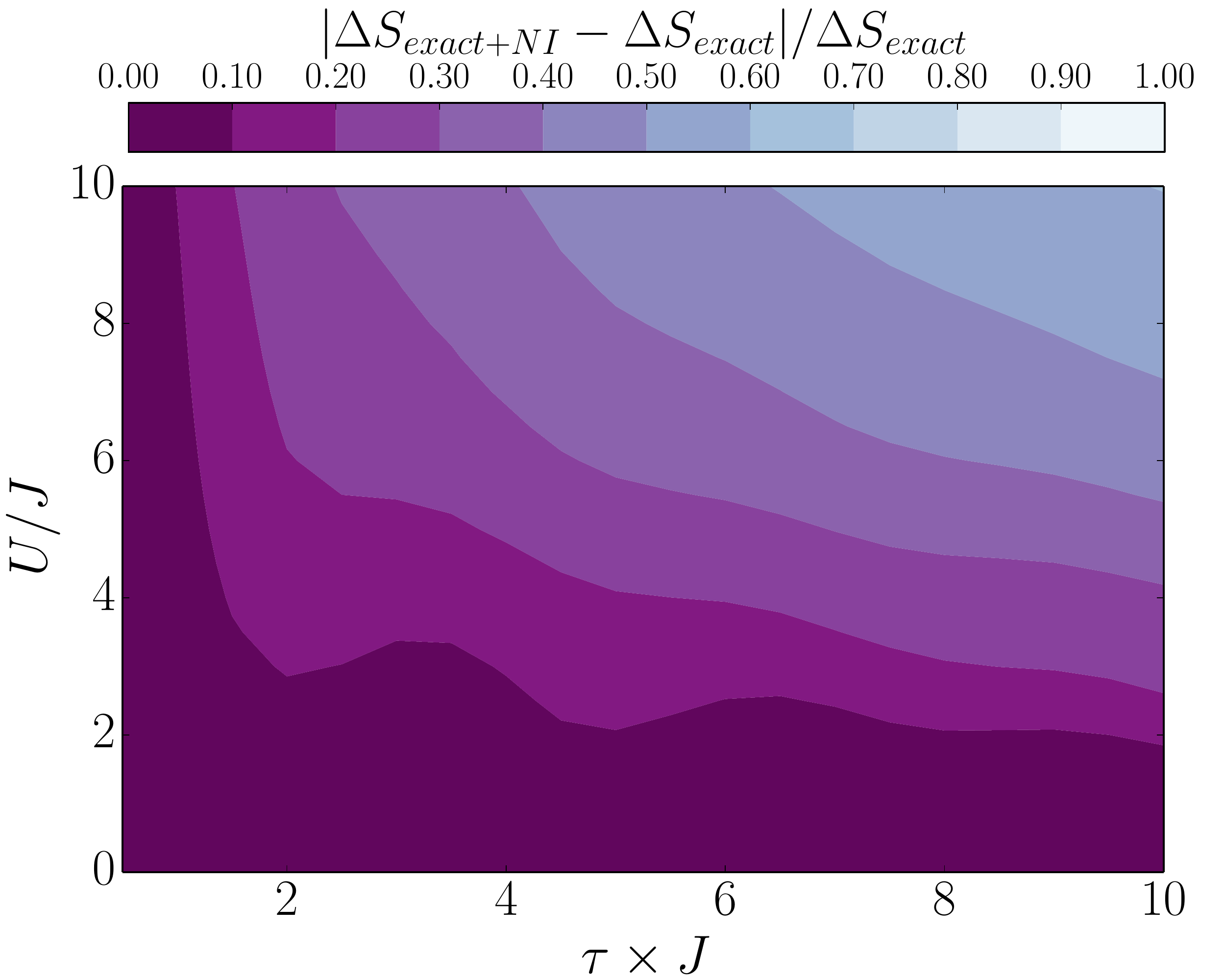}}
\subfloat[$T = 2.5J/k_B$]{\includegraphics[width=0.3\textwidth]{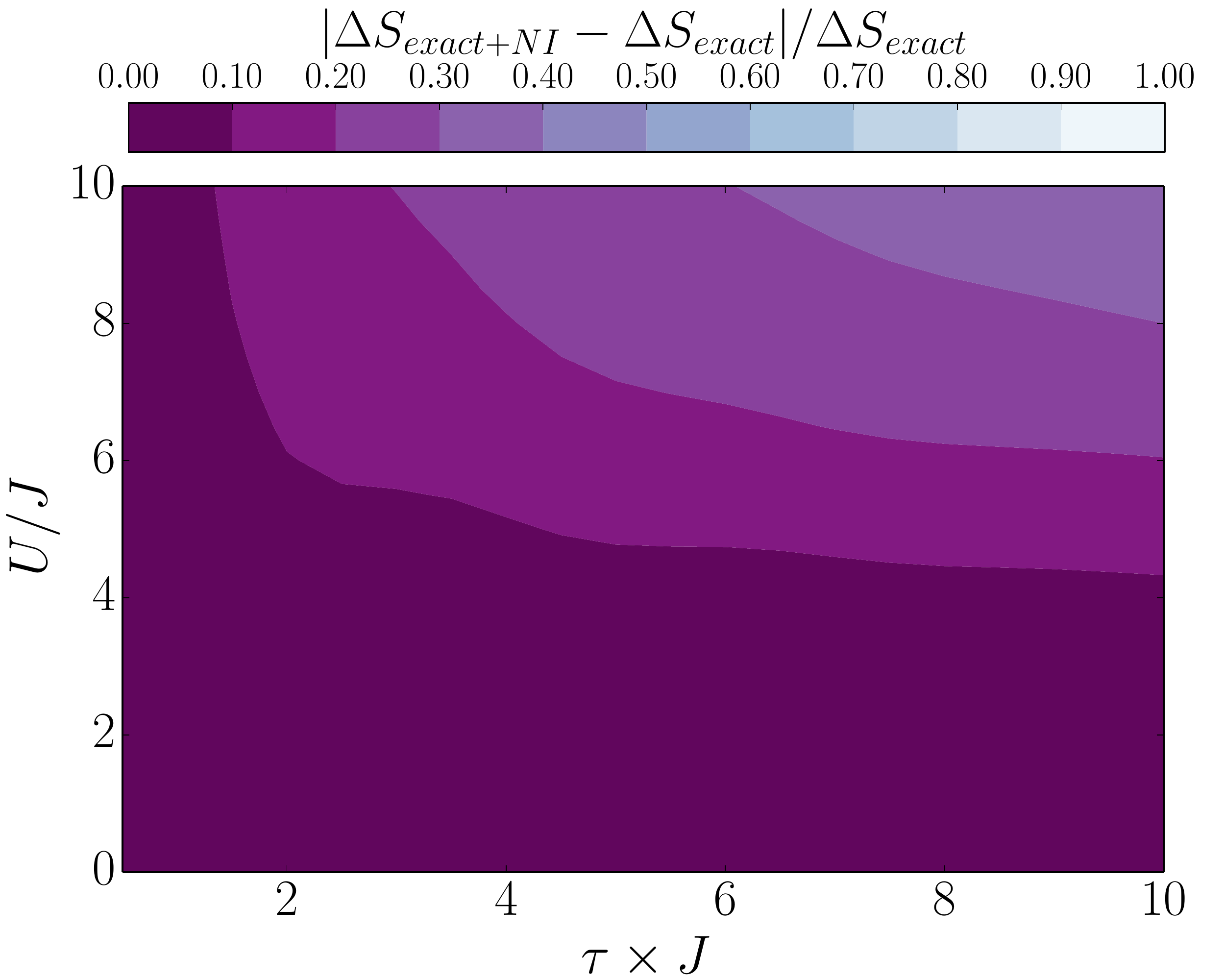}}
\subfloat[$T = 20J/k_B$]{\includegraphics[width=0.3\textwidth]{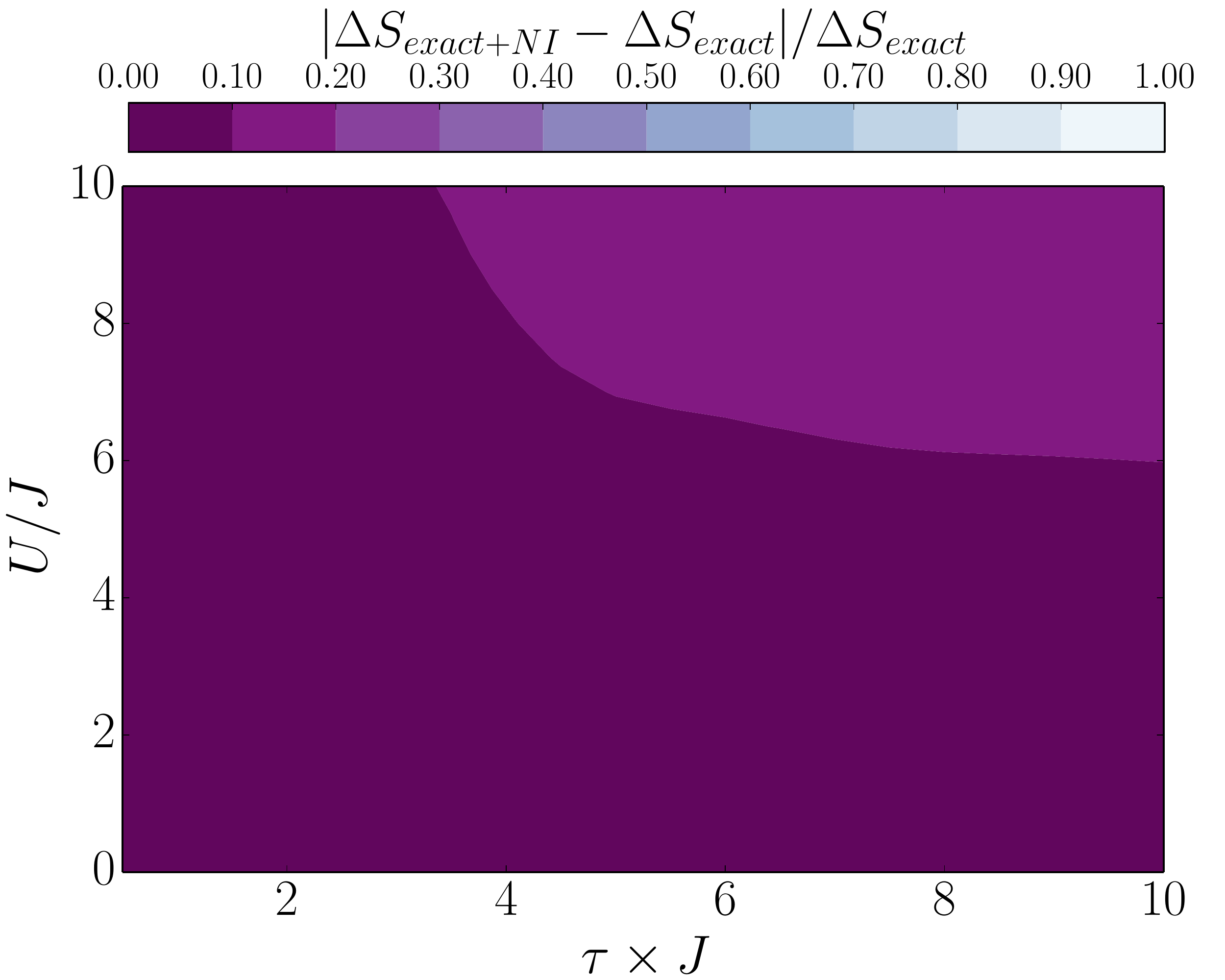}}
\caption{Upper panels: Estimate of the entropy production in the exact + NI approximation versus $\tau$ (x-axis) and $U$ (y-axis), for 6 site chains and AEF driving potential. Temperatures as indicated. Lower panels: relative difference between exact and exact + NI entropy production for AEF driving potential. Parameters as for the upper panels.}
\label{fig:ex+NI_entropy_AEF}
\end{figure*}

Estimates for the entropy variation in the `exact+NI' approximation for MI and AEF driving potentials and six-site chains are shown in figures~\ref{fig:ex+NI_entropy_MI} and \ref{fig:ex+NI_entropy_AEF}, respectively.

Similarly to the `comb' potential, the `exact+NI' approximation with the MI and AEF driving potentials recover to a good extent the qualitatively the behaviour of $\Delta S^{exact}$ for low and intermediate temperatures. For high temperatures the qualitative behaviour is recovered only for $U\lesssim2$.

Quantitatively, the areas of worse performance are related to the areas of worse performance for the corresponding $\langle W_{ext}^{exact+NI}\rangle$, however the approximation performs worse for the entropy than for the extracted work for MI, and better for AEF. Overall the approximation improves its quantitative performance with temperature, as it reproduces well the limits of the entropy variation range, and especially so at high temperature.

%
%

%

\bibliography{ref}
\end{document}